\documentclass[12pt]{article}
\usepackage{a4wide,epsfig,psfrag,amsmath,amssymb,cite,scalefnt}
\usepackage{color}
\usepackage{slashed}

\usepackage{fancyvrb}

\graphicspath{ {figs/} }

\newcommand{\ep}{\epsilon}
\newcommand{\nc}{N_C}
\newcommand{\nl}{n_l}

\newcommand{\ii}{\mathrm{i}}

\newcommand{\nf}{n_f}

\newcommand{\alphas}{\alpha_\mathrm{s}}

\allowdisplaybreaks

\begin{document}

\title{\vskip-3cm{\baselineskip14pt
    \begin{flushleft}
     \normalsize CERN-TH-2023-012, P3H-23-006, TTP23-002, ZU-TH 05/23
    \end{flushleft}} \vskip1.5cm
  Massive three-loop form factors: anomaly contribution
  }

\author{
  Matteo Fael$^{a}$,
  Fabian Lange$^{b,c}$,
  Kay Sch\"onwald$^{d}$,
  Matthias Steinhauser$^{b}$
  \\
  {\small\it (a) Theoretical Physics Department, CERN,}\\
  {\small\it 1211 Geneva, Switzerland}
  \\
  {\small\it (b) Institut f{\"u}r Theoretische Teilchenphysik,
    Karlsruhe Institute of Technology (KIT),}\\
  {\small\it 76128 Karlsruhe, Germany}
  \\
  {\small\it (c) Institut f{\"u}r Astroteilchenphysik,
  Karlsruhe Institute of Technology (KIT),}\\
  {\small\it 76344 Eggenstein-Leopoldshafen, Germany}
  \\
  {\small\it (d) Physik-Institut, Universit\"at Z\"urich, Winterthurerstrasse 190,}\\
  {\small\it 8057 Z\"urich, Switzerland}
}

\date{}

\maketitle

\thispagestyle{empty}

\begin{abstract}

\noindent
We compute three-loop corrections to the singlet form factors for
massive quarks using a semi-analytic method which provides precise results
over the whole kinematic range. Particular emphasis is put on the anomaly
contribution originating from an external axial-vector current. We also
discuss in detail the contribution for a pseudoscalar current and
verify the chiral Ward identity to three-loop order.
Explicit results are presented for the low- and high-energy regions and the
expansions around threshold.

\end{abstract}

\thispagestyle{empty}


\newpage


\section{Introduction}
\label{sec::introduction}

Form factors are important building blocks in any quantum field theory.  In
QED and QCD they constitute the virtual corrections for many important
processes both at lepton and hadron colliders such as Higgs boson production
and decay, lepton pair production via the Drell-Yan process, and electron-muon
scattering at low energies.

In this paper we consider QCD corrections to heavy-quark form factors of an
external current. At one- and two-loop order such calculations have been
performed already some time
ago~\cite{Mastrolia:2003yz,Bonciani:2003ai,Bernreuther:2004ih,Bernreuther:2004th,Bernreuther:2005rw,Bernreuther:2005gw,Gluza:2009yy,Henn:2016tyf,Ahmed:2017gyt,Ablinger:2017hst,Lee:2018nxa}.
Recently we computed the three-loop
corrections for the so-called non-singlet contributions, where the external
current couples to the same fermion line as the external quarks~\cite{Fael:2022rgm,Fael:2022miw} (see also
Refs.~\cite{Henn:2016tyf,Lee:2018rgs,Ablinger:2018yae,Ablinger:2018zwz,Lee:2018nxa,Blumlein:2019oas}
for partial results of simpler subsets).\footnote{Recently the total cross
  section for heavy-quark production at lepton colliders has been computed at
  next-to-next-to-next-to-leading order~\cite{Chen:2022vzo}. In this
  calculation the vector form factor enters as building block.}
In Ref.~\cite{Fael:2022miw} we also considered those singlet contributions
where the external current couples to
massive internal quarks, but only for vector, scalar, and pseudoscalar currents, i.e.\ omitting the axial-vector current.
In this work we close this gap, compute all
contributions for all four currents coupling to massless and massive quarks, and
provide complete results for the singlet contributions. This requires a
detailed discussion of the anomaly contribution for the axial-vector
current following the line of the corresponding two-loop calculation of Ref.~\cite{Bernreuther:2005rw}.

For completeness we want to mention that completely massless form factors are
available up to four-loop order~\cite{Lee:2022nhh} (see
Refs.~\cite{Baikov:2009bg,Lee:2010cga,Gehrmann:2010ue} for the corresponding
three-loop results).  Three-loop corrections to massless form factors where
the external current couples to massive quarks have been considered in
Ref.~\cite{Chen:2021rft}. This reference also contains a detailed
discussion of the renormalization of the axial-vector current
contribution. However, at three-loop order there are further subtleties
for massive final-state quarks.

\begin{figure}[t]
    \begin{center}
      \includegraphics[width=0.15\textwidth]{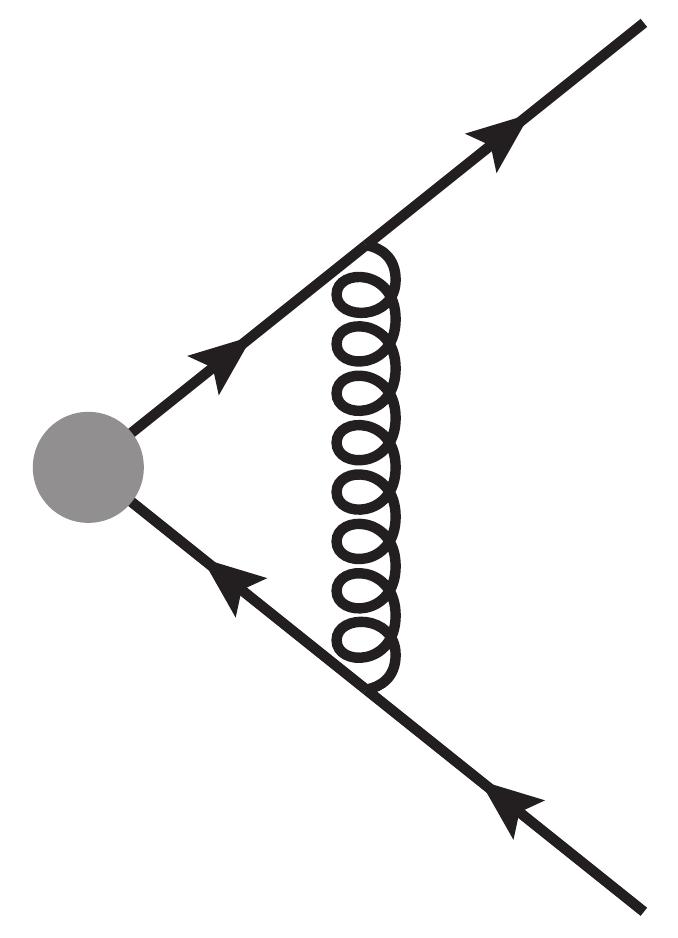}
      \hspace{0.9cm}
      \includegraphics[width=0.15\textwidth]{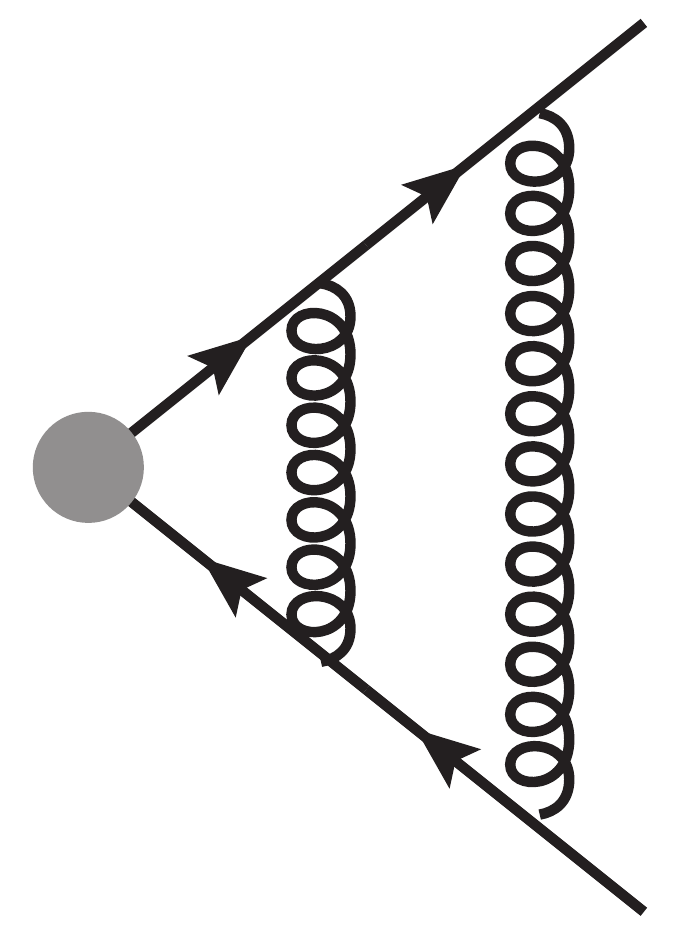}
      \hspace{0.9cm}
      \includegraphics[width=0.15\textwidth]{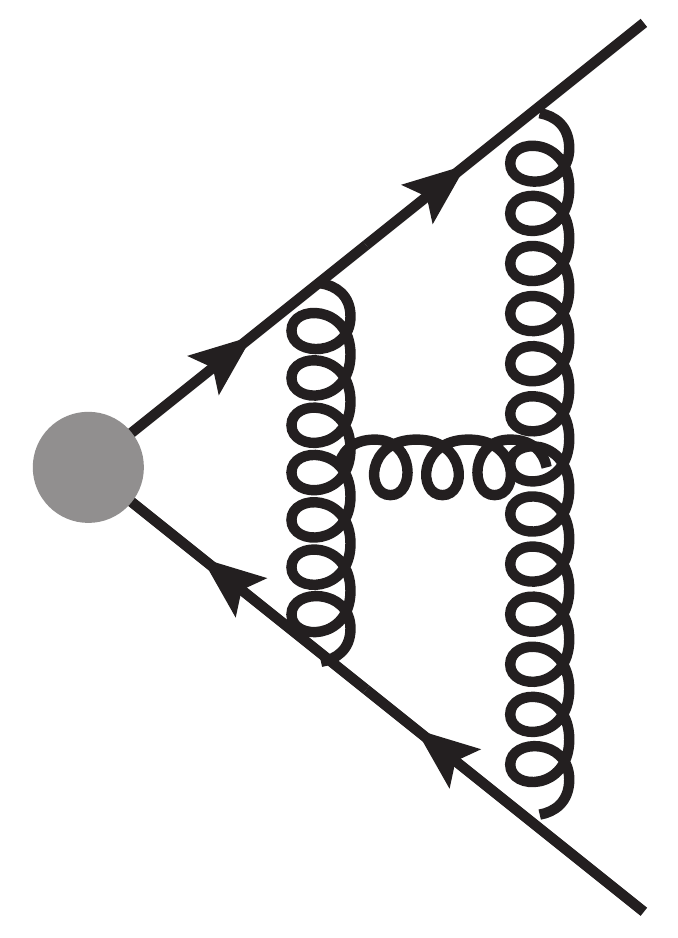}
      \\ \vspace{0.2cm}
      \includegraphics[width=0.2\textwidth]{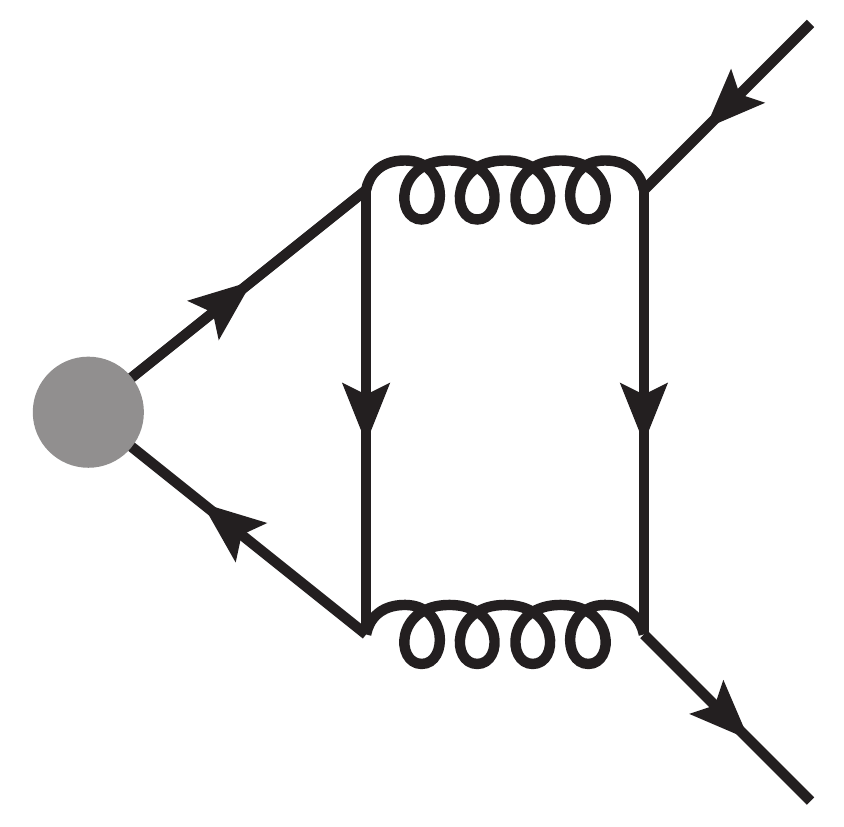}
      \hspace{0.5cm}
      \includegraphics[width=0.2\textwidth]{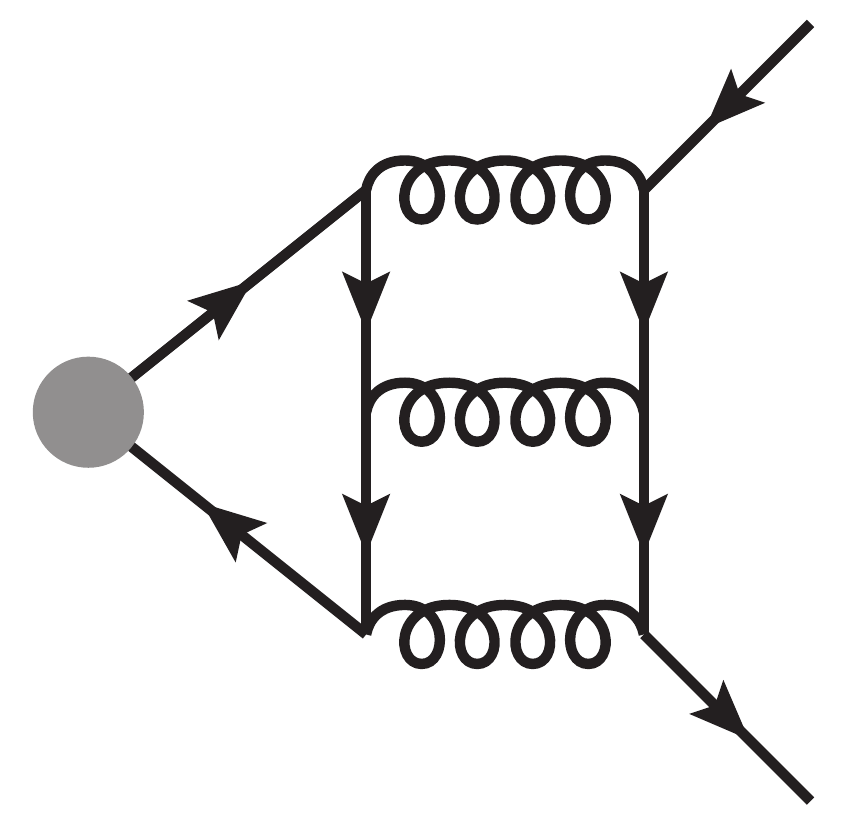}
      \hspace{0.5cm}
      \includegraphics[width=0.2\textwidth]{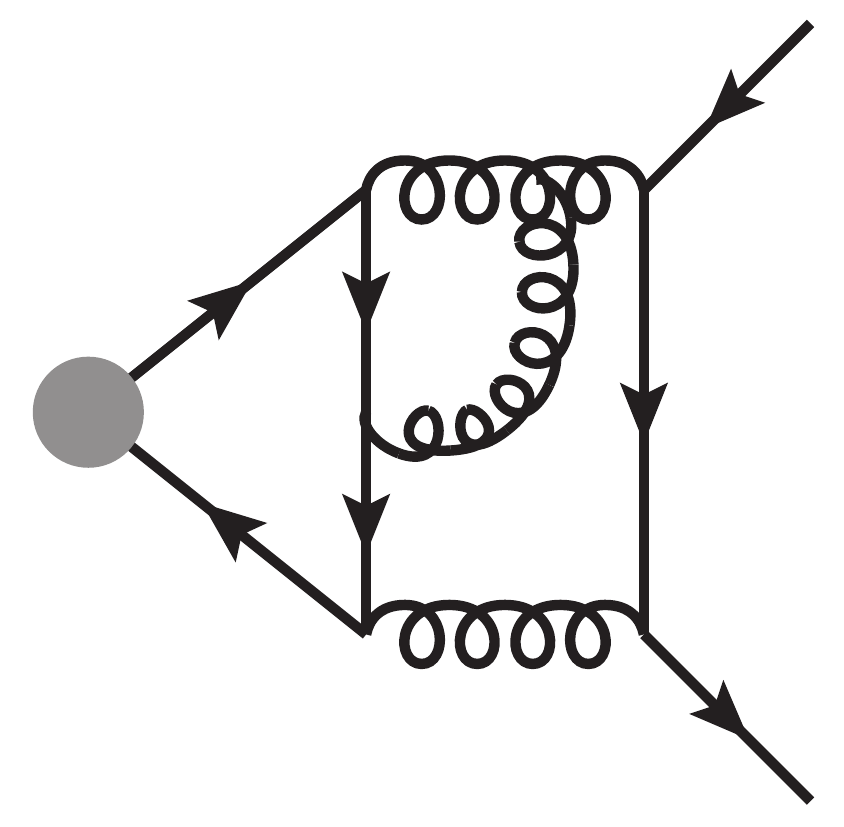}
    \caption{\label{fig::diags}Sample Feynman diagrams contributing to the
      heavy-quark form factors. The top row displays non-singlet and
      the bottom row singlet diagrams up to three loops in QCD.
      The gray blob represents the coupling to the external current.}
    \end{center}
\end{figure}

In the literature (see, e.g.,
Refs.~\cite{Larin:1991tj,Larin:1993tq,Bernreuther:2005rw}) one often finds the
terms ``flavour-singlet'' and ``flavour-non-singlet'' which refer to certain
combinations of (axial-vector) currents (and not to Feynman diagrams). The
former is simply the sum of the axial-vector currents of the quarks involved
in the theory. On the other hand, ``flavour-non-singlet'' refers to the
difference of the axial-vector currents of the two quarks of a generation.
The flavour-non-singlet current, which corresponds to the Z boson
coupling in the SM, is conserved while the flavour-singlet current is anomalous.

In this paper we define ``singlet'' and ``non-singlet'' at the level of
Feynman diagrams and use the notion:
\begin{itemize}
\item Singlet: The external current does not couple to the fermion line of the
  final-state quarks.
\item Non-singlet: The external current couples directly to the fermion line of
  the final-state quarks.
\end{itemize}
This is illustrated by the Feynman diagrams shown in Fig.~\ref{fig::diags},
where the first and second row contain non-singlet and singlet contributions,
respectively.

For the non-singlet contributions it is possible to use anti-commuting
$\gamma_5$. On the other hand, the singlet contributions require
a definition of $\gamma_5$ in which traces of the form
$\mbox{Tr}(\gamma_5\gamma^\mu\gamma^\nu\gamma^\rho\gamma^\sigma)$
do not vanish. In this work we adopt the approach from Ref.~\cite{Larin:1993tq}, which is often called Larin scheme.
Of course this prescription can also
be applied to the non-singlet contributions. As a cross check we repeat the
calculation of Ref.~\cite{Fael:2022miw} and
show that the final results for the finite form factors
are identical in both prescriptions.

In the following we refer to massless and massive singlet
contributions depending on whether the external current
couples to massless or massive quarks, respectively.
Results for the massive singlet form factors with external vector and scalar currents
have already been presented in Ref.~\cite{Fael:2022miw}.
In this work we complete the massive and provide the massless singlet contributions.

For vector and scalar currents $\gamma_5$ is absent and thus these contributions can be treated in analogy to the non-singlet contributions.
Since the vector current contribution vanishes at two-loop order due to Furry's theorem, it is finite at three-loop level.
The scalar and pseudoscalar form factors only receive massive singlet
contributions because the triangles vanish due to the Dirac algebra in the massless case.  The results can be found in Ref.~\cite{Fael:2022miw}.
Note, however, that in Ref.~\cite{Fael:2022miw} the finite renormalization constant
for the pseudoscalar current has not been taken into account.
In this work we correct this deficit.
We consider
the combination of the singlet and non-singlet contributions
and introduce in both parts a non-anti-commuting $\gamma_5$.
As already mentioned above, the main focus of the present work is
on the axial-vector contribution.

The outline of this paper is as follows: In the next Section we introduce our
notation and briefly mention our treatment of $\gamma_5$. Afterwards we
discuss the renormalization of the singlet form factors with special emphasis
on the axial-vector current and the subtraction of infrared divergences.
In Section~\ref{sec::chiral} we discuss the chiral Ward identity and we
dedicate Section~\ref{sec::comp} to the computation of the two- and three-loop
vertex integrals and a discussion of the various cross checks which we have
performed.  Section~\ref{sec::results} contains our results. Finally we
conclude in Section~\ref{sec::concl}.  In Appendix~\ref{app::projectors} we
explicitly state the projectors for the form factors and in
Appendix~\ref{app::ren_const} we provide all relevant renormalization
constants related to the treatment of $\gamma_5$. Appendix~\ref{app::resFF}
contains explicit results for the massive singlet form factors and in
Appendix~\ref{app::GGtil} analytic results for the one- and two-loop
expressions of the form factor induced by the pseudoscalar gluonic
operator are presented.  In Appendix~\ref{app:formfactors3l} we provide a
description of the package \texttt{FF3l} where all results for the three-loop
massive form factors are implemented.



\section{Notation}
\label{sec::notation}

We consider the vector ($v$), axial-vector ($a$), scalar ($s$), and pseudoscalar ($p$) currents
\begin{eqnarray}
  j_\mu^v &=& \bar{\psi}\gamma_\mu\psi\,,\nonumber\\
  j_\mu^a &=& \bar{\psi}\gamma_\mu\gamma_5\psi\,,\nonumber\\
  j^s &=& m \,\bar{\psi}\psi\,,\nonumber\\
  j^p &=& \ii m \,\bar{\psi}\gamma_5\psi\,.
  \label{eq::currents}
\end{eqnarray}
The factor $m$ is introduced such that the scalar and pseudoscalar currents
have vanishing anomalous dimensions.

It is convenient to decompose the three-point functions with an external
quark-anti-quark pair into scalar form factors which we denote by
\begin{eqnarray}
  \Gamma_\mu^v(q_1,q_2) &=&
  F_1^v(q^2)\gamma_\mu - \frac{\ii}{2m}F_2^v(q^2) \sigma_{\mu\nu}q^\nu
  \,, \nonumber\\
  \Gamma_\mu^a(q_1,q_2) &=&
  F_1^a(q^2)\gamma_\mu\gamma_5 {- \frac{1}{2m}F_2^a(q^2) q_\mu }\gamma_5
  \,, \nonumber\\
  \Gamma^s(q_1,q_2) &=& {m} F^s(q^2)
  \,, \nonumber\\
  \Gamma^p(q_1,q_2) &=& {\ii m} F^p(q^2) {\gamma_5}
  \,,
  \label{eq::Gamma}
\end{eqnarray}
where the momenta $q_1$ and $q_2$ are the momenta of the incoming quark and
the outgoing anti-quark, respectively, which are on-shell, i.e.\
$q_1^2=q_2^2=m^2$.  Furthermore, $q = q_1 - q_2$ is the outgoing momentum of
the current with $q^2 = s$ and
$\sigma_{\mu\nu} = \ii [\gamma_\mu, \gamma_\nu]/2$.  The form factors $F_i^k$
are conveniently obtained by applying appropriate projectors which we show in
Appendix~\ref{app::projectors}.  We denote the non-singlet and singlet
contributions to the form factors by\footnote{In
  Refs.~\cite{Fael:2022rgm,Fael:2022miw} no subscript has been used for the
  non-singlet contribution.}
\begin{eqnarray}
  F^{k}_{\rm non-sing} \mbox{ and  } F^{k}_{\mathrm{sing}, h/l}
  \,.
\end{eqnarray}
The second subscript $h$ or $l$ is used to distinguish the contributions where the external current couples to a massive or massless internal quark loop.

The colour structure of the two-loop singlet form factors is $C_FT_F$.
For the three-loop singlet contributions we have altogether five colour
structures: $C_F^2T_F$, $C_FC_AT_F$, $C_FT_F^2n_h$, $C_FT_F^2n_l$ and
$(d^{abc})^2/N_C$ where $C_F=T_F (N_C^2-1)/N_C$ and $C_A=2 T_F N_C$ are the
quadratic Casimir operators of the $\mathrm{SU}(N_C)$ gauge group in the
fundamental and adjoint representation, respectively, $n_l$ is the number of
massless quark flavors, and $T_F=1/2$.  For convenience we introduce $n_h=1$
for closed quark loops which have the same mass as the external quarks.
We then denote the total number of quark flavors by $n_f = n_l + n_h$.
Furthermore we have $(d^{abc})^2 = T_F^3 (N_C^2-1)(N_C^2-4)/(2N_C)$.
This colour structure only appears
for the vector current, whereas the remaining four colour factors only appear for the
axial-vector, scalar, and pseudoscalar currents.

For later convenience we introduce the perturbative expansion of the
various (bare, renormalized, finite, \ldots) quantities as
\begin{eqnarray}
  F &=& \sum_{i\ge0} \left(\frac{\alpha_s(\mu)}{\pi}\right)^i F^{(i)}
        \,,
        \label{eq::Fas}
\end{eqnarray}
where $\alpha_s$ depends on the number of active flavours.  We perform the
calculation of the bare diagrams and the renormalization of the ultraviolet
counterterms in $n_f$-flavour QCD with $n_f=n_l+n_h$.
We decouple the heavy quark from the running of $\alpha_s$
before subtracting the infrared poles (cf.\ Subsection~\ref{sub::IR}) such
that our final finite result for the form factors is parameterized in terms
of $\alpha_s^{(n_l)}$.  Note that in Eq.~(\ref{eq::Fas}) the singlet diagrams
start to contribute to $F^{(2)}$.

In case we implement the definition of $\gamma_5$ from
Ref.~\cite{Larin:1993tq} we replace it both in the Feynman rule for
the current and in the projector for the axial-vector and
pseudoscalar current according to
\begin{eqnarray}
  \gamma^\mu\gamma^5 & \to & \frac{\ii}{3!} \varepsilon^{\mu\nu\rho\sigma} \gamma_{[\nu}\gamma_\rho\gamma_{\sigma]}
  \,,\nonumber\\
  \gamma^5 & \to & \frac{\ii}{4!} \varepsilon^{\mu\nu\rho\sigma} \gamma_{[\mu}\gamma_\nu\gamma_\rho\gamma_{\sigma]}
  \,.
                   \label{eq::gamma5}
\end{eqnarray}
The square brackets on the r.h.s.\ denote anti-symmetrization of the
corresponding indices. After applying the projectors we obtain
products of two $\varepsilon$ tensors which we replace by
\begin{eqnarray}
  \varepsilon_{\alpha_1\alpha_2\alpha_3\alpha_4}
  \varepsilon_{\beta_1\beta_2\beta_3\alpha_4}
  = |(\delta_{\alpha_i\beta_j})|
  \,.
\end{eqnarray}
The determinant on the r.h.s.\ of this equation is interpreted in
$d=4-2\epsilon$ dimensions.



\section{Renormalization and infrared subtraction}
\label{sec::renormalization}

In order to obtain the UV-renormalized form factors we perform a parameter
renormalization for $\alpha_s$ in the $\overline{\rm MS}$ and for the heavy-quark
mass $m$ in the on-shell scheme.  In addition, we take into account the wave
function renormalization for the external quarks in the on-shell scheme.  For the
scalar and pseudoscalar current we renormalize the factor $m$ in the
definition of the currents (see Eq.~(\ref{eq::currents})) in the
$\overline{\rm MS}$ scheme.\footnote{Note that in Ref.~\cite{Fael:2022miw} the
  factor $m$ has been renormalized in the $\overline{\rm MS}$ scheme for the
  non-singlet current. However, for the singlet currents the on-shell
  scheme has been used.}

For the pseudoscalar and axial-vector currents there are additional
renormalization constants which depend on the considered current and on the
treatment of $\gamma_5$.  In the following we discuss in detail the
renormalization of the corresponding form factors.

After renormalization the form factors still contain infrared poles.
We discuss their subtraction in Subsection~\ref{sub::IR}.


\subsection{Pseudoscalar form factor $F^p$}

The two-loop singlet diagram contributing to the pseudoscalar form factor does not develop
sub-divergences and thus the form factor is finite. Similarly, at three-loop
order the counterterm contributions from the quark wave function, $\alpha_s$,
$m$, and the overall renormalization constant related to the non-vanishing
anomalous dimension of $j^p$ are sufficient to render the three-loop singlet
contributions ultraviolet finite. As a consequence it is not necessary to
separate singlet and non-singlet contributions and we can consider the proper sum
\begin{eqnarray}
  F^{p,\rm bare} &=& F^{p,\rm bare}_{\rm non-sing} + F^{p,\rm bare}_{\rm sing}
                     \,,
\end{eqnarray}
and adopt the $\gamma_5$ prescription of Ref.~\cite{Larin:1993tq} in all
contributions. This leads to
\begin{eqnarray}
  F^{p} &=& Z_{p}^{\rm fin} Z_{p}^{\overline{\rm MS}}
  Z_2^{\rm OS} F^{p,\rm bare}
  \Bigg|_{m^{\rm bare}=Z_m^{\rm OS}m^{\rm OS}, \, \alpha_s^{\rm bare}=Z_{\alpha_s}\alpha_s}
  \,,
              \label{eq::Fpf}
\end{eqnarray}
where $Z_2^{\rm OS}$ is the on-shell wave function renormalization constant for
the external quarks.

In case we drop the singlet contributions and use anti-commuting
$\gamma_5$ we have $Z_{p}^{\rm fin}=1$ and
$Z_{p}^{\overline{\rm MS}}=Z_m^{\overline{\rm MS}}$ in the above formula,
where $Z_m^{\overline{\rm MS}}$ is the $\overline{\rm MS}$ renormalization
constant of the quark mass.  For the $\gamma_5$ prescription of
Ref.~\cite{Larin:1993tq} explicit results for $Z_{p}^{\rm fin}$ and
$Z_{p}^{\overline{\rm MS}}$ can be found in Eq.~(\ref{eq::Zp}).  It is a
welcome cross check of our calculation that the non-singlet contribution of
$F^{p}$ agrees in the two approaches up to three-loop order.

The results for $F^{p}_{\rm sing}$ have already been shown in Ref.~\cite{Fael:2022miw}.
However, in this reference $Z_p^{\rm fin}=1$ has
been chosen and $Z_m^{\rm OS}$ has been used instead of $Z_p^{\overline{\rm MS}}$.
This has, of course, no influence on the finiteness of the form factor (after infrared subtraction), but the finite terms differ.


\subsection{Axial-vector form factors $F_1^a$ and $F_2^a$}

The singlet diagram contributions to the axial-vector form factor develop the
famous Adler-Bell-Jackiw anomaly~\cite{Adler:1969gk,Bell:1969ts} which leads
to a rather non-trivial renormalization.
In our derivation we assume that all $\nf = n_l + n_h$ quarks are grouped into doublets and $n_h = 1$.
We then introduce the ``flavour-non-singlet'' current
\begin{equation}
  J^a_{\mathrm{NS}, \mu} = \sum_{i=1}^{\nf} a_i \bar\psi_i \gamma_\mu \gamma_5 \psi_i \,,
  \label{eq::JaNS}
\end{equation}
where $a_i$ is the coupling of the quarks to the $Z$ boson in the SM.  For us
it is sufficient to assume $a_i = \pm 1$ depending on the weak isospin of the
quark.  The sum in Eq.~(\ref{eq::JaNS}) is to be understood such that the
massive form factors of quark flavour $i$ originating from $J^a_{\mathrm{NS}, \mu}$
can be written as
\begin{equation}
  F^a_{i,\mathrm{NS}} = F^a_{i,\mathrm{non-sing}} + F^a_{i,\mathrm{sing},h} - F^a_{i,\mathrm{sing},l}\,,
  \label{eq::FaNS}
\end{equation}
where $F^a_{i,\mathrm{sing},h}$ and $F^a_{i,\mathrm{sing},l}$ denote the
massive and massless singlet contributions as introduced in Section~\ref{sec::notation},
respectively.  The relative sign between $F^a_{i,\mathrm{sing},h}$
and $F^a_{i,\mathrm{sing},l}$ guarantees the anomaly cancellation in the SM.
It is well known that $J^a_{\mathrm{NS}, \mu}$ renormalizes multiplicatively
which also holds for the form factors
\begin{equation}
  F^a_{i,\mathrm{NS}} = Z_\mathrm{NS} Z_2^{\rm OS} F^{a,\mathrm{bare}}_{i,\mathrm{NS}}
  \,,
  \label{eq::FaNS_ren}
\end{equation}
where parameter renormalization on the r.h.s.\ in analogy to
Eq.~(\ref{eq::Fpf}) is understood. The renormalization constant $Z_\mathrm{NS}$
can be decomposed into
\begin{eqnarray}
  Z_{\mathrm{NS}} &=& Z_{a,\rm NS}^{\rm fin} Z_{a,\rm NS}^{\overline{\rm MS}}
  \label{eq::ZaNS}
\end{eqnarray}
with the $\overline{\rm MS}$ renormalization constant $Z_{a,\rm NS}^{\overline{\rm MS}}$ and the finite renormalization $Z_{a,\rm NS}^{\rm fin}$.
Up to the required order $Z_{a,\rm NS}^{\rm fin}$ and $Z_{a,\rm NS}^{\overline{\rm MS}}$ can be found in Eq.~(\ref{eq::Za}) in Appendix~\ref{app::ren_const}.

We also define the ``flavour-singlet'' current
\begin{equation}
  J^a_{\mathrm{S}, \mu} = \sum_{i=1}^{\nf} \bar\psi_i \gamma_\mu \gamma_5 \psi_i \,,
\end{equation}
where all quarks couple to the axial-vector current with the same sign.  Hence the form
factors decompose into
\begin{equation}
  F^a_{i,\mathrm{S}} = F^a_{i,\mathrm{non-sing}} + F^a_{i,\mathrm{sing},h} +
  \sum_{j=1}^{\nl} F^a_{i,\mathrm{sing},j}
  \,.
\end{equation}
Again the current and the form factors renormalize multiplicatively, i.e.
\begin{equation}
  F^a_{i,\mathrm{S}} = Z_\mathrm{S} Z_2^{\rm OS} F^{a,\mathrm{bare}}_{i,\mathrm{S}} ,
\end{equation}
where
\begin{eqnarray}
  Z_{\mathrm{S}} &=& Z_{a,\rm S}^{\rm fin} Z_{a,\rm S}^{\overline{\rm MS}}
\end{eqnarray}
can be decomposed in the same manner as $Z_{\mathrm{NS}}$ in Eq.~(\ref{eq::ZaNS}).
We again refer to Eq.~(\ref{eq::Za}) for the explicit renormalization constants.

With these definitions one can derive the renormalization for the non-singlet
and singlet axial-vector form factors $F^a_{i,\mathrm{sing}}$ and
$F^a_{i,\mathrm{non-sing}}$.
In the non-singlet case we have a multiplicatively renormalization
without any interference of the singlet diagram contributions. It is
given by
\begin{equation}
  F^a_{i,\mathrm{non-sing}} = Z_\mathrm{NS} Z_2^{\rm OS} F^{a,\mathrm{bare}}_{i,\mathrm{non-sing}}
  \,.
\end{equation}

On the other hand, for the renormalized singlet diagram contributions we have
to consider the difference
$\frac{1}{\nf}(F^a_{i,\mathrm{S}} - F^a_{i,\mathrm{NS}})$.  Since the SM is
anomaly free, $F^{a,\mathrm{bare}}_{i,\mathrm{sing},h}$ and
$F^{a,\mathrm{bare}}_{i,\mathrm{sing},l}$ have to renormalize in the same
way and one finds (see also Ref.~\cite{Chen:2021rft})
\begin{equation}
  F^a_{i,\mathrm{sing},j} = Z_\mathrm{NS} Z_2^{\rm OS}
  F^{a,\mathrm{bare}}_{i,\mathrm{sing},j}
  + \frac{1}{\nf}(Z_\mathrm{S} - Z_\mathrm{NS}) Z_2^{\rm OS}
  \left( F^{a,\mathrm{bare}}_{i,\mathrm{non-sing}} + \sum_{k=1}^{\nf}
    F^{a,\mathrm{bare}}_{i,\mathrm{sing},k} \right)
  \,,
\end{equation}
where $j\in\{h,l\}$ and
\begin{eqnarray}
  \frac{1}{\nf}(Z_\mathrm{S} - Z_\mathrm{NS}) &=& Z_{a,\rm S}^{\rm fin} Z_{a,\rm S}^{\overline{\rm MS}} -
                                 Z_{a,\rm NS}^{\rm fin} Z_{a,\rm NS}^{\overline{\rm MS}}
                                                  \nonumber\\
  &=&
  \left(\frac{\alpha_s}{\pi}\right)^2 C_F T_F
  \biggl(
    \frac{3}{8 \epsilon }
    +\frac{3}{16}
  \biggr)
  + \left(\frac{\alpha_s}{\pi}\right)^3 C_F T_F
  \biggl(
    \frac{1}{\epsilon^2}
    \biggl[
       \frac{1}{12} T_F \bigl( n_h + n_l \bigr)
       \nonumber \\ &&
       -\frac{11}{48} C_A
    \biggr]
    + \frac{1}{\epsilon}
    \biggl[
        \frac{109}{288} C_A
        -\frac{9}{16} C_F
        + \frac{1}{72} T_F \bigl( n_h + n_l \bigr)
    \biggr]
    +\biggl[
        \frac{13}{16} \zeta _3
        -\frac{163}{864}
    \biggr] C_A
    \nonumber \\ &&
    -\biggl[
         \frac{3}{4} \zeta_3
         -\frac{23}{64}
    \biggr] C_F
    + \frac{11}{54} T_F \bigl(n_h + n_l \bigr)
  \biggr)
  + {\cal O}(\alpha_s^4)
                         \,.
  \label{eq::Za_sing_nonsing}
\end{eqnarray}
Again we implicitly assume parameter renormalization in analogy to Eq.~(\ref{eq::Fpf}).

Since $(Z_\mathrm{S} - Z_\mathrm{NS})$ starts at $\mathcal{O}(\alphas^2)$, we
need $F^{a,\mathrm{bare}}_{i,\mathrm{non-sing}}$ only to one-loop order and
the last term on the right-hand-side can be neglected.  It is crucial to use
the same prescription for $\gamma_5$ both in the calculation of
$F^{a,\mathrm{bare}}_{i,\mathrm{sing},j}$ and
$F^{a,\mathrm{bare}}_{i,\mathrm{non-sing}}$.
Furthermore, it is important to keep the higher-order terms
in $\ep$ in the tree-level expression
$F^{a,\mathrm{bare},(0)}_{i,\mathrm{non-sing}}$.  For example, the
$\alphas^3/\epsilon^2$ term of $(Z_\mathrm{S} - Z_\mathrm{NS})$ multiplies the
$\mathcal{O}(\ep)$ term of $F^{a,\mathrm{bare},(0)}_{i,\mathrm{non-sing}}$
and produces a term
proportional to $\alphas^3/\ep$ which is necessary to cancel all poles for the
$C_AC_FT_F$ and $C_FT_F^2$ colour factors.  The finiteness of the $C_F^2T_F$
colour factor is guaranteed through the ${\cal O}(\alpha_s^2)$ term of
$(Z_\mathrm{S} - Z_\mathrm{NS})$ which multiplies
$F^{a,(1),\rm bare}_{1,\rm non-sing}$.

\subsection{\label{sub::IR}Infrared divergences}

After the ultraviolet renormalization we still have infrared poles which we treat via
\begin{eqnarray}
  F^f &=& Z^{-1} F\,,
\end{eqnarray}
where $F$ is the UV renormalized form factor and
$F^f$ is finite, i.e., the limit $\epsilon\to0$ can be taken.
$Z$ can be constructed from the cusp anomalous dimension
which has been computed to three-loop order in Refs.~\cite{Polyakov:1980ca,Korchemsky:1987wg,Grozin:2014hna,Grozin:2015kna}.
In our calculation we express $F^f$ in terms of $\alpha_s^{(n_l)}$.




\section{\label{sec::chiral}Chiral Ward identity}

For the axial-vector current the non-renormalization of the
Adler-Bell-Jackiw (ABJ) anomaly implies that the
equation
\begin{eqnarray}
  \label{eq::ABJ}
  (\partial^\mu j_\mu^a)_\mathrm{R} &=& 2 (j^p)_\mathrm{R} + \frac{\alpha_s}{4\pi} T_F (G\tilde G)_\mathrm{R}
\end{eqnarray}
holds at the level of renormalized operators (indicated by the subscript
$\mathrm{R}$)~\cite{Adler:1969er}.  It relates the derivative of the
axial-vector current to the pseudoscalar current and the pseudoscalar gluonic
operator
\begin{equation}
  G\tilde G = \epsilon_{\mu\nu\rho\sigma} G^{a,\mu\nu} G^{a,\rho\sigma} \,,
\end{equation}
where $G^{a,\mu\nu}$ is the field strength tensor of the gluon.  In analogy to
Eq.~(\ref{eq::Gamma}) the three-point functions of $\partial^\mu j_\mu^a$ and
$G\tilde G$ with a massive quark-anti-quark pair can be decomposed as
\begin{eqnarray}
  \Gamma_{\partial J}^a(q_1,q_2) &=& {2 \ii m} F_{\partial J} (q^2) {\gamma_5} \,,
  \nonumber\\
  \Gamma_{G\tilde G}(q_1,q_2) &=& {2 \ii m} F_{G\tilde G} (q^2) {\gamma_5}\,,
\end{eqnarray}
with the form factors $F_{\partial J}$ and $F_{G\tilde G}$.
This allows us to rewrite Eq.~(\ref{eq::ABJ}) at the level of form factors as
\begin{eqnarray}
  F_{\partial J, \mathrm{non-sing}}  &=& F^{p,f}_{\rm non-sing}
                         \label{eq::WI2}
\end{eqnarray}
for the non-singlet and
\begin{eqnarray}
  F_{\partial J, \mathrm{sing}}  &=& F^{p,f}_{\rm sing} + \frac{\alpha_s}{4\pi} T_F F^f_{G\tilde G}
                         \label{eq::WI}
\end{eqnarray}
for the singlet contributions.  Equations~(\ref{eq::WI2}) and~(\ref{eq::WI})
are usually referred to as chiral Ward identities, the latter especially as
the anomalous chiral Ward identity.  In this work we use them as non-trivial
cross checks of our results. This is particularly interesting for
Eq.~(\ref{eq::WI}) which involves finite renormalization constants related to
the treatment of $\gamma_5$.

\begin{figure}[t]
  \begin{center}
      \includegraphics[width=0.15\textwidth]{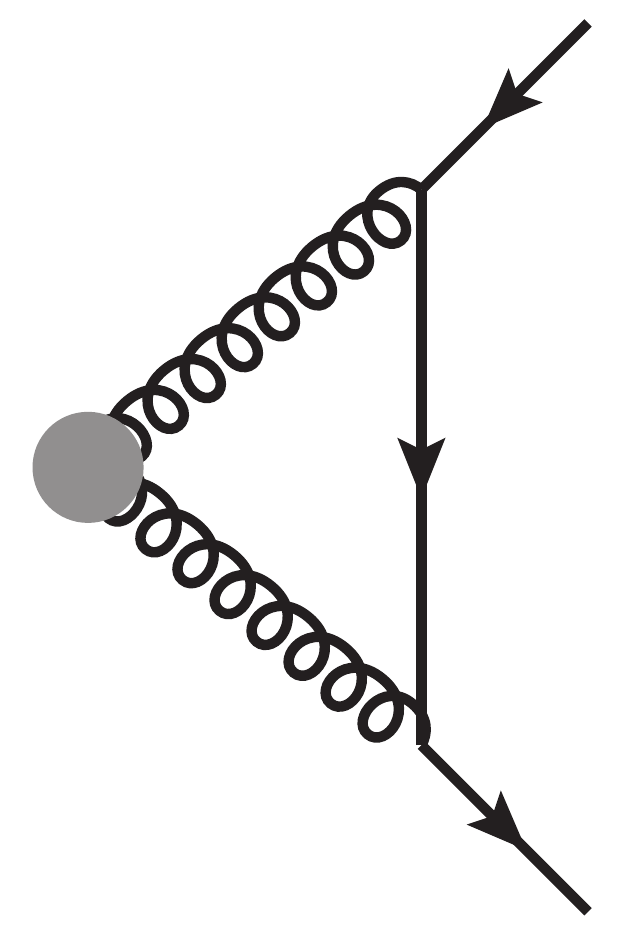}
      \hspace{0.6cm}
      \includegraphics[width=0.15\textwidth]{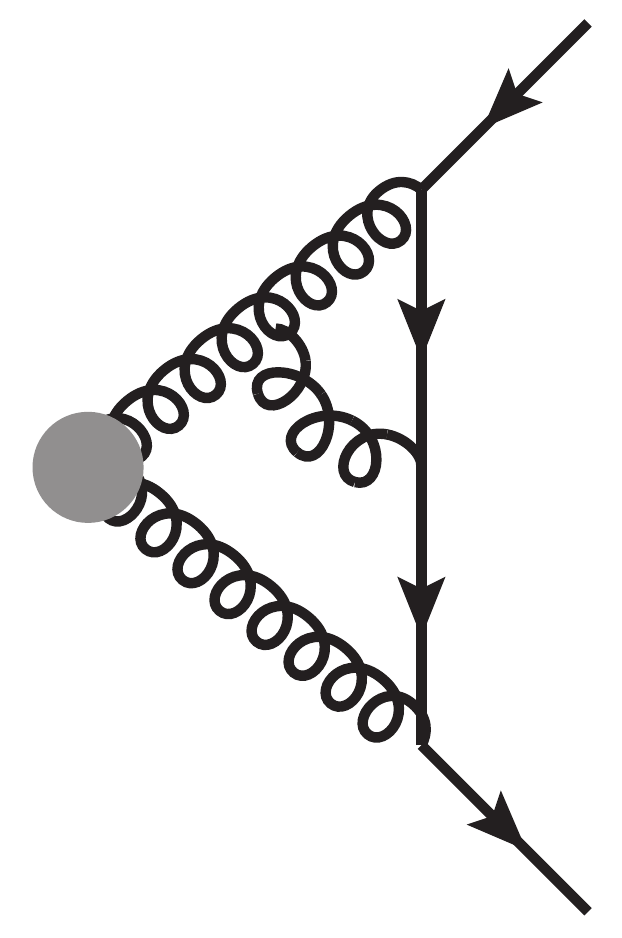}
      \hspace{0.6cm}
      \includegraphics[width=0.15\textwidth]{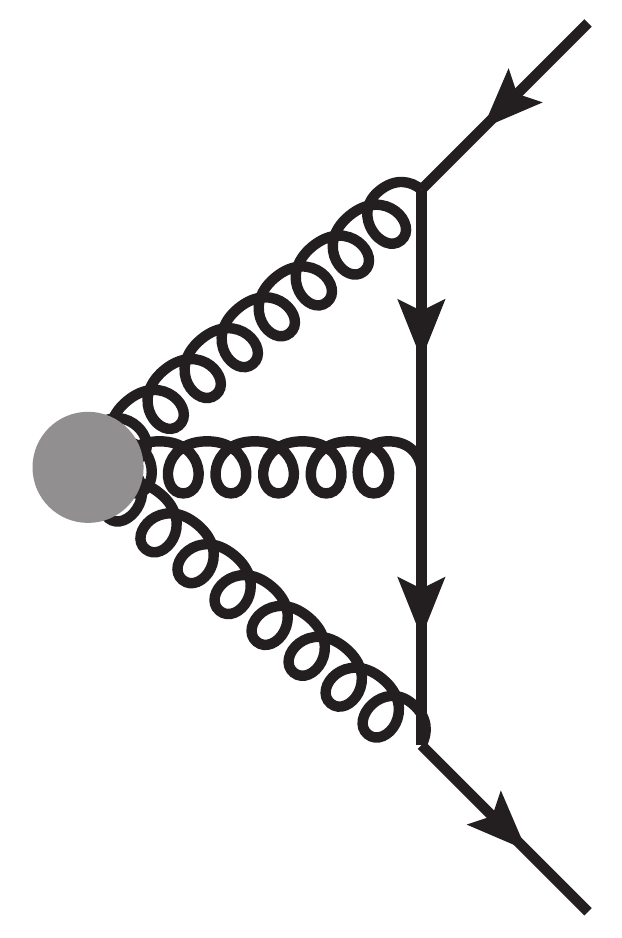}
    \caption{\label{fig::GGtil}One- and two-loop sample Feynman diagrams
      contributing to $F_{G\tilde{G}}$.}
  \end{center}
\end{figure}

For this check we require $F^f_{G\tilde{G}}$ to ${\cal O}(\alpha_s^2)$.
Since the operators $G\tilde{G}$ and $\partial^\mu j_\mu^a$ mix under
renormalization, the finite expression is given by
\begin{eqnarray}
  F_{G\tilde{G}}^f &=& Z^{-1} Z_2^{\rm OS}
                       \left( Z_{G\tilde{G}} F_{G\tilde{G}}^\mathrm{bare} + Z_{GJ} F_{\partial J}^\mathrm{bare} \right)
                       \Big|_{m^{\rm bare}=Z_m^{\rm OS}m^{\rm OS}, \, \alpha_s^{\rm bare}=Z_{\alpha_s}\alpha_s}
                       \,.
                       \label{eq::FGGtil}
\end{eqnarray}
The renormalization constants $Z_{G\tilde{G}}$ and $Z_{GJ}$ have been computed in Refs.~\cite{Larin:1993tq,Zoller:2013ixa,Ahmed:2015qpa,Ahmed:2021spj,Luscher:2021bog,Chen:2021gxv}.
To the required orders they read
\begin{eqnarray}
  Z_{G\tilde{G}} &=& Z_{\alpha_s}
                     = 1 + \frac{\alpha_s}{\pi}
                     \frac{1}{\epsilon} \left (-
                     \frac{11}{12}C_A + \frac{1}{3} T_F n_f
                     \right)
                     + {\cal O}(\alpha_s^2)
                     \,,
                     \nonumber\\
  Z_{GJ} &=&     \frac{\alpha_s}{\pi} \frac{3 C_F}{\epsilon }
                +\left(\frac{\alpha_s}{\pi}\right)^2
                \biggl(
                    \frac{1}{\epsilon ^2}
                    \biggl[
                        C_F n_f T_F
                        -\frac{11}{4} C_A C_F
                    \biggr]
                    +\frac{1}{\epsilon }
                    \biggl[
                        \frac{71}{24} C_A C_F
                        \nonumber \\ &&
                        -\frac{21}{8} C_F^2
                        -\frac{1}{6} C_F T_F n_f
                    \biggr]
                \biggr)
                + {\cal O}(\alpha_s^3)
                \,.
\end{eqnarray}
The factor $Z^{-1}$ again subtracts infrared poles, cf.\ Subsection~\ref{sub::IR}.
We thus have to compute one- and
two-loop corrections for $F_{G\tilde{G}}$ and
one-loop corrections to $F_{\partial J}$ since
$Z_{GJ}$ starts at ${\cal O}(\alpha_s)$.

We compute $F_{G\tilde{G}}^\mathrm{bare}$ using the same setup as for the other form factors which is described in Section~\ref{sec::comp}.
Sample Feynman diagrams contributing to it are shown in Fig.~\ref{fig::GGtil}.
We apply the same projector as
for the pseudoscalar current and use the prescription of $\gamma_5$ from
Eq.~(\ref{eq::gamma5}).

To compute $F_{\partial J}$ we follow two different strategies:
First we treat $\partial^\mu j_\mu^a$ as an independent operator, implement its Feynman rule
using Eq.~(\ref{eq::gamma5}),
and then apply again the projector to the pseudoscalar current.
Secondly we apply the derivative to $\Gamma_\mu^a$ in its decomposed form of Eq.~(\ref{eq::Gamma}) and employ the Dirac equation as well as an anti-commuting $\gamma_5$ to find
\begin{equation}
  F_{\partial J} = F^{a,f}_{1} + \frac{s}{4m^2} F^{a,f}_{2} .
\end{equation}
Thus we can simply use the expressions for $F^{a,f}_{1}$ and $F^{a,f}_{2}$ directly instead of computing $F_{\partial J}$.
Both approaches lead to identical results.
It is interesting to mention that the higher order $\epsilon$ terms
of the tree-level expression
\begin{eqnarray}
  F_{\partial J}^{(0)}  =  {1 - \frac{11\ep}{3} + 4 \ep^2 - \frac{4\ep^3}{3}\,,}
\end{eqnarray}
are crucial to obtain the correct result, at least with our choice of projectors, cf.\ Appendix~\ref{app::projectors}.

After inserting the bare results and counterterms into Eq.~(\ref{eq::FGGtil}) we
obtain a finite result for $F_{G\tilde{G}}^f$ which we present in Appendix~\ref{app::GGtil}.
This then allows us to check the anomalous Ward identity~(\ref{eq::WI2}) in Subsection~\ref{sub::checks}.



\section{\label{sec::comp}Computational details}

For our calculation we use the same automated setup as for the calculation of
the non-singlet form factors in Refs.~\cite{Fael:2022rgm,Fael:2022miw}.  We generate the diagrams
with {\tt qgraf}~\cite{Nogueira:1991ex} and process them with {\tt q2e}
and~{\tt exp}~\cite{Harlander:1997zb,Seidensticker:1999bb,q2eexp} to obtain
{\tt FORM}~\cite{Kuipers:2012rf} code for each individual amplitude. After
applying the projectors and taking the traces each amplitude is written as a
linear combination of scalar functions which belong to certain integral
families. The reduction to master integrals is performed with
\texttt{Kira}~\cite{Maierhofer:2017gsa,Klappert:2020nbg} with \texttt{Fermat}~\cite{fermat}. At this step it is
important to choose a good basis where the dependence on the kinematic variable
and the space-time factorizes. For this step we use the program
\texttt{ImproveMaster.m} developed in Ref.~\cite{Smirnov:2020quc} in an improved version.
Once we know the master integrals for each individual integral family we use
\texttt{Kira} to find a minimal set which reduces the number of master
integrals from $1995$ to $316$ for the massive and from $698$ to $158$ for massless singlet contributions. Next we establish the differential
equations with the help of \texttt{LiteRed}~\cite{Lee:2012cn,Lee:2013mka}.  At
this point only the boundary conditions of all master integrals at some
initial value for $s/m^2$ are needed such that the method of
Ref.~\cite{Fael:2021kyg} can be applied to obtain results for all master
integrals in the whole kinematic range.
We already computed the master integrals for the massive singlet contributions in Ref.~\cite{Fael:2022miw}.
Thus we only describe the calculation of the massless singlet contributions in Subsection~\ref{sec::comp3l}.

The calculation of $F_{G\tilde{G}}$ follows the same general setup.
However, instead of \texttt{q2e} we use \texttt{tapir}~\cite{Gerlach:2022qnc}.
The different mass patterns require the introduction of new integral families
which lead to 3 and 24 master integrals at one- and two-loop order,
respectively. 9 of the two-loop master integrals are known from the two-loop
calculation of the non-singlet form factors.
We describe the analytical computation of the remaining 15 master integrals in Subsection~\ref{sec::masters}.


\subsection{\label{sec::comp3l}Computation of massive vertex integrals at three loops}

The method for our calculation of the three-loop master integrals is described
in detail in Ref.~\cite{Fael:2022miw}.  We deviate slightly from the steps
outlined in this reference by not computing analytical boundary conditions in
the asymptotic limit $s \to 0$. Instead we use numerical boundary conditions
obtained with \texttt{AMFlow}~\cite{Liu:2022chg}\footnote{See
  Refs.~\cite{Liu:2017jxz,Liu:2020kpc,Liu:2021wks,Liu:2022tji,Liu:2022mfb} for
  more details on the auxiliary mass flow method.} at $s/m^2=-1$ which
corresponds to a regular point.  More precisely, we use
\texttt{AMFlow} with \texttt{Kira}~\cite{Maierhofer:2017gsa,Klappert:2020nbg}
as reduction back-end to compute all master integrals as expansions up to
$\ep^6$ at $s/m^2=-1$.  The coefficients of these expansions are floating
point numbers which we obtain with $100$ significant digits within a few days
of runtime for all integral families except one for which we obtain only $85$
significant digits.  From there we derive symbolic expansion at
\begin{align}
    s/m^2 &= \{ -\infty, -32, -28, -24, -16, -12, -8, -4, -3, -2, -1, -3/4,
            -1/2, -1/4, 0,
    \nonumber \\
    &  1/4, 1/2, 1, 2, 3, 7/2, 4, 9/2, 5, 6, 8, 10, 14, 20, 26, 32, 40, 52 \}
\end{align}
and match subsequent expansions in between where the radii of convergence
overlap.  In this way we find a semianalytic expression for the master
integrals over the whole range of $s/m^2$.  In practice we do the following:
We start from the expansion at $s/m^2=-1$ where we can directly match to the
numerical boundary conditions provided by \texttt{AMFlow}.  From there we can
move with the expansions either to smaller or larger values of $s/m^2$.  On
the one hand, we match along the negative axis to $s/m^2 \to - \infty$ and
obtain the expansion for $s/m^2 \to + \infty$ by analytic continuation.  Then
we match down to smaller positive values of $s/m^2$ until $s/m^2=1$.  On the
other hand, we move from $s/m^2=-1$ to larger values where we stop at the
two-particle treshold at $s/m^2=4$.  We check that both ways of expanding and
matching agree in the overlap region of $1 < s/m^2 < 4$ within the expected
accuracy.  This constitues a non-trivial cross check on the calculation of the
master integrals.  We additionally cross check the expansion at $s=0$ where a
subset of master integrals have been computed analytically.  Furthermore, all
master integrals have been computed at $s/m^2=2$ and $s/m^2=6$ with $30$ digit
precision using \texttt{AMFlow} to check the results obtained through the
differential equations. We find agreement within the expected uncertainty.



\subsection{\boldmath Calculation of master integrals for $F_{G\tilde{G}}$}
\label{sec::masters}

Let us briefly describe the calculation for the master integrals needed for
 $F_{G\tilde{G}}$ at two-loop order.
First, we establish a system of differential equations in the variable $x$
defined by
\begin{eqnarray}
 s &=& - \frac{(1-x)^2}{x} ~.
\end{eqnarray}
The system of differential equations is subsequently solved with the methods
described in Ref.~\cite{Ablinger:2018zwz}.  In practice this means that we do
not bring the system to canonical form, but we decouple coupled systems of
differential equations into one higher-dimensional one using the package
\texttt{OreSys}~\cite{ORESYS} (which is based on
\texttt{Sigma}~\cite{Schneider:2007}) and solve this equation order-by-order
in $\epsilon$ with \texttt{HarmonicSums}~\cite{HarmonicSums}.  The largest
coupled system we encounter here is a $3\times3$ system.  For the complete
solution we have to provide boundary conditions.  To do this, we choose to
compute the master integrals in the limit $s \to 0$ ($x \to 1$).  However,
since the diagrams can have cuts through only massless lines, the limit
$s \to 0$ needs an asymptotic expansion.  While the asymptotic expansion for
some integrals can be constructed by direct integration or via simple
Mellin-Barnes representations, we apply the method of
regions~\cite{Beneke:1997zp} as implemented in
\texttt{asy.m}~\cite{Jantzen:2012mw} to the more involved master integrals.
It turns out that there are three different regions, which scale as
$\chi^{-0\epsilon}$, $\chi^{-2\epsilon}$ and $\chi^{-4\epsilon}$ in the variable
$\chi=\sqrt{-s/m^2}$.  The hard region $\propto \chi^{-0\epsilon}$ leads to massive
propagators which are well studied in the literature (see, e.g.,
Ref.~\cite{Fleischer:1999hp}).  The integrals in the second region
$\propto \chi^{-2\epsilon}$ can be calculated in closed form in terms of $\Gamma$
functions.  In the region $\propto \chi^{-4\epsilon}$ we encountered one integral
which could not be calculated in terms of $\Gamma$ functions.  For this
integral we used \texttt{HyperInt}~\cite{Panzer:2014caa} to obtain the result
expanded in $\epsilon$.  It turns out that the solutions of all master
integrals can be written in terms of harmonic
polylogarithms~\cite{Remiddi:1999ew}.  We provide these results in an
ancillary file~\cite{progdata}.

The analytic results have been cross checked against numerical evaluations with
\texttt{FIESTA5}~\cite{Smirnov:2021rhf} in the euclidean region ($0<x<1$).



\subsection{Cross checks}
\label{sub::checks}

There are a number of checks which support the correctness of our result
which we summarize in the following.

At two-loop order we reproduce the massless and massive
singlet axial-vector and pseudoscalar results presented in Ref.~\cite{Bernreuther:2005rw}.  We
also agree with the one-loop corrections to $F_{G\tilde{G}}$.

Furthermore, we have performed our calculation for general QCD gauge parameter
$\xi$ and have checked that it drops out in the final result. This is a
non-trivial check at three loops where $\xi$ cancels only after including
the counterterm contribution from mass renormalization.

At three loops we have cross checked the results for the massless singlet
master integrals by evaluating them numerically with AMFlow~\cite{Liu:2022chg}
at $s/m^2=2$ and $s/m^2=6$.  This is an important consistency check for the
method which we use to compute the master integrals.  We chose these points
because they are separated by at least one special point like the thresholds
and the high-energy expansion from our boundary conditions.  Crossing these
special points is the most difficult step in our approach.

A further check is the use of naive $\gamma_5$ and non-anti-commuting
$\gamma_5$ for the non-singlet contributions of the axial-vector and
pseudoscalar currents. Both calculations agree after taking into account the
proper $\overline{\rm MS}$ and finite renormalization constants, see Section~\ref{sec::results}.

\begin{figure}[t]
  \begin{center}
    \begin{tabular}{cc}
      \includegraphics[width=0.47\textwidth]{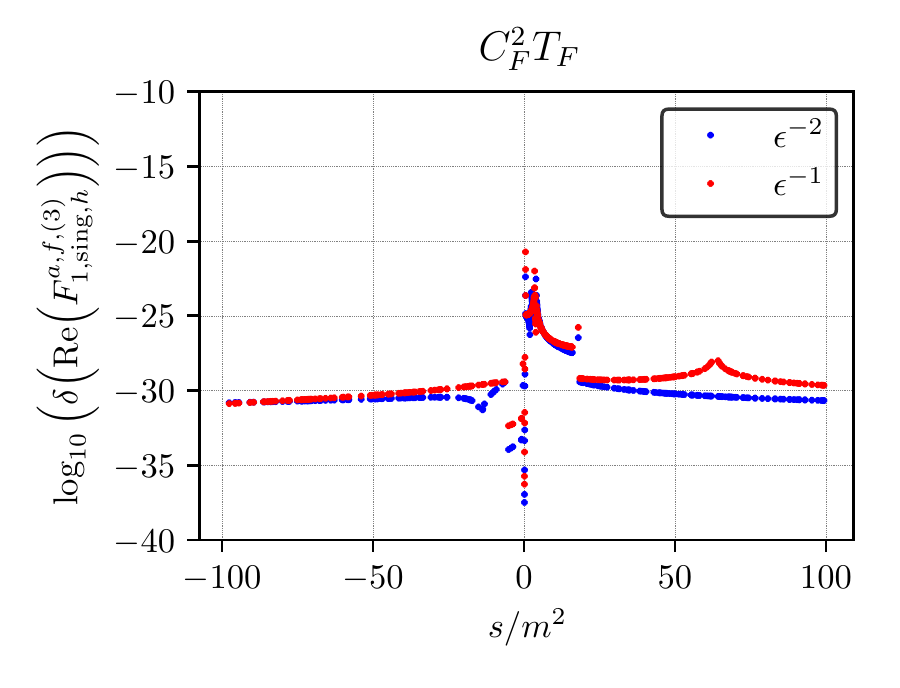}
      &
      \includegraphics[width=0.47\textwidth]{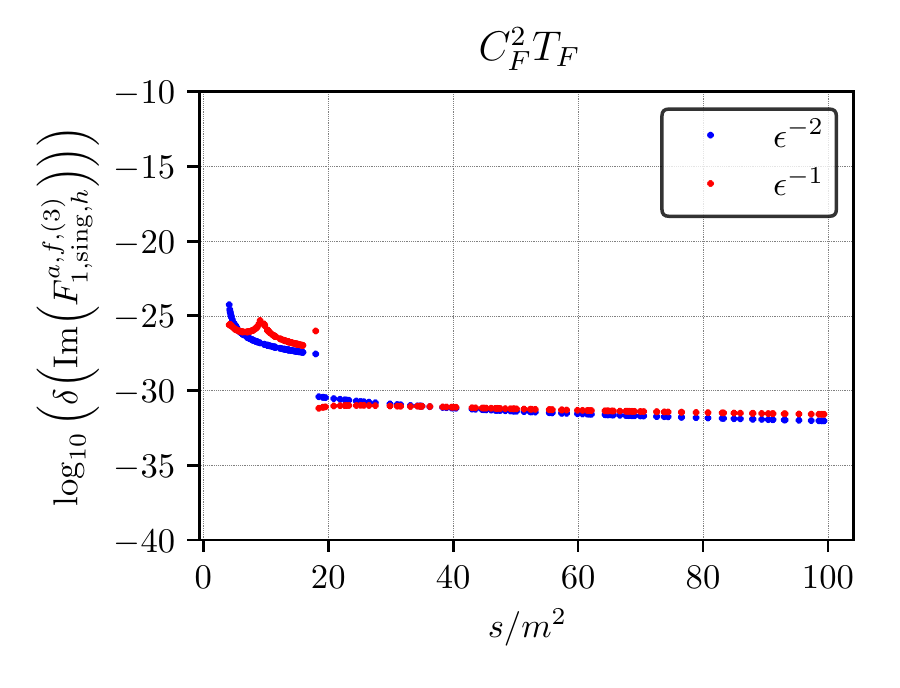}
      \\
      \includegraphics[width=0.47\textwidth]{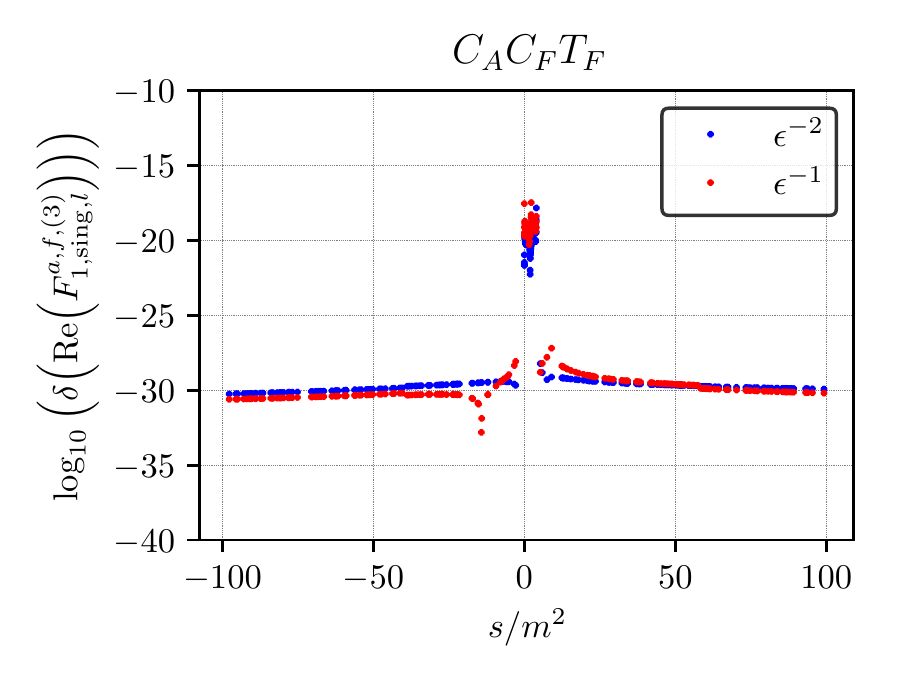}
      &
      \includegraphics[width=0.47\textwidth]{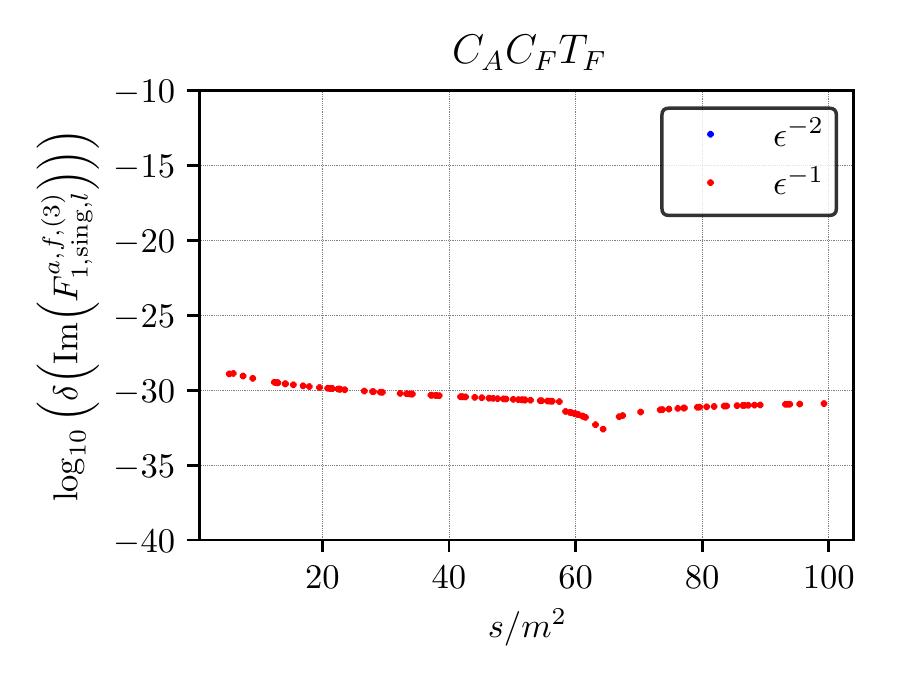}
      \\
    \end{tabular}
    \caption{\label{fig::pole-cancellation}
      Relative cancellation of the poles for the $C_F^2 T_F$ colour factor of $F^{a,f,(3)}_{1,\mathrm{sing},h}$ and the $C_A C_F T_F$ colour factor of $F^{a,f,(3)}_{1,\mathrm{sing},l}$.
    }
  \end{center}
\end{figure}

Since our three-loop results are mainly floating point numbers, the poles also only cancel numerically against the analytically known counterterms.
We can therefore use the precision of these cancellations as cross check and estimate of the uncertainty.
As in Ref.~\cite{Fael:2022miw} we define
\begin{equation}
  \label{eq::delta-def}
  \delta \Big(F^{f,(3)}\big|_{\ep^i}\Big) = \frac{F^{(3)}\big|_{\ep^i}+F^{(\text{CT}+{Z})}\big|_{\ep^i}}{F^{(\text{CT}+{Z})}\big|_{\ep^i}}\,,
\end{equation}
which represents the number of correct digits for the poles of order $\ep^i$.
As representative examples we show the $C_F^2 T_F$ colour factor of $F^{a,f,(3)}_{1,\mathrm{sing},h}$ and the $C_A C_F T_F$ colour factor of $F^{a,f,(3)}_{1,\mathrm{sing},l}$ in Fig.~\ref{fig::pole-cancellation}.

It is clearly visible that the poles cancel with at least around 20 digits for the massive singlet and at least around 15 digits for the massless singlet contributions.
In both cases we obtain this worst precision in the region $0 \leq s < 4 m^2$, while it is around 30 digits over large ranges of $s$.
Since the precision is similar or better for the other colour factors and form factors, we refrain from showing more plots.

\begin{figure}[t]
  \begin{center}
    \begin{tabular}{cc}
      \includegraphics[width=0.47\textwidth]{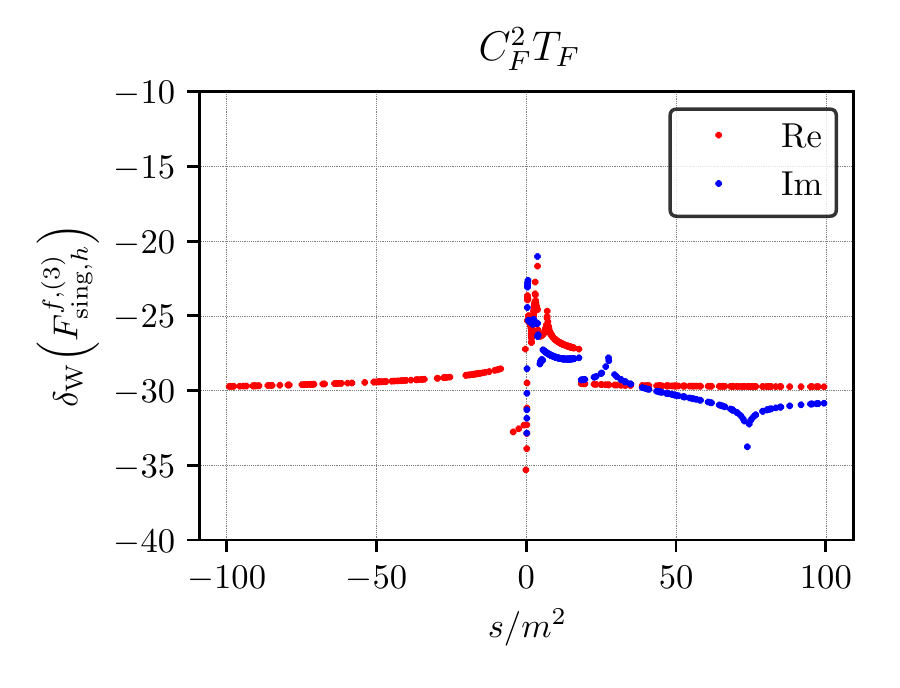}
      &
      \includegraphics[width=0.47\textwidth]{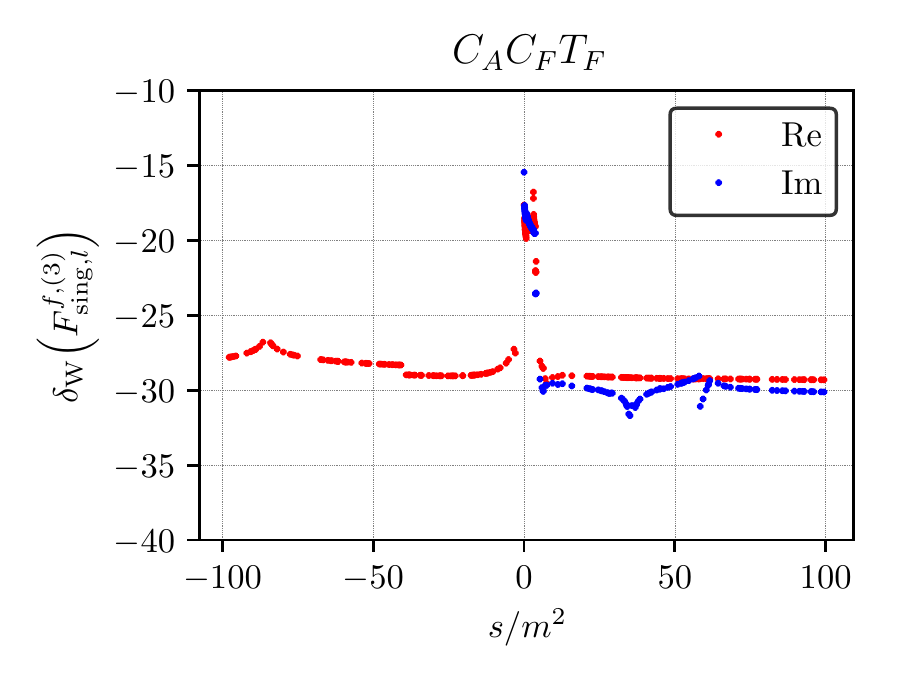}
    \end{tabular}
    \caption{\label{fig::anomalous-Ward} Precision to which the anomalous Ward
      identity in Eq.~(\ref{eq::WI}) is fulfilled for the $C_F^2 T_F$ colour
      factor of the massive and the $C_A C_F T_F$ colour factor for the
      massless singlet contributions. The quantity $\delta_W$ is defined
        in Eq.~(\ref{eq::deltaW}). }
  \end{center}
\end{figure}

Finally, we can explicitly check the chiral Ward identities of Eqs.~(\ref{eq::WI2}) and~(\ref{eq::WI}) which relate $F^a_1$, $F^a_2$, $F_p$, and $F_{G\tilde{G}}$.
Since they hold on the level of finite form factors, they allow us to check their finite terms.
This is especially interesting for the singlet contributions with their nontrivial renormalization including finite pieces, cf.\ Section~\ref{sec::renormalization}.
We define the relative precision with respect to the analytically computed $F_{G\tilde{G}}$ as
\begin{equation}
  \delta_\mathrm{W}\Big(F^{f,(3)}_\mathrm{sing}\Big) = \frac{F^{a,f,(3)}_{1,\mathrm{sing}} + \frac{s}{4m^2} F^{a,f,(3)}_{2,\mathrm{sing}} - F^{p,f,(3)}_{\rm sing} - \big(\frac{\alpha_s}{4\pi} T_F F^f_{G\tilde G}\big)^{(3)}}{ \big(\frac{\alpha_s}{4\pi} T_F F^f_{G\tilde G}\big)^{(3)}} \,.
  \label{eq::deltaW}
\end{equation}
In Fig.~\ref{fig::anomalous-Ward} we show it for two colour factors of the massive and massless singlet contributions.

The precision is similar compared to the pole cancellation discussed before and we again refrain from showing more than two representative examples.




\section{\label{sec::results}Results for the singlet form factors}

In this Section we discuss our results for the singlet form factors.  We
present expansions for small and large values of $s$ and for $s\to 4m^2$ and
show results for the finite form factors in the whole $s$ range.  For better
readability we concentrate in the main text to the contributions where the
external current couples to massless quarks and relegate the formulae and plots
for the massive singlet contributions to Appendix~\ref{app::resFF}.

We provide all results obtained in this paper as well as the non-singlet and singlet results from Refs.~\cite{Fael:2022rgm,Fael:2022miw} as \texttt{Mathematica} package available in Ref.~\cite{gitmath}.
Furthermore we implemented these results in the \texttt{Fortran}
library \texttt{FF3l} available in Ref.~\cite{gitfortran} which allows for a fast numerical evaluation of
all form factors.
The library is described in more detail in Appendix~\ref{app:formfactors3l}.


\subsection{Comparison of naive and Larin $\gamma_5$ prescription for
  axial-vector and pseudoscalar non-singlet form factors}

It is interesting to discuss the tree-level results for the two
$\gamma_5$ prescriptions. For the axial-vector and
pseudoscalar current we find
\begin{eqnarray}
  F_{1,\rm naive}^{a,(0)} &=& 1\,,\nonumber\\
  F_{2,\rm naive}^{a,(0)} &=& 0\,,\nonumber\\
  F_{1,\rm larin}^{a,(0)} &=& 
  1
  - \epsilon \frac{32m^2-5s}{3(4m^2-s)}
  + \epsilon^2 \frac{4(4m^2+s)}{3(4m^2-s)}
  - \epsilon^3 \frac{4s}{3(4m^2-s)}
  \,,\nonumber\\
  F_{2,\rm larin}^{a,(0)} &=&
  - \epsilon \frac{8 {m^2}(2m^2-s)}{s(4m^2-s)}
  + \epsilon^2 \frac{64 m^2 (2m^2-s)}{3s(4m^2-s)}
  - \epsilon^3 \frac{32 m^2(2m^2-s)}{3s(4m^2-s)}
  \,,\nonumber\\
  F_{\rm naive}^{p,(0)} &=& 1
  \,,\nonumber\\
  F_{\rm larin}^{p,(0)} &=& 
  1
  - \epsilon \frac{12m^2+19s}{6s}
  + \epsilon^2 \frac{44m^2+13s}{6s}
  - \epsilon^3 \frac{2(12m^2-s)}{3s}
  + \epsilon^4 \frac{2(4m^2-s)}{3 s}\,,
  \label{eq::Ftree}
\end{eqnarray}
with our choice of projectors, cf.\ Appendix~\ref{app::projectors}.
Note that there is a non-trivial $s$ dependence at higher orders in
$\epsilon$. Through renormalization of the quark wave function and the
subtraction of infrared divergences they induce finite terms in $\epsilon$ at
one-loop order.  At two and three loops even poles are generated which are
important to obtain finite expressions for the form factors.

We have used the prescription of Ref.~\cite{Larin:1993tq} for $\gamma_5$
also for the one-, two-, and three-loop form factors. After renormalization and
infrared subtraction we obtain
\begin{eqnarray}
  F_{1,\rm naive}^{a,(i),f} = F_{1,\rm larin}^{a,(i),f} \,,\nonumber\\
  F_{2,\rm naive}^{a,(i),f} = F_{2,\rm larin}^{a,(i),f} \,,\nonumber\\
  F_{\rm naive}^{p,(i),f}   = F_{\rm larin}^{p,(i),f} \,,
\end{eqnarray}
for $i=1,2,3$. Let us stress that it is important to
thoroughly follow the instructions from Section~\ref{sec::renormalization} and
take into account all relevant renormalization constants
from Appendix~\ref{app::ren_const}.



\subsection{\label{sub::expansions}Expansions for $s\to 0$, $s\to-\infty$, and $s\to4m^2$}

In this Section we concentrate on the singlet contributions and present
explicit results for the expansions for small and large values of $s$ and close
to threshold. We choose $\mu^2=m^2$ for the renormalization scale. For
completeness we present both two- and three-loop expressions.

Including terms up to linear order in $\chi = \sqrt{-s/m^2}$ we obtain
for the massless singlet contribution in the limit $s\to0$
\begin{align}
    &F_{1,{\rm sing},l}^{v,f}\Big|_{s\to0} =
      \left(\frac{\alpha_s}{\pi}\right)^3 \frac{d_{abc}d^{abc}}{\nc} \left[ -0.64927 + 0.99711 \chi \right]
      \,,
    \\
    &F_{2,{\rm sing},l}^{v,f}\Big|_{s\to0} =
      \left(\frac{\alpha_s}{\pi}\right)^3 \frac{d_{abc}d^{abc}}{\nc} \left[ -5.7080 - 6.5797 \ln (\chi) + \chi \bigl(8.1838-3.7011 \ln (\chi) \bigr) \right]
      \,,
    \\
    &F_{1,{\rm sing},l}^{a,f}\Big|_{s\to0} =
      \left(\frac{\alpha_s}{\pi}\right)^2 C_F T_F \biggl[
        - \frac{7}{4}
        + \frac{\pi^2}{4} \chi
      \biggr]
      \nonumber\\
    & + \left(\frac{\alpha_s}{\pi}\right)^3 C_F T_F \biggl[
      C_F \biggl(
        -1.4887 + 1.2337 \chi
      \biggr)
      + C_A \biggl(
        -9.0185 + \chi \bigl(6.3166 - 7.8134 \ln(\chi) \bigr)
      \biggr)
      \nonumber \\ &
      +  T_F n_h \biggl(-0.32519\biggr)
      +  T_F n_l \biggl(
        3.6797 + \chi \bigl(-1.4751 + 3.2899 \ln(\chi) \bigr)
      \biggr)
    \biggr]
      \,,
    \\
    &F_{2,{\rm sing},l}^{a,f}\Big|_{s\to0} =
      \left(\frac{\alpha_s}{\pi}\right)^2 C_F T_F
      \biggl[
        \frac{\pi^2}{2 \chi}
        - \frac{2}{3} \ln^2(\chi)
        + \frac{25}{9} \ln(\chi)
        - \frac{95}{54}
        - \frac{\pi^2}{9}
      \biggr]
      \nonumber\\
    & + \left(\frac{\alpha_s}{\pi}\right)^3 C_F T_F \biggl[
      C_F \biggl(
          \frac{2.4674}{\chi }
          +6.3840 \ln (\chi )
          +2.2099
          +\chi  \bigl( 2.8786 \ln(\chi )-5.0719 \bigr)
      \biggr)
      \nonumber \\ &
      + C_A \biggl(
          -\frac{15.627 \ln (\chi )+5.3408}{\chi }
          +0.81481 \ln ^3(\chi )
          -2.6308 \ln ^2(\chi )
          +4.2083 \ln (\chi )
          \nonumber \\ &
          +14.089
          +\chi  (6.0657 \ln (\chi )-5.3476)
      \biggr)
      +  T_F n_h \biggl(
        -0.96834 \ln (\chi )
        +0.15303
        +0.90471 \chi
      \biggr)
      \nonumber \\ &
      +  T_F n_l \biggl(
         \frac{6.5797 \ln (\chi )+4.5177}{\chi }
        -0.29630 \ln ^3(\chi )
        +1.2593 \ln ^2(\chi )
        +0.47451 \ln (\chi )
        \nonumber \\ &
        -5.9964
        +\chi  (-2.4674 \ln (\chi )-0.049185)
      \biggr)
    \biggr]
      \,,
      \label{eq::nlsing_s0}
  \end{align}
where terms of ${\cal O}(\chi^2)$ have been neglected and the analytic continuation
for $s>0$ is given by $\chi = \sqrt{-s/m^2} = -{\rm i}\sqrt{s/m^2}$.  The results for
the massive singlet form factors can be found in Eq.~(\ref{eq::nhsing_s0}).
It is interesting to note that the axial-vector form factor
$F_{2,{\rm sing},l}^{a,f}$ develops $1/\sqrt{-s/m^2}$ terms, both at two and three
loops, which are absent in the massive case.  $F_{2,{\rm sing},l}^{a,f}$ also
has logarithmic contributions up to third order in the $(s/m^2)^0$ term whereas
$F_{2,{\rm sing},l}^{v,f}$ only has linear logarithms.
$F_{1,{\rm sing},l}^{a,f}$ starts to develop logarithms at order $\sqrt{-s/m^2}$
and the vector contribution $F_{1,{\rm sing},l}^{v,f}$ only at order $s/m^2$.

In the high-energy limit the expansions of the massless singlet form
factors are given by
\begin{align}
    &F_{1,{\rm sing},l}^{v,f}\Big|_{s\to-\infty} =
      \left(\frac{\alpha_s}{\pi}\right)^3\biggl[
        -0.334349
        +\frac{m^2 }{-s} \biggl(
          -0.00833333 l_s^5-0.116245 l_s^4-0.639133 l_s^3
          \nonumber \\ &
          -0.484656 l_s^2+13.7669 l_s+46.9765
        \biggr)
      \biggr]
      \,,
    \\
    &F_{2,{\rm sing},l}^{v,f}\Big|_{s\to-\infty} =
      \left(\frac{\alpha_s}{\pi}\right)^3
      \frac{m^2 }{-s} \biggl[
        -4.57974 l_s-7.34102
      \biggr]
      \,,
    \\
    &F_{1,{\rm sing},l}^{a,f}\Big|_{s\to-\infty} =
      \left(\frac{\alpha_s}{\pi}\right)^2 C_F T_F \biggl[
        -\frac{3}{4} l_s
        -\frac{9}{4}
        + \frac{\pi^2}{12}
        + \frac{m^2}{-s}\biggl\{
          \frac{1}{2} l_s^2
          + \frac{3}{2} l_s
          + \frac{1}{2}
          + \frac{\pi^2}{2}
        \biggr\}
      \biggr]
      \nonumber \\
    & + \left(\frac{\alpha_s}{\pi}\right)^3 C_F T_F \biggl[
      C_F \biggl(
        0.1875 l_s^3+0.919383 l_s^2+1.7663 l_s+0.520574
      \biggr)
      \nonumber \\ &
      + C_A \biggl(
        -0.6875 l_s^2-4.09631 l_s-6.70052
      \biggr)
      +  T_F n_h \biggl(
        0.25 l_s^2+1.03502 l_s+2.34309
      \biggr)
      \nonumber \\ &
      +  T_F n_l \biggl(
        0.25 l_s^2+1.03502 l_s+2.34309
      \biggr)
      + \frac{m^2}{-s} \biggl\{
        C_F \biggl(
          -0.0833333 l_s^4-0.529589 l_s^3
          \nonumber \\ &
          -5.50593 l_s^2-17.2508 l_s-32.6278
        \biggr)
        + C_A \biggl(
          -0.00208333 l_s^5-0.0751055 l_s^4+0.141666 l_s^3
          \nonumber \\ &
          +3.33973 l_s^2+15.0217 l_s+36.7552
        \biggr)
        +  T_F n_h \biggl(
          -0.166667 l_s^3-1.59058 l_s^2-3.29888 l_s
          \nonumber \\ &
          -7.38784
        \biggr)
        +  T_F n_l \biggl(
          -0.166667 l_s^3-1.09058 l_s^2-3.50612 l_s-6.4258
        \biggr)
      \biggr\}
    \biggr]
      \,,
    \\
    &F_{2,{\rm sing},l}^{a,f}\Big|_{s\to-\infty} =
      \left(\frac{\alpha_s}{\pi}\right)^2 C_F T_F
      \frac{m^2}{-s}
      \biggl[
        - \frac{1}{2} l_s^2
        - 3 l_s
        - 2
        - \frac{\pi^2}{3}
      \biggr]
      \nonumber\\
    & + \left(\frac{\alpha_s}{\pi}\right)^3 C_F T_F \frac{m^2}{-s} \biggl[
      C_F \biggl(
        0.104167 l_s^4+1. l_s^3+6.68117 l_s^2+22.4839 l_s+34.67
      \biggr)
      \nonumber \\ &
      + C_A \biggl(
        0.0208333 l_s^4-0.611111 l_s^3-7.80858 l_s^2-30.0535 l_s-49.2293
      \biggr)
      \nonumber \\ &
      +  T_F n_h \biggl(
        0.222222 l_s^3+2.05556 l_s^2+6.33333 l_s+8.54753
      \biggr)
      \nonumber \\ &
      +  T_F n_l \biggl(
        0.222222 l_s^3+2.05556 l_s^2+6.33333 l_s+10.147
      \biggr)
    \biggr]
      \,,
      \label{eq::nlsing_sinf}
  \end{align}
where $l_s = \log(m^2/(-s-i\delta))$ and we neglect terms which are suppressed by $m^4/s^2$.
In the leading term there are at most cubic logarithms which
are present for $F_{1,{\rm sing},l}^{a,f}$. In the subleading term
$l_s^5$ terms appear for $F_{1,{\rm sing},l}^{a,f}$
and $F_{1,{\rm sing},l}^{v,f}$ whereas the leading logarithm
for $F_{2,{\rm sing},l}^{a,f}$ is $l_s^4$ and
$F_{2,{\rm sing},l}^{v,f}$ only has linear subleading logarithms.
The corresponding results for the massive singlet form factors
can be found in Eq.~(\ref{eq::nhsing_sinf}).

For some of the coefficients in the high-energy expansion our method provides
a numerical accuracy of several ten digits for the massless and several hundred digits for the massive singlet contributions.
The accuracy for the massless contributions is of course limited by the numerical boundary conditions while we have analytic boundary conditions for the massive contributions.
The high accuracy allows for the
application of the {\tt PSLQ} algorithm~\cite{PSLQ} to reconstruct the
analytic expressions.
For example we find
\begin{align}
  F_{1,\mathrm{sing},l}^{a,f,(3)} \Big|_{m^0/(-s)^0,\,l_s^3} &= \frac{3 C_F^2 T_F}{16} , \nonumber\\
  F_{1,\mathrm{sing},l}^{a,f,(3)} \Big|_{m^0/(-s)^0,\,l_s^2} &= C_F^2 T_F \left( \frac{9}{8} - \frac{\pi^2}{48} \right) - \frac{11 C_A C_F T_F}{16} + \frac{C_F T_F^2 n_h}{4} + \frac{C_F T_F^2 n_l}{4}
\end{align}
for the leading and subleading logarithms of $F_{1,\mathrm{sing},l}^{a,f,(3)}$.

Close to threshold it is convenient to parameterize the
form factors in terms of the velocity of the produced quarks,
$\beta = \sqrt{1 - 4 m^2/s}$. We observe that the two-loop
and the three-loop vector corrections start with $\beta^0$. The three-loop
axial-vector form factors develop $1/\beta$ terms which read
\begin{align}
  &F_{1,{\rm sing},l}^{a,f}\Big|_{s\to4m^2} =
    \left(\frac{\alpha_s}{\pi}\right)^3 {C_F^2} T_F \frac{1}{\beta}
    \biggl[
      \bigl( 2.6544-0.4750 {\rm i} \bigr) l_{2 \beta }-3.4005-3.6946 {\rm i}
    \biggr]
    \,,
  \\
  &F_{2,{\rm sing},l}^{a,f}\Big|_{s\to4m^2} =
    \left(\frac{\alpha_s}{\pi}\right)^3 {C_F^2} T_F \frac{1}{\beta}
    \biggl[
      - \bigl(0.18704+1.18515 {\rm i} \bigr) l_{2 \beta } + 0.79281-0.18115 {\rm i}
    \biggr]
    \,,
    \label{eq::nlsing_sthr}
\end{align}
where $l_{2\beta}=\log(2\beta)$.
The $1/\beta$ terms for the massive singlet form factors are provided in
Eq.~(\ref{eq::nhsing_sthr}).



\subsection{\label{sub::plots}Finite form factors}

\begin{figure}[h]
  \begin{center}
    \begin{tabular}{ccc}
      \includegraphics[width=0.3\textwidth]{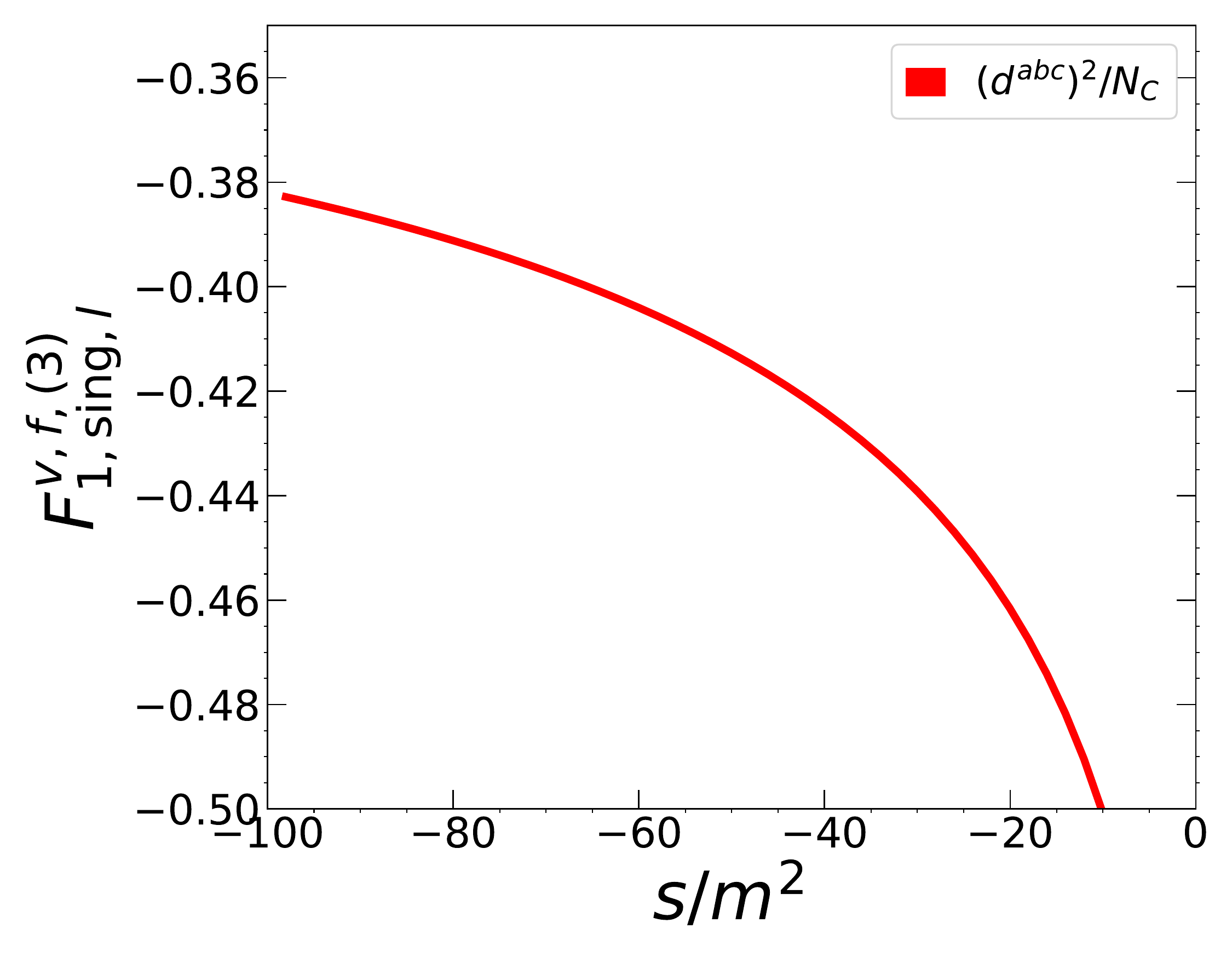} &
      \includegraphics[width=0.3\textwidth]{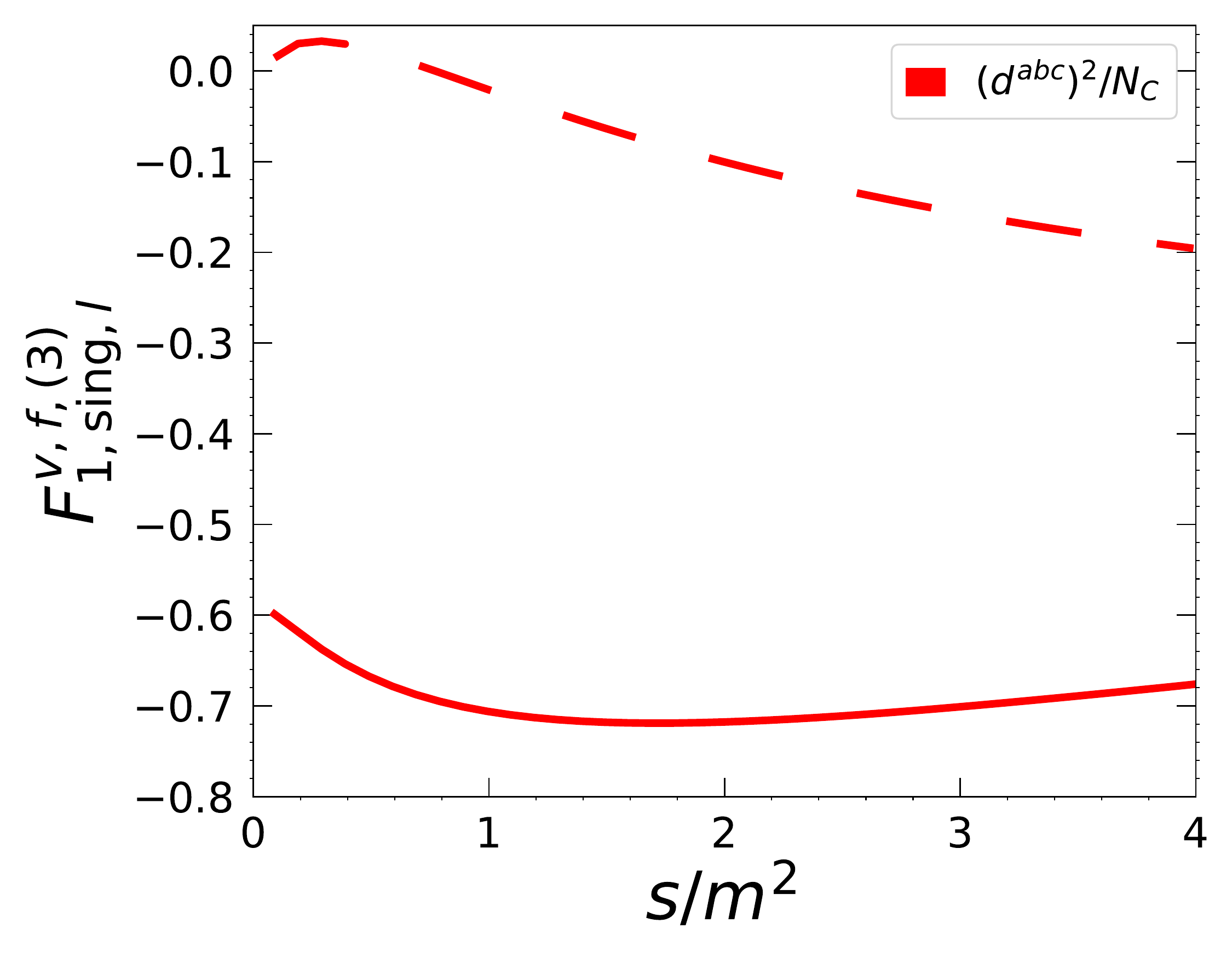} &
      \includegraphics[width=0.3\textwidth]{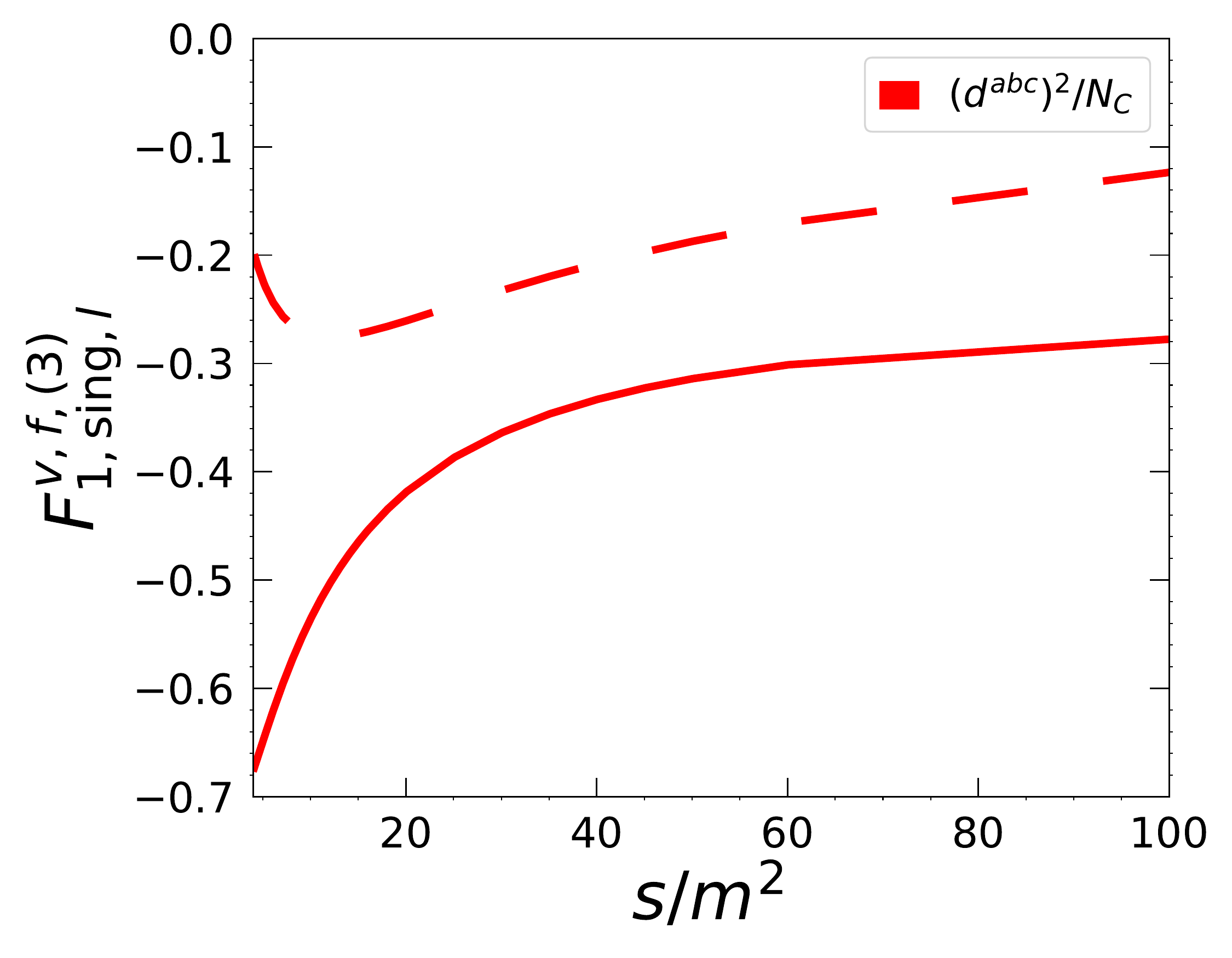}
      \\ (a) & (b) & (c) \\
      \includegraphics[width=0.3\textwidth]{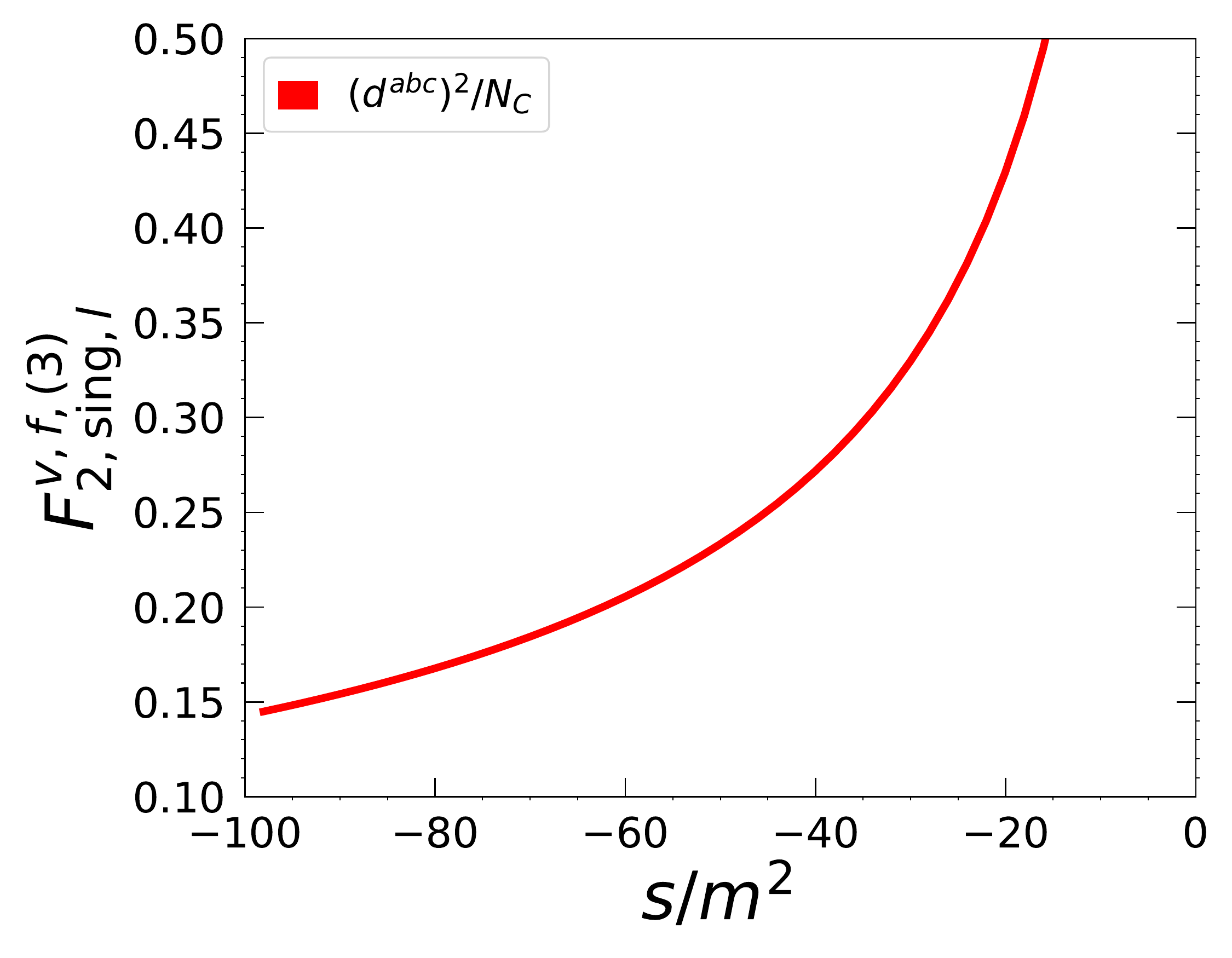} &
      \includegraphics[width=0.3\textwidth]{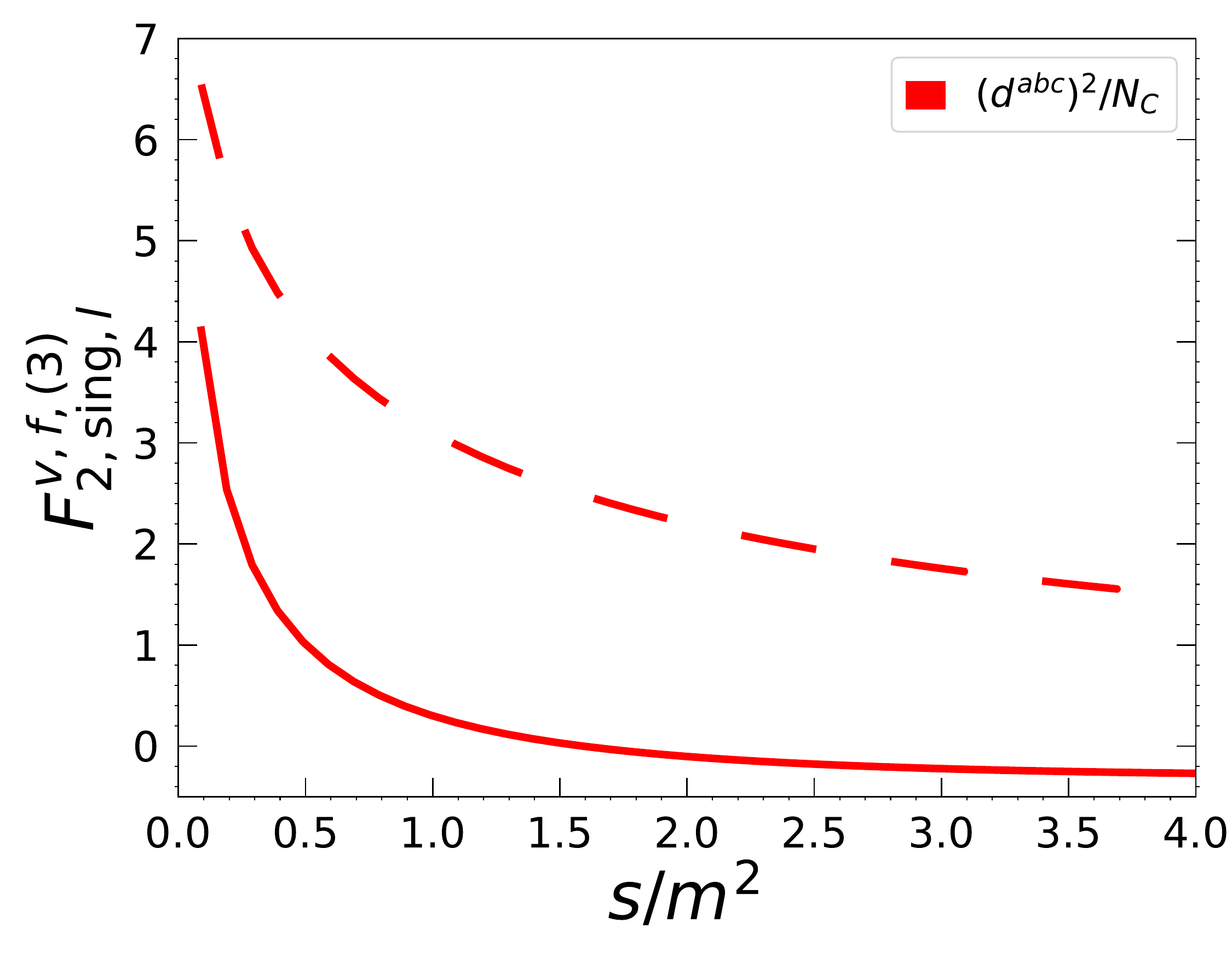} &
      \includegraphics[width=0.3\textwidth]{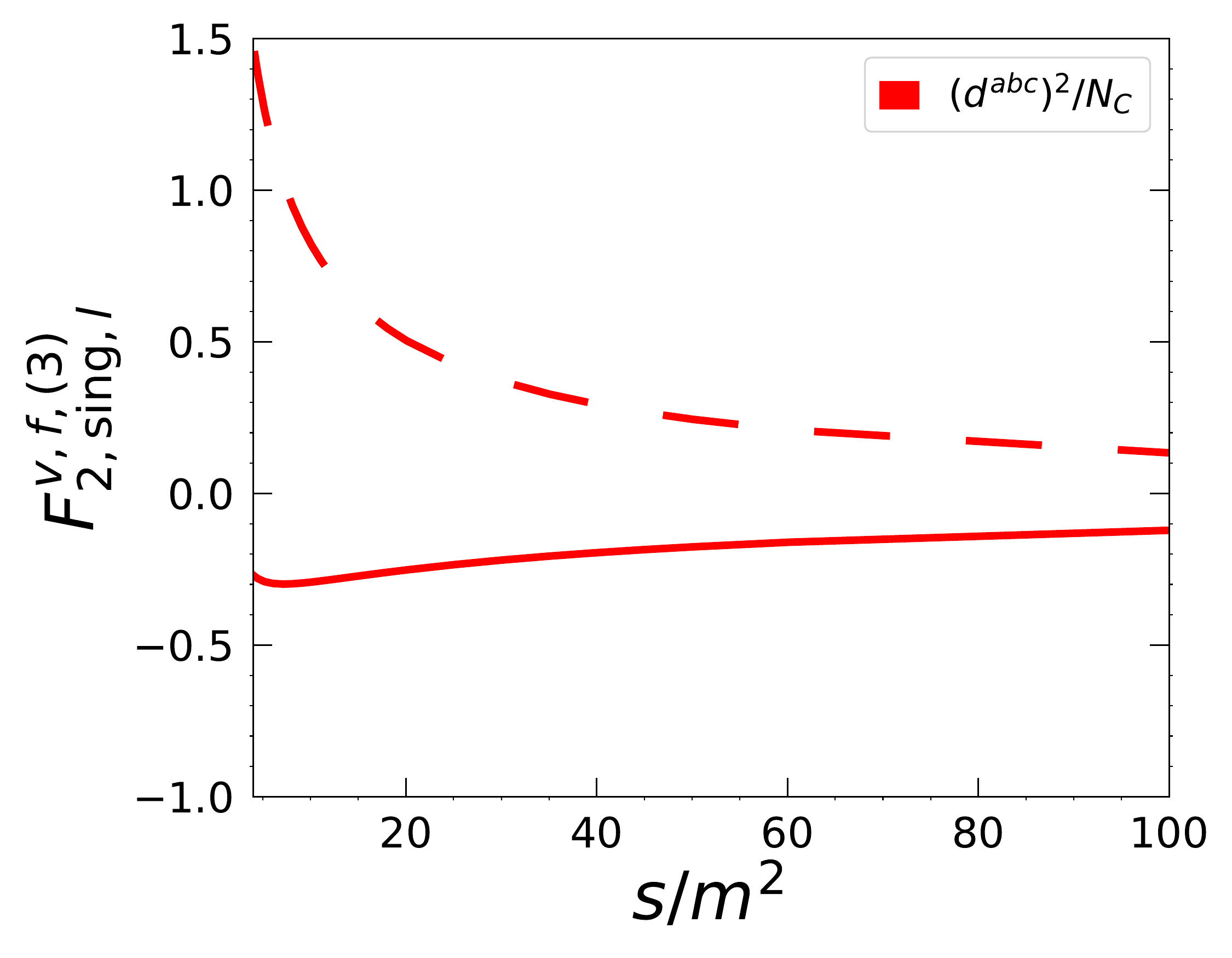}
      \\ (d) & (e) & (f) \\
      \includegraphics[width=0.3\textwidth]{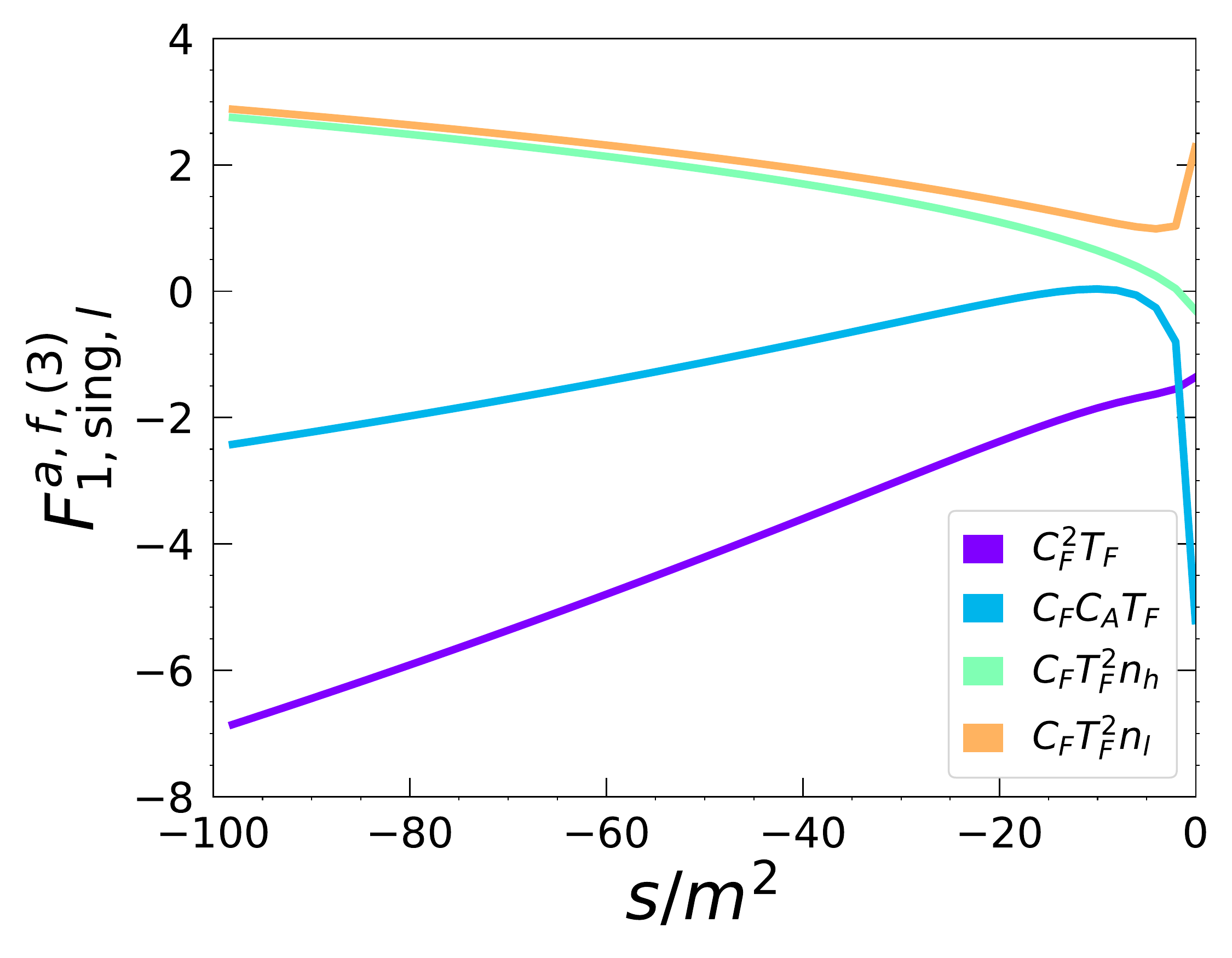} &
      \includegraphics[width=0.3\textwidth]{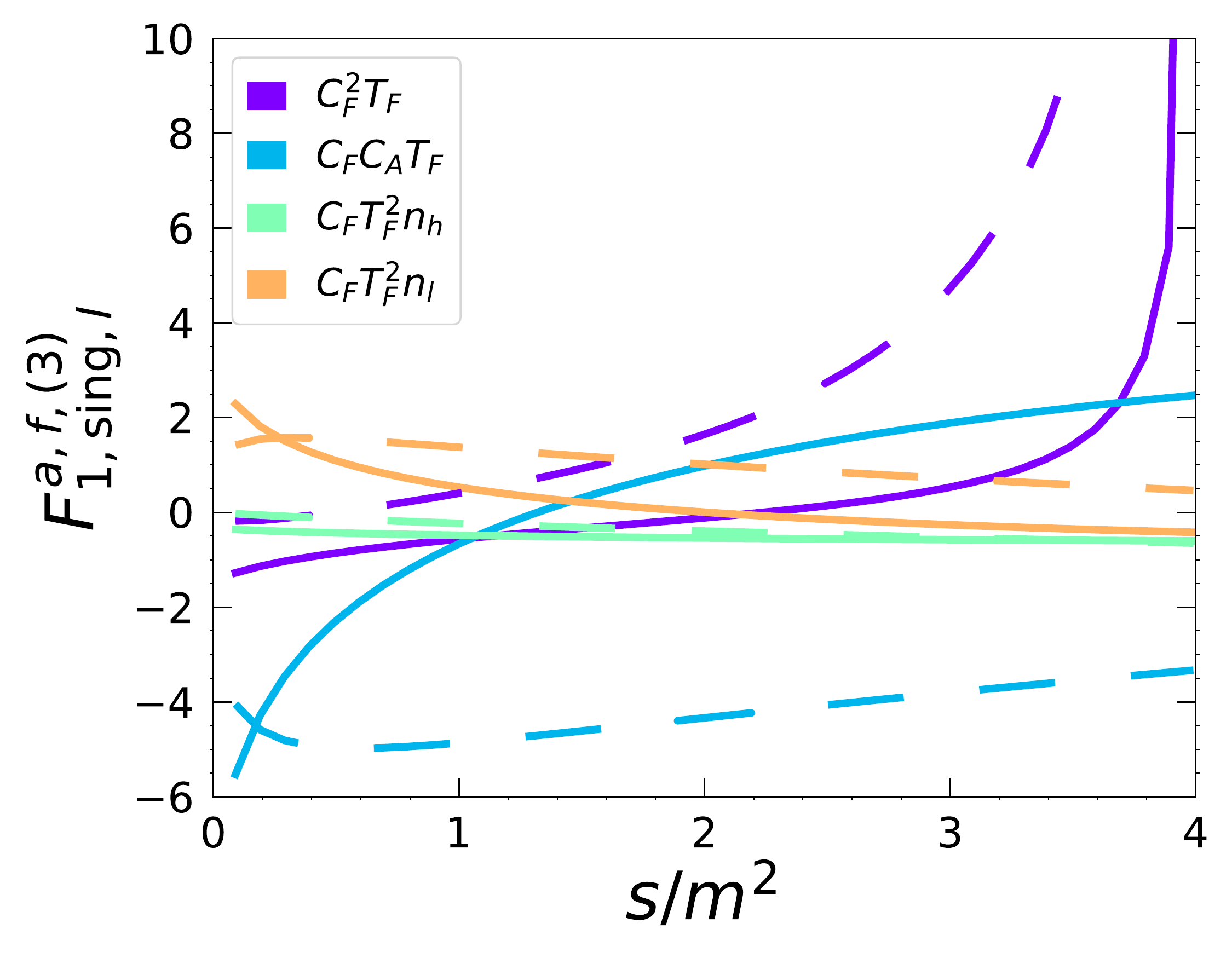} &
      \includegraphics[width=0.3\textwidth]{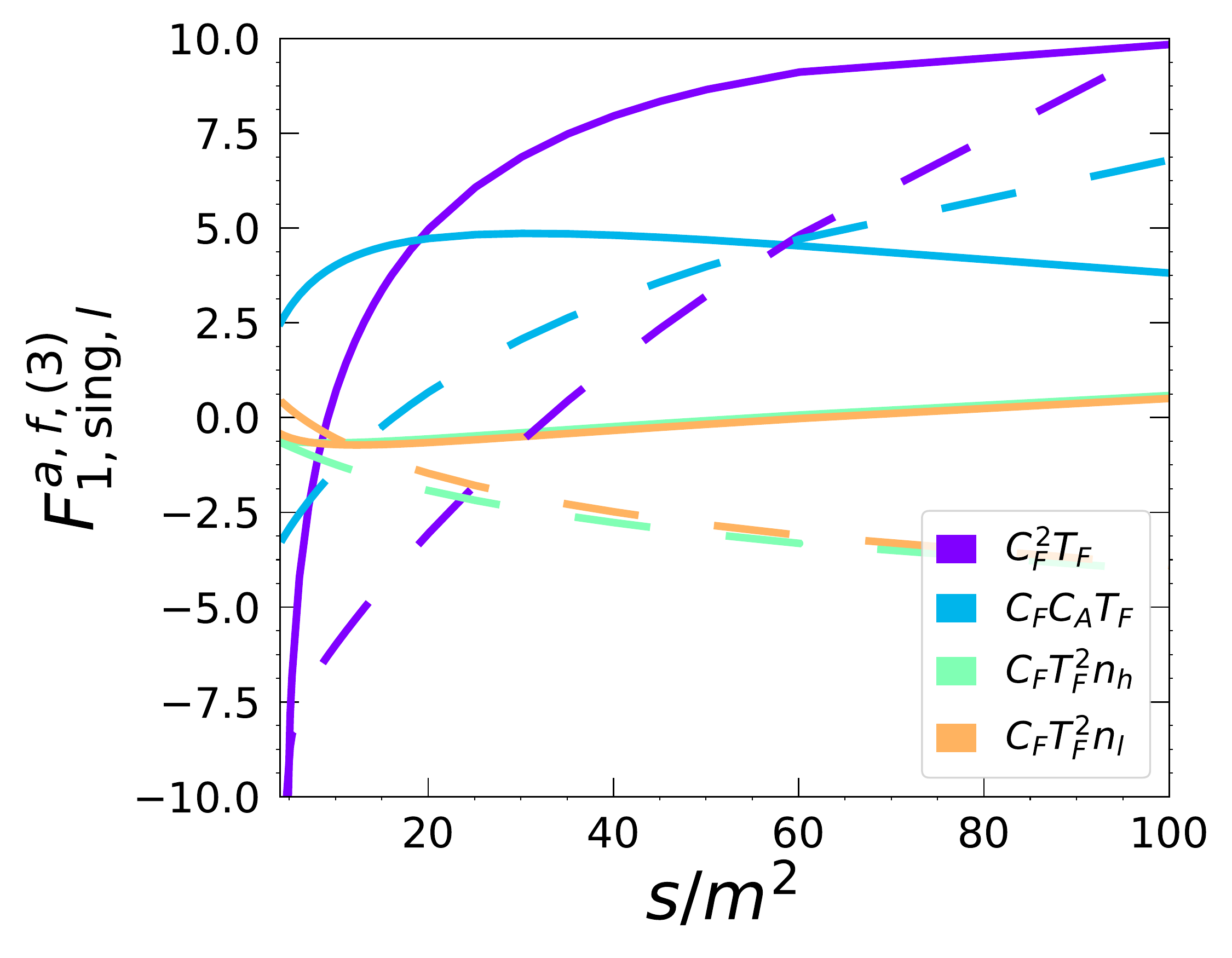}
      \\ (g) & (h) & (i) \\
      \includegraphics[width=0.3\textwidth]{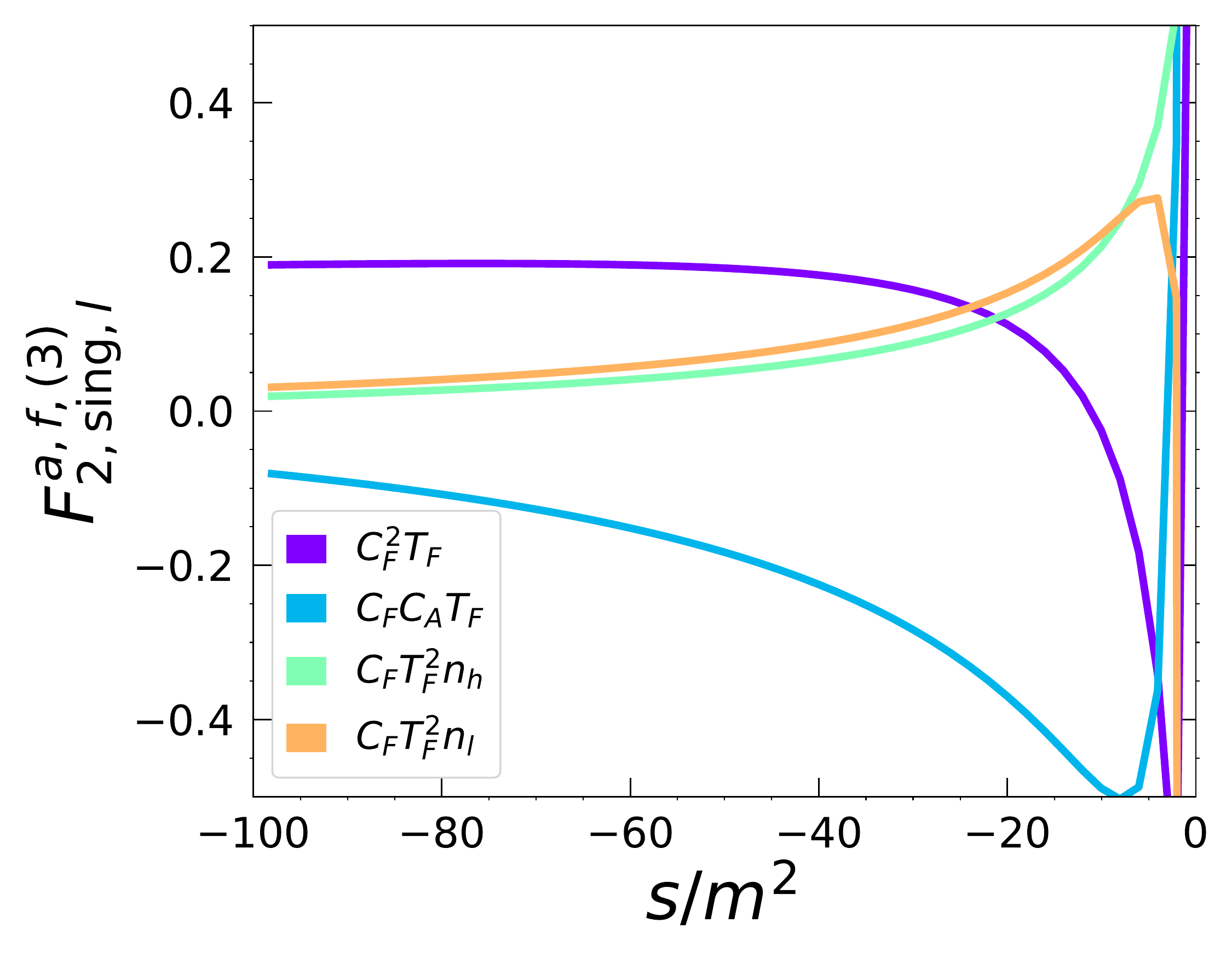} &
      \includegraphics[width=0.3\textwidth]{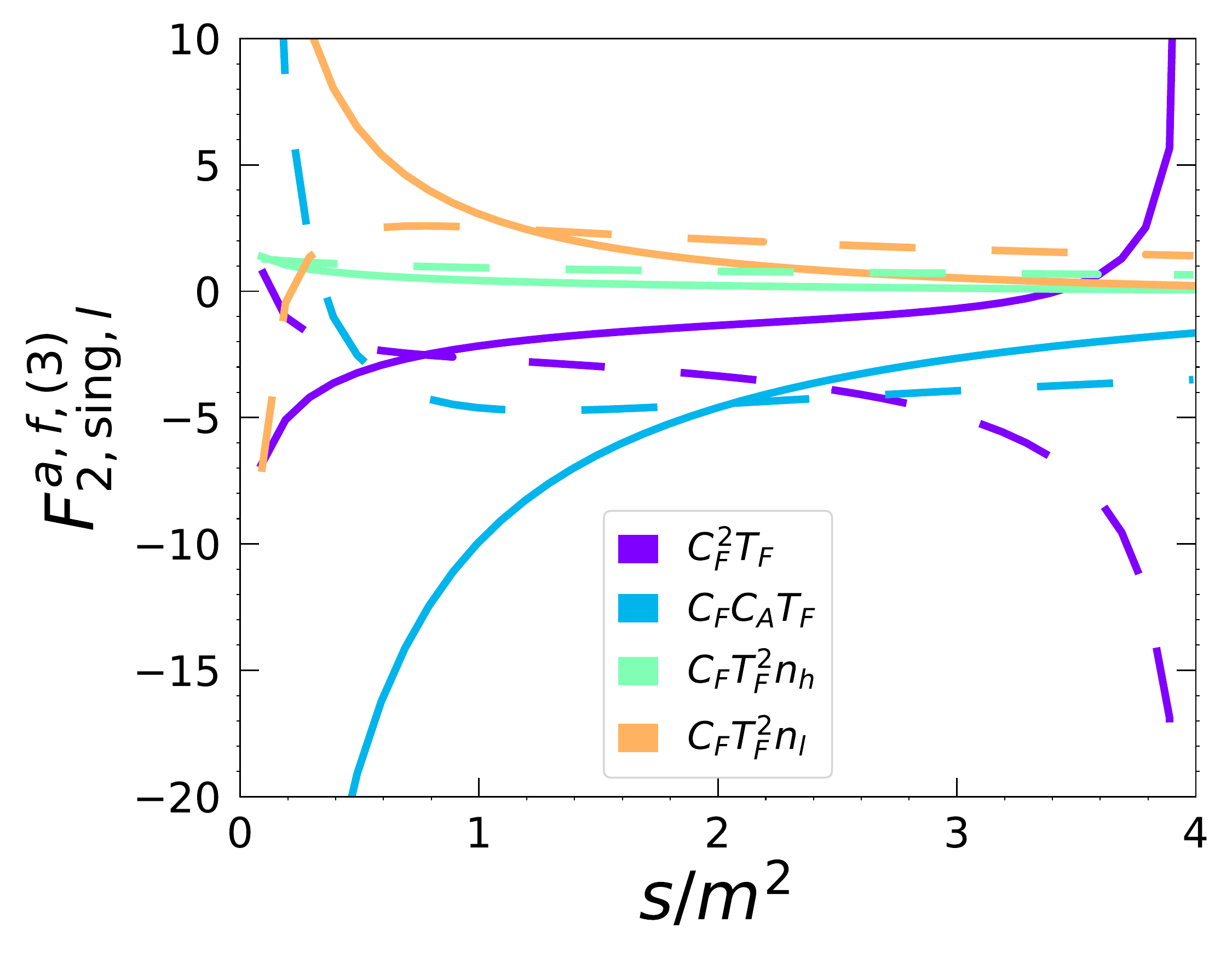} &
      \includegraphics[width=0.3\textwidth]{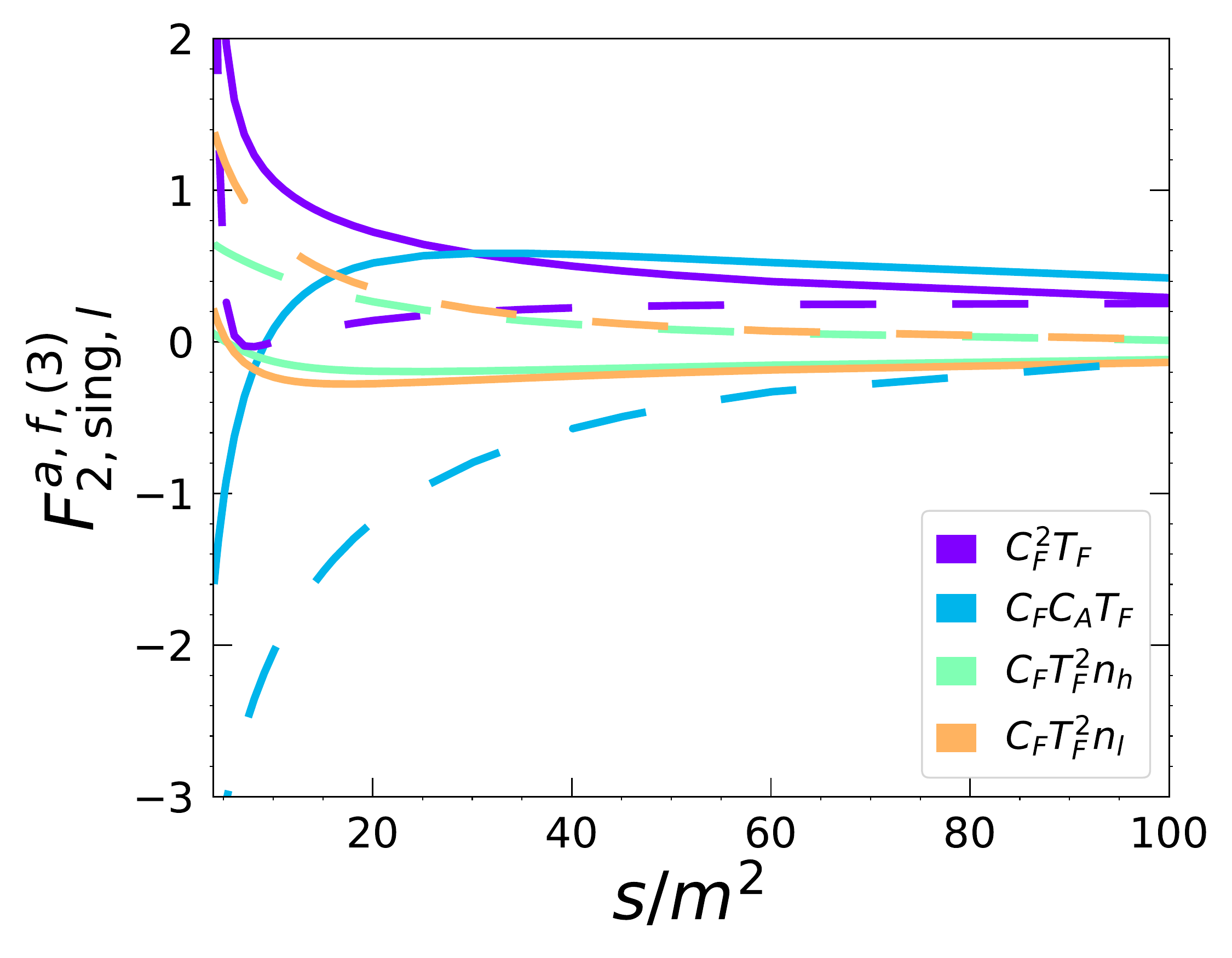}
      \\ (j) & (k) & (l)
    \end{tabular}
    \caption{\label{fig::FF_nlsing_1} Massless singlet form factors as a
      function of $s$ for $\mu^2=m^2$.}
  \end{center}
\end{figure}

In Fig.~\ref{fig::FF_nlsing_1} we show the finite results for the massless
singlet form factors as a function of $s$. We subdivide the energy
range into three parts corresponding to negative values of $s$, the region
between $s=0$ and the threshold $s=4m^2$, and above threshold and show results
for all individual colour factors.  We present both real (solid) and imaginary
(dashed) parts.  In contrast to the non-singlet contributions the singlet form
factors develop an imaginary part also for $s\in[0,4m^2]$ since there are cuts
through the gluons and in the massless singlet case in addition through the
massless quarks.  One recognizes the strong power-like divergences for $s\to0$
and $s\to 4m^2$ which are present in some of the form factors. On the other
hand, the logarithmic divergences in the various limits exhibit only a mild
behaviour.




\section{\label{sec::concl}Conclusions}

The main result of this paper are the three-loop corrections to the singlet form
factors with massive external quarks where external vector, axial-vector,
scalar, or pseudoscalar currents couple to a closed massless or massive quark loop.
This complements the non-singlet and massive singlet contributions presented
in Refs.~\cite{Fael:2022rgm,Fael:2022miw}.  We present our results in an
easy-to-use form as \texttt{Mathematica} package and {\tt Fortran} library
with high numerical precision in the whole $s$ range. Our method
allows for a systematic improvement of the accuracy if needed.

For the computation of the master integrals we use the ``expand and match''
approach which has been introduced in Ref.~\cite{Fael:2021kyg} and further
developed in Refs.~\cite{Fael:2022rgm,Fael:2022miw}. It provides analytic
expansions with numerical coefficients for all master integrals around
properly chosen kinematic points leading to precise results for the form
factors in the respective energy region. In the paper we provide expansions
around the physically interesting points $s\to0$, $s\to-\infty$ and
$s\to4m^2$. In some cases the numerical precision is sufficiently high such
that the analytic result of the expansion coefficients can be reconstructed.

In the course of our calculation we obtained a number of further interesting
results. For example, we have applied two different prescriptions for the
treatment of $\gamma_5$ to the non-singlet axial-vector and pseudoscalar form
factors and have checked by an explicit calculation that the final finite
expressions are identical. Furthermore, we have computed analytic two-loop
corrections to the massive pseudoscalar-gluon-heavy-quark vertex which
we needed to check the non-renormalization of the Adler-Bell-Jackiw anomaly at
three-loop order.



\section*{\label{sec::ack}Acknowledgments}

This research was supported by the Deutsche Forschungsgemeinschaft (DFG,
German Research Foundation) under grant 396021762 --- TRR 257 ``Particle
Physics Phenomenology after the Higgs Discovery'' and by the European
Research Council (ERC) under the European Union's Horizon 2020 research and
innovation programme grant agreement 101019620 (ERC Advanced Grant TOPUP).
The work of M.F.\ was supported in part by the European Union’s Horizon 2020
research and innovation program under the Marie Sk\l{}odowska-Curie
grant agreement No.\ 101065445 -- PHOBIDE.  The Feynman diagrams were drawn
with the help of Axodraw~\cite{Vermaseren:1994je} and
JaxoDraw~\cite{Binosi:2003yf}.



\begin{appendix}


\section{Projectors}
\label{app::projectors}

To project onto the form factors given by Eq.~(\ref{eq::Gamma}) we define the projectors
\begin{equation}
  F_i = \mathrm{Tr}[ P_i^\mu (\slashed{q_2} + m) \Gamma_{i,\mu} (\slashed{q_1} + m) ]
\end{equation}
with
\begin{align}
  P_{1}^{v,\mu} &= \frac{(4 m^2 - s) \gamma^\mu - 2(3-2\epsilon) m ( q_1^\mu + q_2^\mu )}
                  {4 (-1 + \ep) (4 m^2 - s)^2}
  , \nonumber \\
  P_{2}^{v,\mu} &= \frac{-m^2(4 m^2 - s) \gamma^\mu + (2m^2+s-\epsilon s) m (q_1^\mu + q_2^\mu )}
                  {(-1 + \ep) (4 m^2 - s)^2 s}
  , \nonumber \\
  P_{1}^{a,\mu} &= \frac{s \gamma^\mu \gamma_5 -2 m \gamma_5 ( q_1^\mu - q_2^\mu )}
                  {4 (-1 + \ep) (4 m^2 - s)s}
  , \nonumber \\
  P_{2}^{a,\mu} &= \frac{-s m^2 \gamma^\mu \gamma_5 + (6m^2-4\epsilon m^2-s+\epsilon s)m\gamma_5( q_1^\mu - q_2^\mu )}
                  {(-1 + \ep) (4 m^2 - s) s^2}
  , \nonumber \\
  P^{s,\mu} &= \frac{1}{2 m (4 m^2-s)}
  , \nonumber \\
  P^{p,\mu} &= - {\rm i} \frac{\gamma_5}{2 m s} 
  .
\end{align}
$\gamma_5$ is replaced using Eq.~(\ref{eq::gamma5}).



\section{\label{app::ren_const}Renormalization constants}

In our calculation there are several (non-standard) renormalization constants
which are needed due to the use of non-anti-commuting $\gamma_5$.
For convenience of the reader we reproduce all of them in the following.
We use the notion for ``singlet'' and ``non-singlet''
as defined in Section~\ref{sec::introduction}.

For the axial-vector contribution we need~\cite{Larin:1991tj,Larin:1993tq}
\begin{eqnarray}
  Z_{a,\rm S}^{\overline{\rm MS}} &=&
  1
  +\left(\frac{\alpha_s}{\pi}\right)^2 \frac{1}{\epsilon}
  \biggl(
        \frac{11}{24} C_A C_F
        +\frac{5}{24} C_F n_f T_F
  \biggr)
  +\left(\frac{\alpha_s}{\pi}\right)^3
  \biggl(
         \frac{1}{\epsilon ^2}
        \biggl[
                -\frac{121}{432} C_A^2 C_F
                \nonumber \\ &&
                -\frac{11}{432} C_A C_F T_F n_f
                +\frac{5}{108} C_F T_F^2 n_f^2
        \biggr]
        +\frac{1}{\epsilon }
        \biggl[
                -\frac{77 }{144} C_A C_F^2
                +\frac{1789}{2592} C_A^2 C_F
                \nonumber \\ &&
                +\frac{149 }{2592} C_A C_F T_F n_f
                -\frac{11}{144} C_F^2 T_F n_f
                +\frac{13}{648} C_F T_F^2 n_f^2
        \biggr]
  \biggr)
  +{\cal O}(\alpha_s^4)
  \,,\nonumber\\
  Z_{a,\rm NS}^{\overline{\rm MS}} &=&
  1
  +\left(\frac{\alpha_s}{\pi}\right)^2 \frac{1}{\epsilon}
  \biggl(
        \frac{11}{24}  C_A C_F
        -\frac{1}{6} C_F T_F n_f
  \biggr)
  +\left(\frac{\alpha_s}{\pi}\right)^3
  \biggl(
        \frac{1}{\epsilon ^2}
        \biggl[
                -\frac{121}{432} C_A^2 C_F
                \nonumber \\ &&
                +\frac{11}{54} C_A C_F T_F n_f
                -\frac{1}{27} C_F T_F^2 n_f^2
        \biggr]
        +\frac{1}{\epsilon }
        \biggl[
                -\frac{77}{144} C_A C_F^2
                -\frac{26}{81} C_A C_F T_F  n_f
                \nonumber \\ &&
                +\frac{1789}{2592}  C_A^2 C_F
                +\frac{1}{9} C_F^2 T_F  n_f
                +\frac{1}{162} C_F T_F^2  n_f^2
        \biggr]
  \biggr)
  +{\cal O}(\alpha_s^4)
  \,,\nonumber\\
  Z_{a,\rm S}^{\rm fin} &=&
  1
  -\frac{\alpha_s}{\pi} C_F
  +\left(\frac{\alpha_s}{\pi}\right)^2
  \biggl(
        -\frac{107}{144} C_A C_F
        +\frac{11}{8} C_F^2
        +\frac{31}{144} C_F T_F n_f
  \biggr)
  \nonumber \\ &&
  + \left(\frac{\alpha_s}{\pi}\right)^3
  \biggl(
    \biggl[
        \frac{55 \zeta _3}{48}
        -\frac{133}{2592}
    \biggr] C_A C_F T_F n_f
    +\biggl[
        \frac{2917}{864}
        -\frac{5 \zeta _3}{2}
     \biggr] C_A C_F^2
     \nonumber \\ &&
   +\biggl[
        \frac{7 \zeta _3}{8}
        -\frac{2147}{1728}
    \biggr] C_A^2 C_F
   +\biggl[
        \frac{497}{1728}
        -\frac{13 \zeta _3}{12}
     \biggr] C_F^2 T_F n_f
   +\frac{79}{324} C_F n_f^2 T_F^2
   \nonumber \\ &&
   +\biggl[
        \frac{3 \zeta_3}{2}
        -\frac{185}{96}
     \biggr] C_F^3
  \biggr)
  +{\cal O}(\alpha_s^4)
  \,,\nonumber\\
  Z_{a,\rm NS}^{\rm fin} &=&
  1
  -\frac{\alpha_s}{\pi} C_F
  +\left(\frac{\alpha_s}{\pi}\right)^2
  \biggl(
        -\frac{107}{144} C_A C_F
        +\frac{11}{8} C_F^2
        +\frac{1}{36} C_F T_F n_f
  \biggr)
  \nonumber \\ &&
  +\left(\frac{\alpha_s}{\pi}\right)^3
  \biggl(
        \biggl[
                \frac{2917}{864}
                -\frac{5}{2} \zeta_3
        \biggr] C_A C_F^2
        +\biggl[
                \frac{89}{648}
                +\frac{\zeta_3}{3}
        \biggr] C_A C_F T_F n_f
        +\biggl[
                -\frac{2147}{1728}
                \nonumber \\ &&
                +\frac{7}{8} \zeta_3
        \biggr] C_A^2 C_F
        +\biggl[
                -\frac{185}{96}
                +\frac{3}{2} \zeta_3
        \biggr] C_F^3
        +\biggl[
                -\frac{31 }{432}
                -\frac{1}{3} \zeta_3
        \biggr] C_F^2 T_F n_f
        \nonumber \\ &&
        +\frac{13}{324} C_F T_F^2  n_f^2
  \biggr)
  +{\cal O}(\alpha_s^4)
  \,.
  \label{eq::Za}
\end{eqnarray}
$Z_{a,\rm S}^{\overline{\rm MS}}$ is taken from Eq.~(19) of the arXiv
version of Ref.~\cite{Larin:1993tq} and $Z_{a,\rm S}^{\rm fin}$ from
Eq.~(5.4) of Ref.~\cite{Ahmed:2021spj}.
$Z_{a,\rm NS}^{\overline{\rm MS}}$ and $Z_{a,\rm NS}^{\rm fin}$ are
obtained from Eqs.~(8) and~(11) of Ref.~\cite{Larin:1991tj}.

For the pseudoscalar contribution we have
\begin{eqnarray}
  Z_{p}^{\overline{\rm MS}} &=&
  1
  - \frac{\alpha_s}{\pi} \frac{3 C_F}{4 \epsilon }
  +\left(\frac{\alpha_s}{\pi}\right)^2
  \biggl(
        \frac{1}{\epsilon^2}
        \biggl[
            \frac{11}{32} C_A C_F
            +\frac{9}{32} C_F^2
            -\frac{1}{8} C_F T_F n_f
        \biggr]
        + \frac{1}{\epsilon }
        \biggl[
            \frac{79}{192} C_A C_F
            \nonumber \\ &&
            -\frac{3}{64} C_F^2
            -\frac{11}{48} C_F T_F n_f
        \biggr]
  \biggr)
  +\left(\frac{\alpha_s}{\pi}\right)^3
  \biggl(
        \frac{1}{\epsilon ^3}
        \biggl[
                -\frac{33}{128}  C_A C_F^2
                +\frac{11}{72} C_A C_F T_F n_f
                \nonumber \\ &&
                -\frac{121}{576} C_A^2 C_F
                -\frac{9}{128} C_F^3
                +\frac{3}{32} C_F^2 T_F n_f
                -\frac{1}{36} C_F T_F^2 n_f^2
        \biggr]
        +\frac{1}{\epsilon ^2}
        \biggl[
                -\frac{215}{768} C_A C_F^2
                \nonumber \\ &&
                +\frac{55}{432} C_A C_F T_F n_f
                -\frac{257}{3456} C_A^2 C_F
                +\frac{9}{256} C_F^3
                +\frac{19}{192} C_F^2 T_F n_f
                -\frac{11}{216} C_F T_F^2 n_f^2
        \biggr]
        \nonumber \\ &&
        +\frac{1}{\epsilon }
        \biggl[
                \frac{3203}{2304} C_A C_F^2
                +\bigl(
                        -\frac{29}{144}
                        +\frac{1}{4} \zeta_3
                \bigr) C_A C_F n_f T_F
                -\frac{599}{6912} C_A^2 C_F
                -\frac{43}{128} C_F^3
                \nonumber \\ &&
                +\bigl(
                        -\frac{107}{288}
                        -\frac{1}{4} \zeta_3
                \bigr) C_F^2 T_F n_f
                +\frac{17}{432} C_F T_F^2 n_f^2
        \biggr]
  \biggr)
  +{\cal O}(\alpha_s^4)
  \,,\nonumber\\
  Z_{p}^{\rm fin} &=&
  1
  -2 \frac{\alpha_s}{\pi} C_F
  +\left(\frac{\alpha_s}{\pi}\right)^2
  \biggl(
        \frac{1}{72} C_A C_F
        +\frac{1}{18} C_F T_F n_f
  \biggr)
  +\left(\frac{\alpha_s}{\pi}\right)^3
  \biggl(
        \biggl[
                -\frac{25}{54}
                \nonumber \\ &&
                +\frac{19}{2} \zeta_3
        \biggr] C_A C_F^2
        +\biggl[
                \frac{107}{324}
                +\frac{2}{3} \zeta_3
        \biggr] C_A C_F T_F n_f
        +\biggl[
                -\frac{479}{864}
                -\frac{13}{4} \zeta_3
        \biggr] C_A^2 C_F
        \nonumber \\ &&
        +\biggl[
                \frac{19}{12}
                -6 \zeta_3
        \biggr] C_F^3
        +\biggl[
                -\frac{145}{216}
                -\frac{2}{3} \zeta_3
        \biggr] C_F^2 T_F n_f
        +\frac{13}{162} C_F T_F^2 n_f^2
  \biggr)
  +{\cal O}(\alpha_s^4)\,,
  \label{eq::Zp}
\end{eqnarray}
which corresponds to Eqs.~(13) and~(15) of the arXiv version of
Ref.~\cite{Larin:1993tq}.

Note that $Z_{p}^{\overline{\rm MS}}$ is the renormalization constant
associated to the factor $m$ on the r.h.s.\ of $j^p$ in
Eq.~(\ref{eq::currents}). It replaces the usual $\overline{\rm MS}$
renormalization constant $Z_{m}^{\overline{\rm MS}}$ for the heavy-quark mass
which is used for anti-commuting $\gamma_5$, e.g.\ for the non-singlet
contribution.  In case only the singlet contribution is considered only the
${\cal O}(\alpha_s)$ terms are needed from $Z_{p}^{\overline{\rm MS}}$.  Up to
this order $Z_{p}^{\overline{\rm MS}}$ agrees with
$Z_{m}^{\overline{\rm MS}}$.  Note that in Ref.~\cite{Fael:2022miw} the factor
$m$ in Eq.~(\ref{eq::currents}) has been renormalized on-shell.

We refrain from providing explicit expressions for the wave function, strong
coupling constant, and heavy-quark mass renormalization constants, which have
already been used in Refs.~\cite{Fael:2022rgm,Fael:2022miw}.



\section{\label{app::resFF} Results for the massive singlet contribution}

In this Section we collect the expansions around $s=0$, around the threshold $s=4m^2$, and in the high-energy limit for the massive singlet contributions in the spirit of those for the massless singlet contributions shown in Subsection~\ref{sub::expansions}.
We also show plots over the whole range of $s$.

In the limit $s\to0$ we obtain for the six form factors
\begin{align}
    &F_{1,{\rm sing},h}^{v,f}\Big|_{s\to0} = 0
      \,,
    \\
    &F_{2,{\rm sing},h}^{v,f}\Big|_{s\to0} =
    \left(\frac{\alpha_s}{\pi}\right)^3 \frac{d_{abc}d^{abc}}{n_c}
    \left[ 0.371005 \right]
      \,,
    \\
    &F_{1,{\rm sing},h}^{a,f}\Big|_{s\to0} =
    \left(\frac{\alpha_s}{\pi}\right)^2 C_F T_F \biggl[
      - \frac{19}{12}
      + \frac{2}{9} \pi^2
    \biggr]
    \nonumber\\
  & + \left(\frac{\alpha_s}{\pi}\right)^3 C_F T_F
    \biggl[
      0.79884 C_A
      -4.3999 C_F
      +0.66880 T_F n_h
      +1.2009 T_F n_l
    \biggr]
      \,,
    \\
    &F_{2,{\rm sing},h}^{a,f}\Big|_{s\to0} =
    \left(\frac{\alpha_s}{\pi}\right)^2 C_F T_F
    \biggl[
      \frac{2}{3}
      + \frac{1}{90} \pi^2
      - \frac{1}{24} \pi^2 {\chi}
    \biggr]
    \nonumber\\
    & + \left(\frac{\alpha_s}{\pi}\right)^3 C_F T_F
    \biggl[
       2.4737 C_A
      +3.1457 C_F
      +0.36848 T_F n_h
      -0.73194 T_F n_l
      \nonumber \\ &
      + {\chi}
      \biggl\{
        C_A \bigl(
          1.3022 \ln(\chi)
          -2.2422
        \bigr)
        -1.8506 C_F
        + T_F n_l \bigl(
          -0.54831 \ln(\chi)
          +0.86816
        \bigr)
      \biggr\}
    \biggr]
      \,,
    \\
    &F_{{\rm sing},h}^{s,f}\Big|_{s\to0} =
    \left(\frac{\alpha_s}{\pi}\right)^2 C_F T_F
    \biggl[
      - 2
      + \frac{1}{3} \pi^2
      - \frac{1}{12} \pi^2 {\chi}
    \biggr]
    \nonumber\\
    & + \left(\frac{\alpha_s}{\pi}\right)^3 C_F T_F
    \biggl[
       7.2423 C_A
      -2.1288 C_F
      +0.47311 T_F n_h
      -1.4332 T_F n_l
      \nonumber \\ &
      + \chi
      \left(
        C_A \bigl(
          3.0157 \ln(\chi)
          -4.6557
        \bigr)
        -0.20562 C_F
        + T_F n_l \bigl(
          -1.0966 \ln(\chi)
          +1.4622
        \bigr)
      \right)
    \biggr]
    \,,
    \\
    &F_{{\rm sing},h}^{p,f}\Big|_{s\to0} =
    \left(\frac{\alpha_s}{\pi}\right)^2 C_F T_F
    \biggl[
      \frac{1}{6}
      + \frac{2}{9} \pi^2
      - \frac{1}{8} \pi^2 {\chi}
    \biggr]
    \nonumber\\
    & + \left(\frac{\alpha_s}{\pi}\right)^3 C_F T_F
    \biggl[
      +9.8173 C_A
      -0.55128 C_F
      +0.99399 T_F n_h
      -2.4788 T_F n_l
      \nonumber \\ &
      + {\chi}
      \left(
        C_A \bigl(
          3.9067 \ln(\chi)
          -7.6518
        \bigr)
        -1.8506 C_F
        + T_F n_l \bigl(
          -1.6449 \ln(\chi)
          +2.6045
        \bigr)
        \right)
    \biggr]\,,
    \label{eq::nhsing_s0}
  \end{align}
where again terms of ${\cal O}(\chi^2)$ have been neglected.
Logarithmic contributions appear only at order
$\chi=\sqrt{-s/m^2}$ and thus the limit $s=0$ exists for all form factors.

In the high-energy limit we have
\begin{align}
    &F_{1,{\rm sing},h}^{v,f}\Big|_{s\to-\infty} =
      \left(\frac{\alpha_s}{\pi}\right)^3 \frac{d_{abc}d^{abc}}{n_c}
      \biggl[
        -0.33435 + \frac{m^2}{-s} \big(-0.0083333 l_s^5-0.11624 l_s^4
      \nonumber \\
        & -0.63913 l_s^3-0.83260
     l_s^2+15.749 l_s+66.917\big)
      \biggr]
      \,,
    \\
    &F_{2,{\rm sing},h}^{v,f}\Big|_{s\to-\infty} =
      \left(\frac{\alpha_s}{\pi}\right)^3 \frac{d_{abc}d^{abc}}{n_c}
      \biggl[
        \frac{m^2}{-s} \left(-4.5797 l_s-7.3410\right)
      \biggr]
      \,,
    \\
    &F_{1,{\rm sing},h}^{a,f}\Big|_{s\to-\infty} =
      \left(\frac{\alpha_s}{\pi}\right)^2 C_F T_F \biggl[
        - \frac{3}{4} l_s \!
        - \frac{9}{4}
        + \frac{\pi^2}{12}
        + \frac{m^2}{-s} \biggl\{
          \frac{1}{2} l_s^2
          + \bigl(
            \frac{3}{2}
            - \frac{\pi^2}{3}
          \bigr) l_s \!
          + \frac{1}{2}
          - \frac{\pi^2}{2}
          + 4 \zeta_3
        \biggr\}
      \biggr]
      \nonumber\\
    & + \left(\frac{\alpha_s}{\pi}\right)^3 C_F T_F
      \biggl[
        C_F
        \biggl(
           0.18750 l_s^3
          +0.91938 l_s^2
          +1.7663 l_s
          +0.52057
        \biggr)
        \nonumber \\ &
        + C_A
        \biggl(
          -0.68750 l_s^2
          -4.0963 l_s
          -6.7005
        \biggr)
        +  T_F n_h
        \biggl(
           0.25000 l_s^2
          +1.0350 l_s
          +2.3431
        \biggr)
        \nonumber \\ &
        +  T_F n_l
        \biggl(
           0.25000 l_s^2
          +1.0350 l_s
          +2.3431
        \biggr)
        + \frac{m^2}{-s}
        \biggl\{
          C_F
          \biggl(
              -0.083333 l_s^4
              +0.29288 l_s^3
              \nonumber \\ &
              -3.2283 l_s^2
              -0.27409 l_s
              +2.5837
          \biggr)
          + C_A
          \biggl(
              -0.0020833 l_s^5
              -0.075106 l_s^4
              +0.14167 l_s^3
              \nonumber \\ &
              -0.48000 l_s^2
              -6.8168 l_s
              +9.9023
          \biggr)
          +  T_F n_h
          \biggl(
              -0.16667 l_s^3
              +0.054357 l_s^2
              \nonumber \\&
              +3.1403 l_s
              +6.4236
          \biggr)
          +  T_F n_l
          \biggl(
              -0.16667 l_s^3
              +0.55436 l_s^2
              +2.9330 l_s
              +5.7888
          \biggr)
        \biggr\}
      \biggr]
      \,,
    \\
    &F_{2,{\rm sing},h}^{a,f}\Big|_{s\to-\infty} =
      \left(\frac{\alpha_s}{\pi}\right)^2 C_F T_F \frac{m^2}{-s}
      \biggl[
        - \frac{1}{2} l_s^2
        - 3 l_s
        - 2
        - \frac{\pi^2}{3}
      \biggr]
      \nonumber\\
    & + \left(\frac{\alpha_s}{\pi}\right)^3 C_F T_F \frac{m^2}{-s}
      \biggl[
        C_F
        \biggl(
            0.10417 l_s^4
            +1.0000 l_s^3
            +6.6812 l_s^2
            +22.484 l_s
            +34.670
        \biggr)
        \nonumber \\ &
        + C_A
        \biggl(
            0.020833 l_s^4
            -0.61111 l_s^3
            -7.8086 l_s^2
            -30.054 l_s
            -49.229
        \biggr)
        \nonumber \\ &
        +  T_F n_h
        \biggl(
            0.22222 l_s^3
            +2.0556 l_s^2
            +6.3333 l_s
            +8.5475
        \biggr)
        \nonumber \\ &
        +  T_F n_l
        \biggl(
            0.22222 l_s^3
            +2.0556 l_s^2
            +6.3333 l_s
            +10.147
        \biggr)
      \biggr]
      \,,
    \\
    &F_{{\rm sing},h}^{s,f}\Big|_{s\to-\infty} =
      \left(\frac{\alpha_s}{\pi}\right)^2 C_F T_F
      \biggl[
        - \frac{1}{48} \l_s^4
        + \bigl( 1-\frac{\pi^2}{12} \bigr) l_s^2
        + \bigl( 4 - 3 \zeta_3 \bigr) l_s
        + \frac{2 \pi^2}{3}
        - \frac{\pi^4}{45}
      \biggr]
      \nonumber\\
    & + \left(\frac{\alpha_s}{\pi}\right)^3 C_F T_F \frac{m^2}{-s}
    \biggl[
      C_F
      \biggl(
          0.0041667 l_s^6
          -0.0062500 l_s^5
          +0.062124 l_s^4
          +1.0817 l_s^3
          +4.8496 l_s^2
          \nonumber \\ &
          +32.500 l_s
          +58.066
      \biggr)
      + C_A
      \biggl(
          0.0010417 l_s^6
          -0.022917 l_s^5
          -0.14492 l_s^4
          +0.46401 l_s^3
          \nonumber \\ &
          +3.6270 l_s^2
          +9.0468 l_s
          +16.307
      \biggr)
      +  T_F n_h
      \biggl(
          0.0083333 l_s^5
          +0.023148 l_s^4
          -0.078904 l_s^3
          \nonumber \\ &
          -0.31219 l_s^2
          -2.1741 l_s
          -1.2446
      \biggr)
      +  T_F n_l
      \biggl(
          0.0083333 l_s^5
          +0.023148 l_s^4
          -0.078904 l_s^3
          \nonumber \\ &
          -0.31219 l_s^2
          -3.8614 l_s
          -6.4797
      \biggr)
    \biggr]
      \,,
    \\
    &F_{{\rm sing},h}^{p,f}\Big|_{s\to-\infty} =
      \left(\frac{\alpha_s}{\pi}\right)^2 C_F T_F
      \biggl[
        - \frac{1}{48} \l_s^4
        + \bigl( 1-\frac{\pi^2}{12} \bigr) l_s^2
        - 3 \zeta_3 l_s
        + \frac{\pi^2}{3}
        - \frac{\pi^4}{45}
      \biggr]
      \nonumber\\
    & + \left(\frac{\alpha_s}{\pi}\right)^3 C_F T_F \frac{m^2}{-s}
      \biggl[
        C_F
        \biggl(
            0.0041667 l_s^6
            -0.0062500 l_s^5
            +0.16629 l_s^4
            +1.5817 l_s^3
            +1.9782 l_s^2
            \nonumber \\ &
            +31.884 l_s
            +61.904
        \biggr)
        + C_A
        \biggl(
            0.0010417 l_s^6
            -0.022917 l_s^5
            -0.12408 l_s^4
            -0.14710 l_s^3
            \nonumber \\ &
            -6.3791 l_s^2
            -25.947 l_s
            -33.440
        \biggr)
        +  T_F n_h
        \biggl(
            0.0083333 l_s^5
            +0.023148 l_s^4
            +0.14332 l_s^3
            \nonumber \\ &
            +2.2434 l_s^2
            +4.0771 l_s
            +3.8620
        \biggr)
        +  T_F n_l
        \biggl(
            0.0083333 l_s^5
            +0.023148 l_s^4
            +0.14332 l_s^3
            \nonumber \\ &
            +2.2434 l_s^2
            +2.3898 l_s
            -0.20928
        \biggr)
      \biggr]
      \,,
      \label{eq::nhsing_sinf}
  \end{align}
where terms of order $m^4/s^2$ have been dropped.
As expected, the scalar and pseudoscalar form factors start at order
$m^2/s$ where both develop leading $l_s^6$ terms.
The vector and axial-vector form factors show a similar behavior as
in the massless case discussed around Eq.~(\ref{eq::nlsing_sinf}).

At threshold the three-loop axial-vector, scalar, and pseudoscalar form
factors develop $1/\beta$ poles which are given by
\begin{align}
    &F_{1,{\rm sing},h}^{a,f}\Big|_{s\to4m^2} =
      \left(\frac{\alpha_s}{\pi}\right)^3 C_F^2 T_F \frac{1}{\beta}
      \left[
        0.062172 {\rm i} l_{2 \beta }+ 0.097660 - 0.062172 {\rm i}
      \right]
      \,,
    \\
    &F_{2,{\rm sing},h}^{a,f}\Big|_{s\to4m^2} =
      \left(\frac{\alpha_s}{\pi}\right)^3 C_F^2 T_F \frac{1}{\beta}
      \left[
        - \bigl(3.6207-1.9240  {\rm i} \bigr) l_{2 \beta }
        + 3.0223+5.7495 {\rm i}
      \right]
      \,,
    \\
    &F_{{\rm sing},h}^{s,f}\Big|_{s\to4m^2} =
      \left(\frac{\alpha_s}{\pi}\right)^3 C_F^2 T_F \frac{1}{\beta}
      \left[
        - \bigl( 2.4674-0.8781 {\rm i} \bigr) l_{2 \beta }
        + 3.8466+2.9977 {\rm i}
      \right]
      \,,
    \\
    &F_{{\rm sing},h}^{p,f}\Big|_{s\to4m^2} =
      \left(\frac{\alpha_s}{\pi}\right)^3 C_F^2 T_F \frac{1}{\beta}
      \left[
        -\bigl(6.0881-3.6463 {\rm i} \bigr) l_{2 \beta }
        + 5.7276+9.5631 {\rm i}
      \right]
      \,.
      \label{eq::nhsing_sthr}
  \end{align}

\begin{figure}[h]
  \begin{center}
    \begin{tabular}{ccc}
      \includegraphics[width=0.3\textwidth]{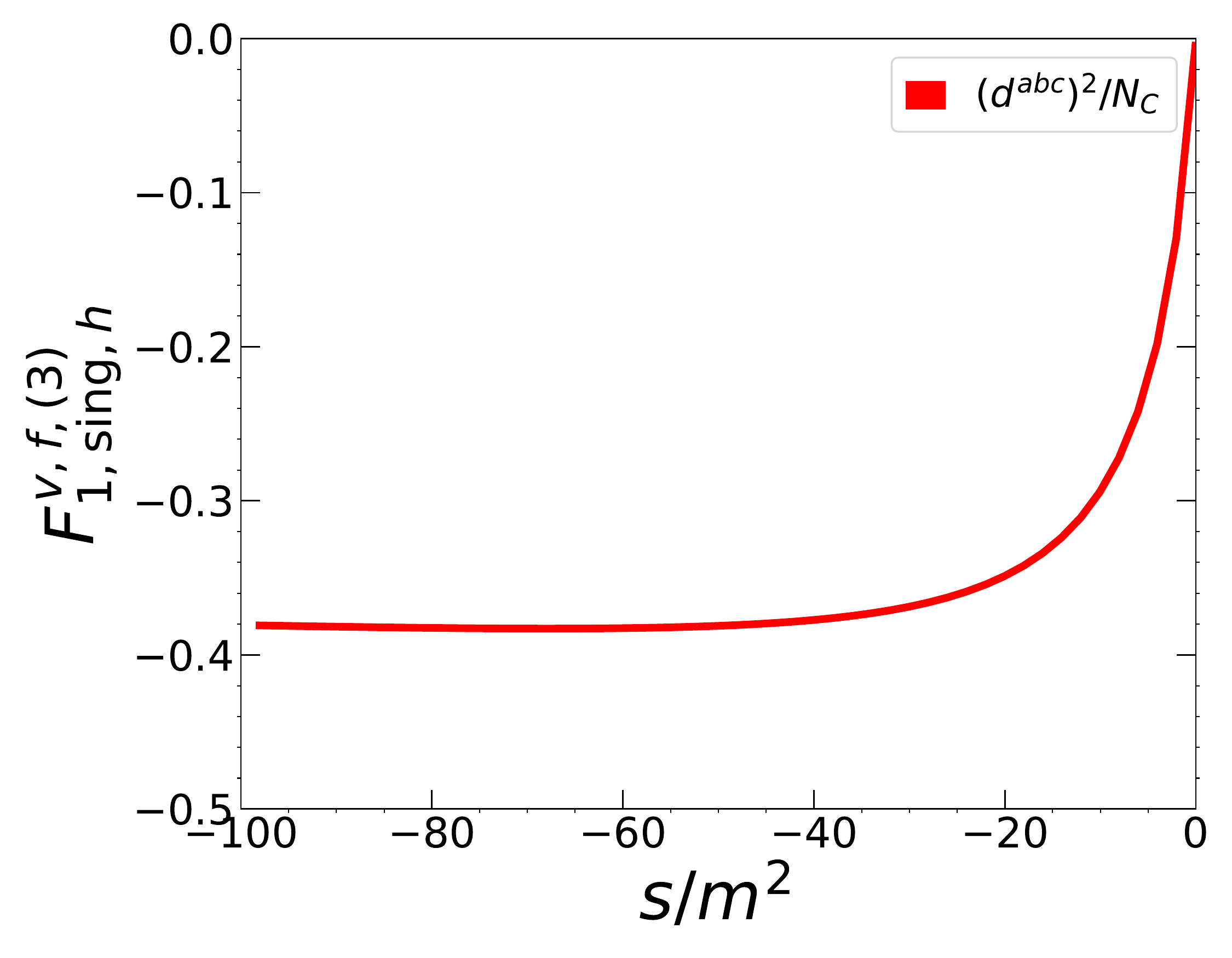} &
      \includegraphics[width=0.3\textwidth]{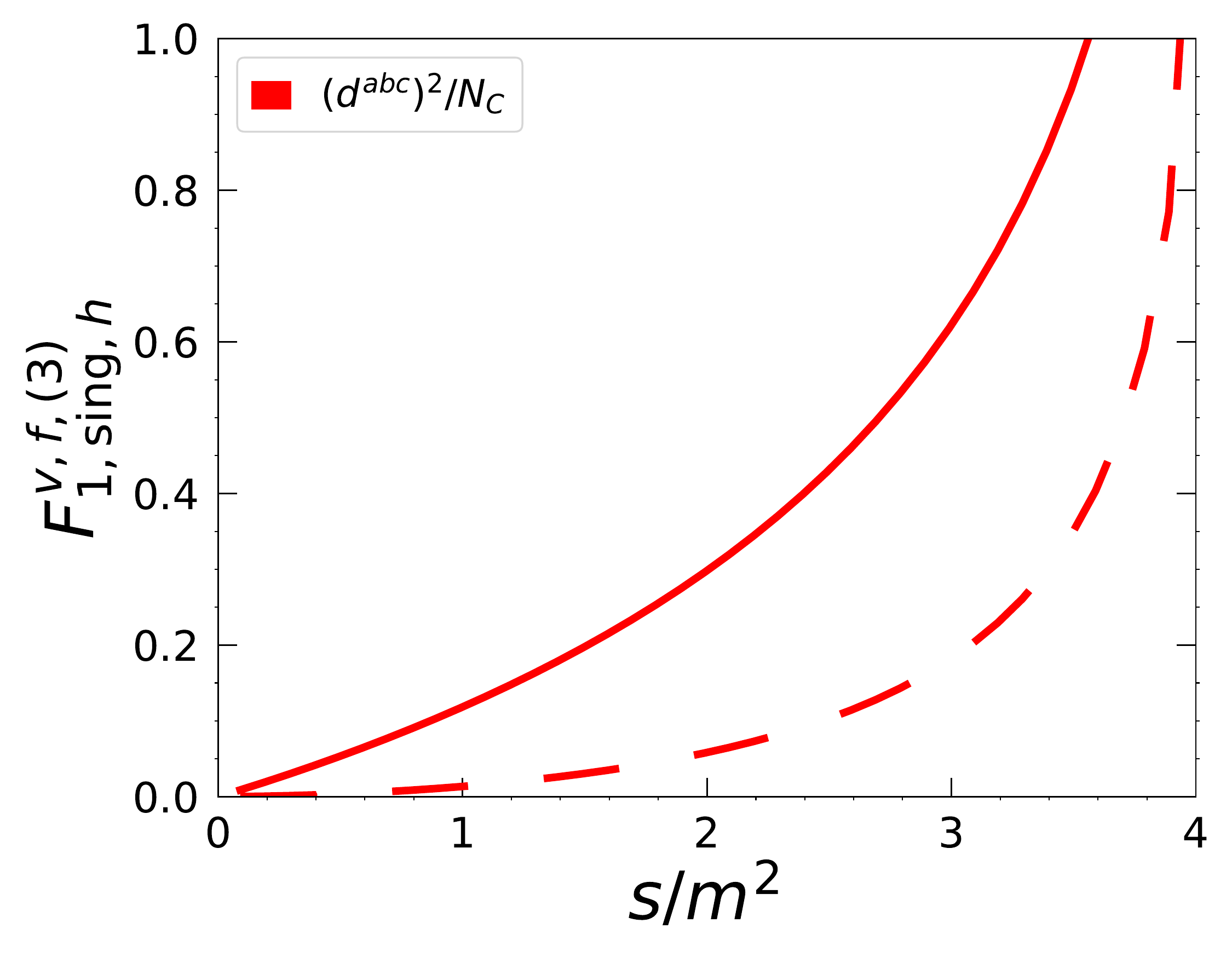} &
      \includegraphics[width=0.3\textwidth]{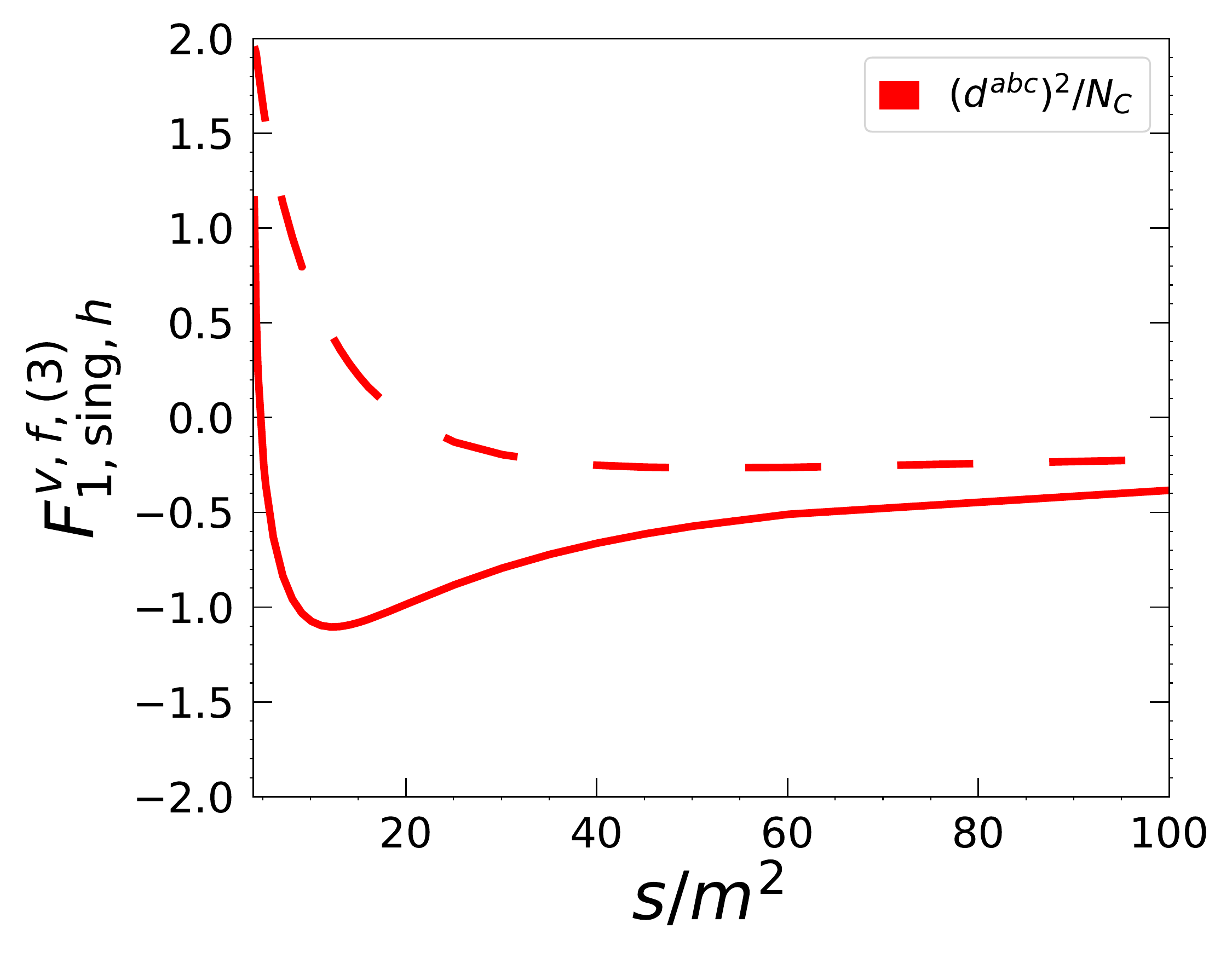}
      \\ (a) & (b) & (c) \\
      \includegraphics[width=0.3\textwidth]{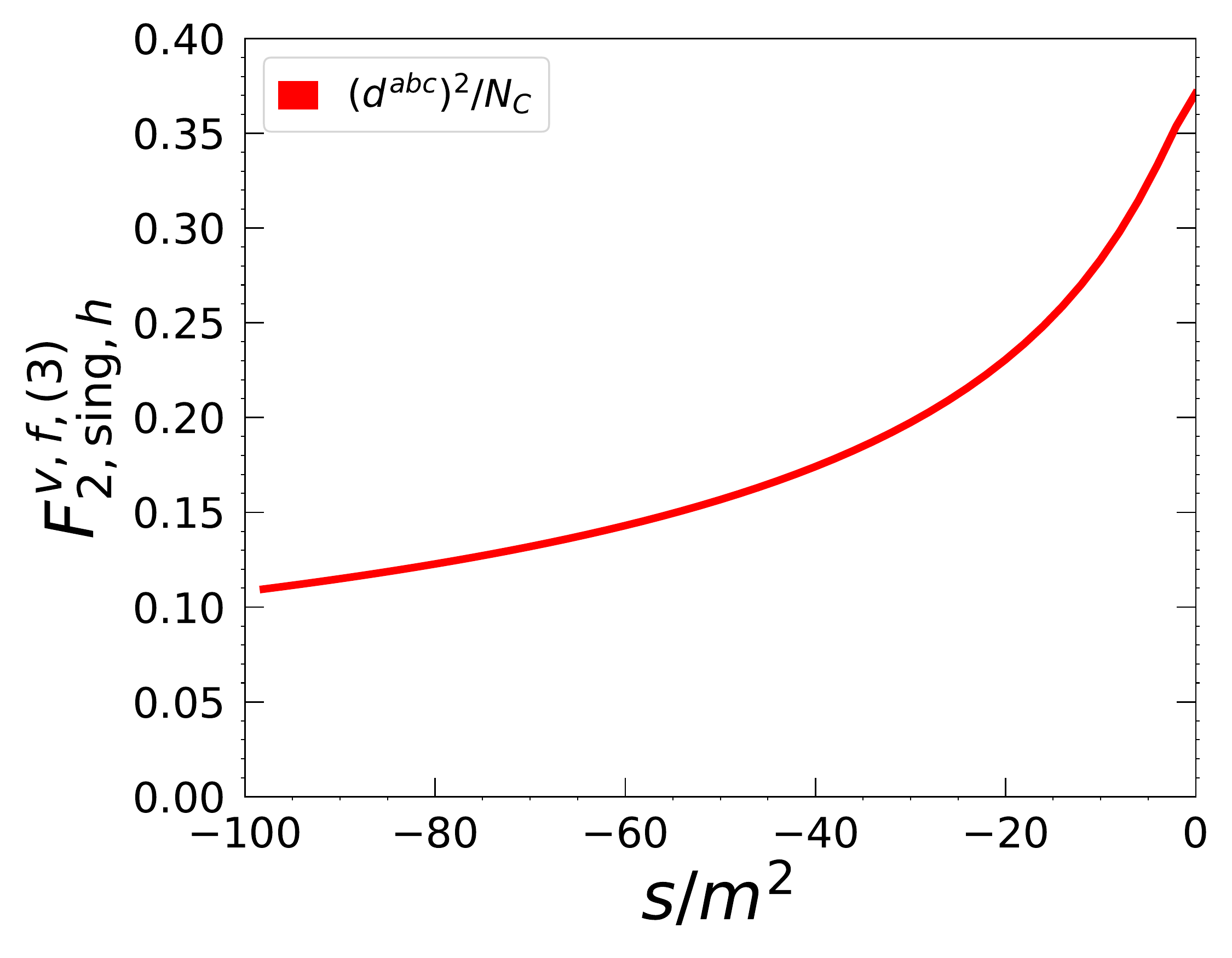} &
      \includegraphics[width=0.3\textwidth]{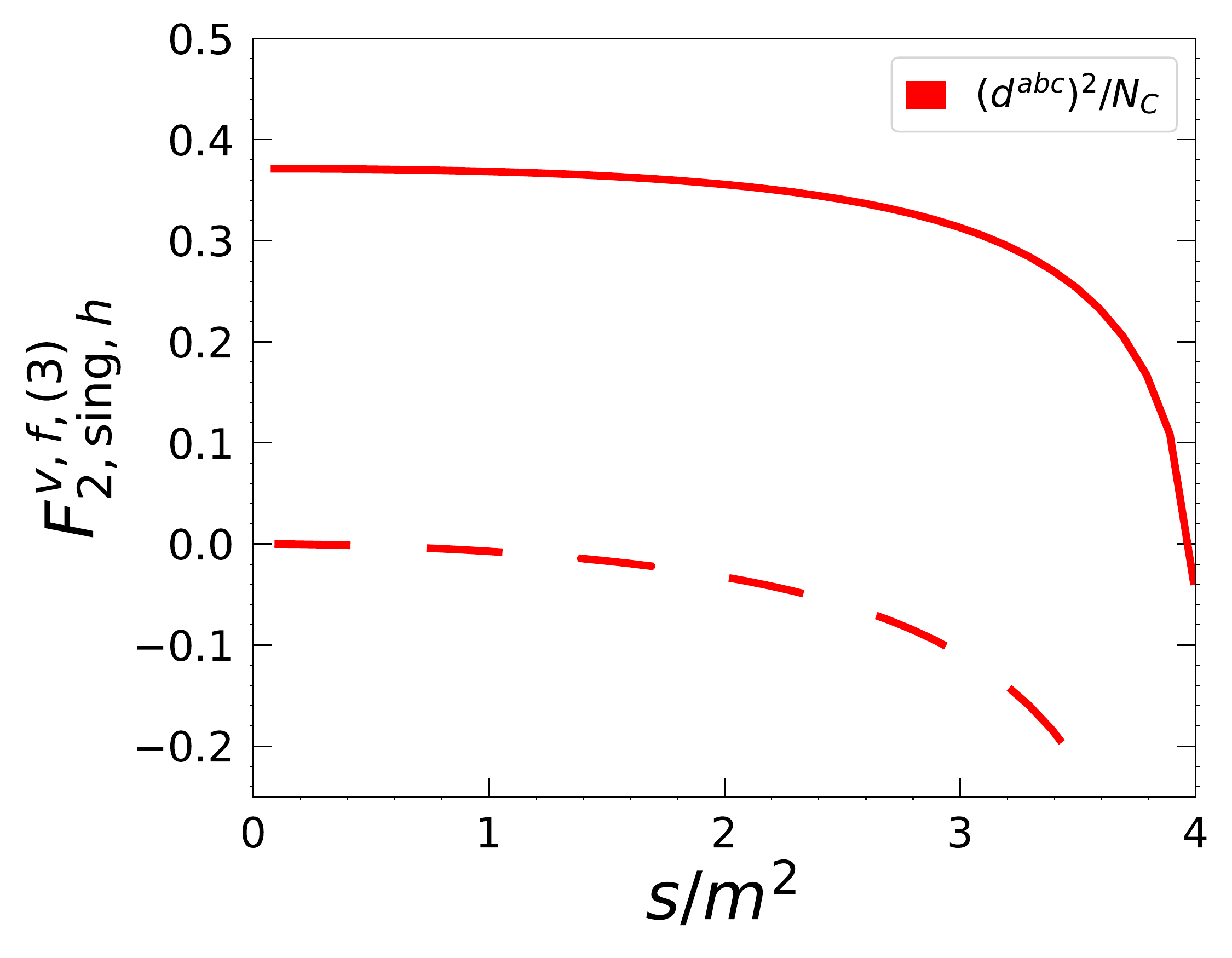} &
      \includegraphics[width=0.3\textwidth]{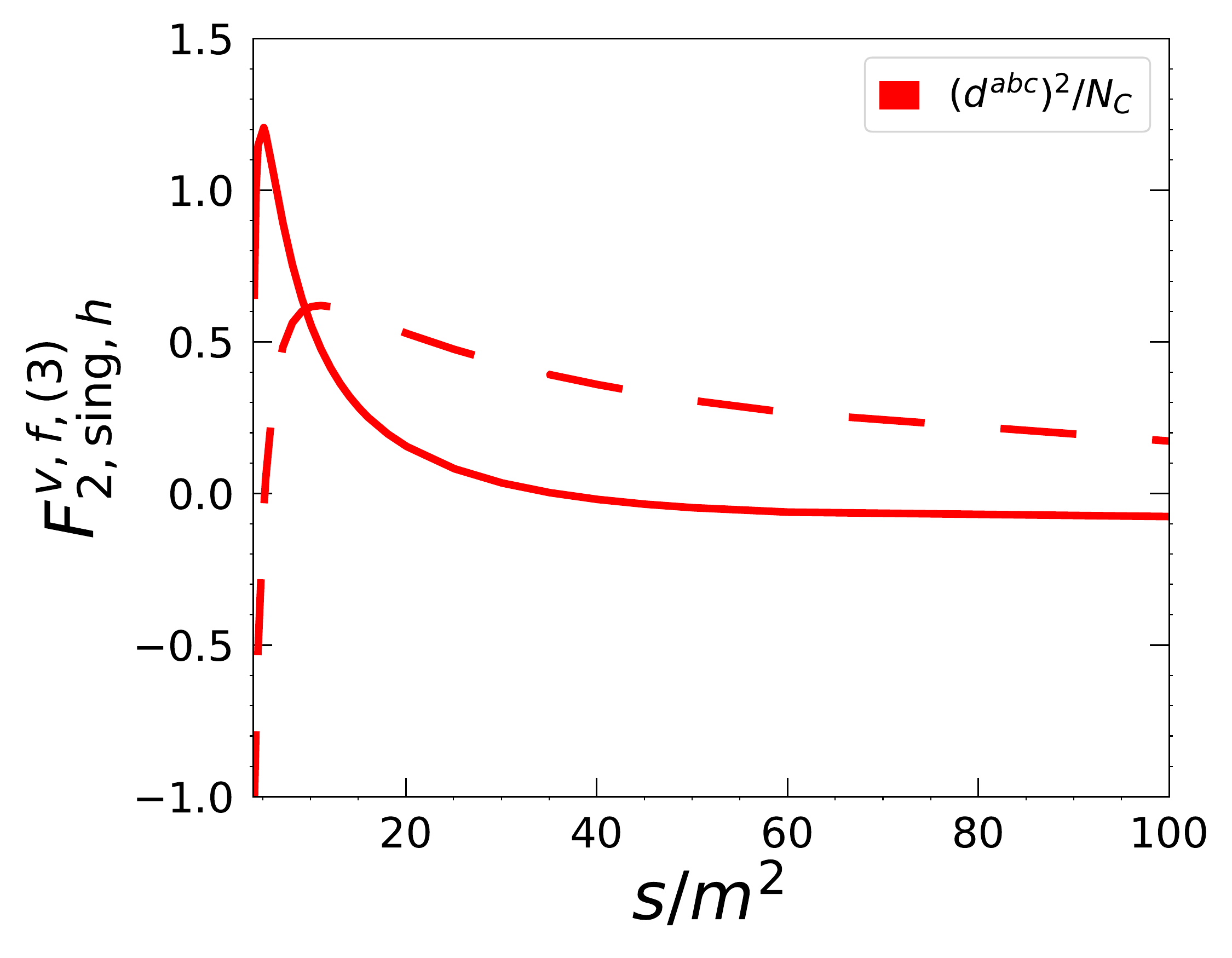}
      \\ (d) & (e) & (f) \\
      \includegraphics[width=0.3\textwidth]{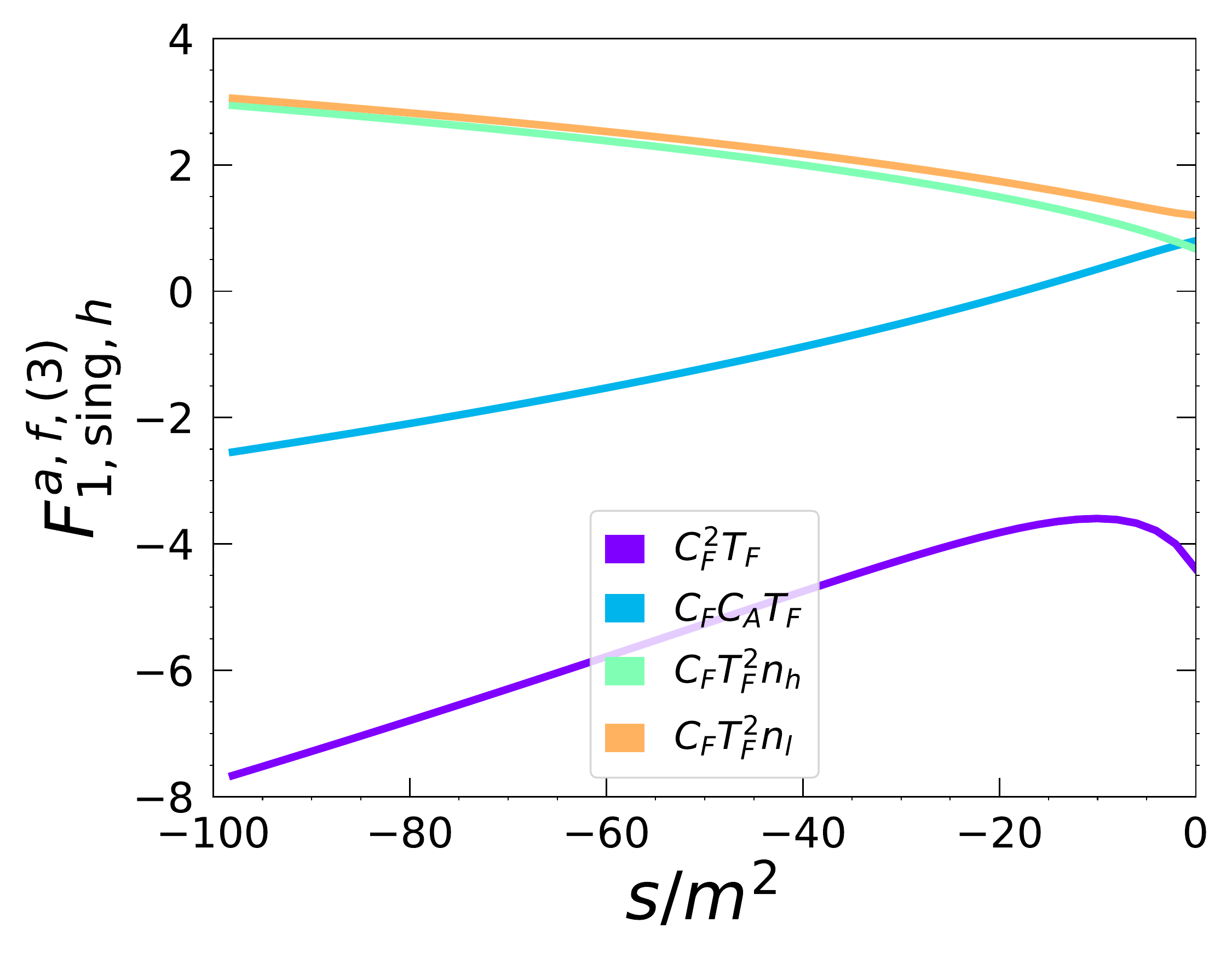} &
      \includegraphics[width=0.3\textwidth]{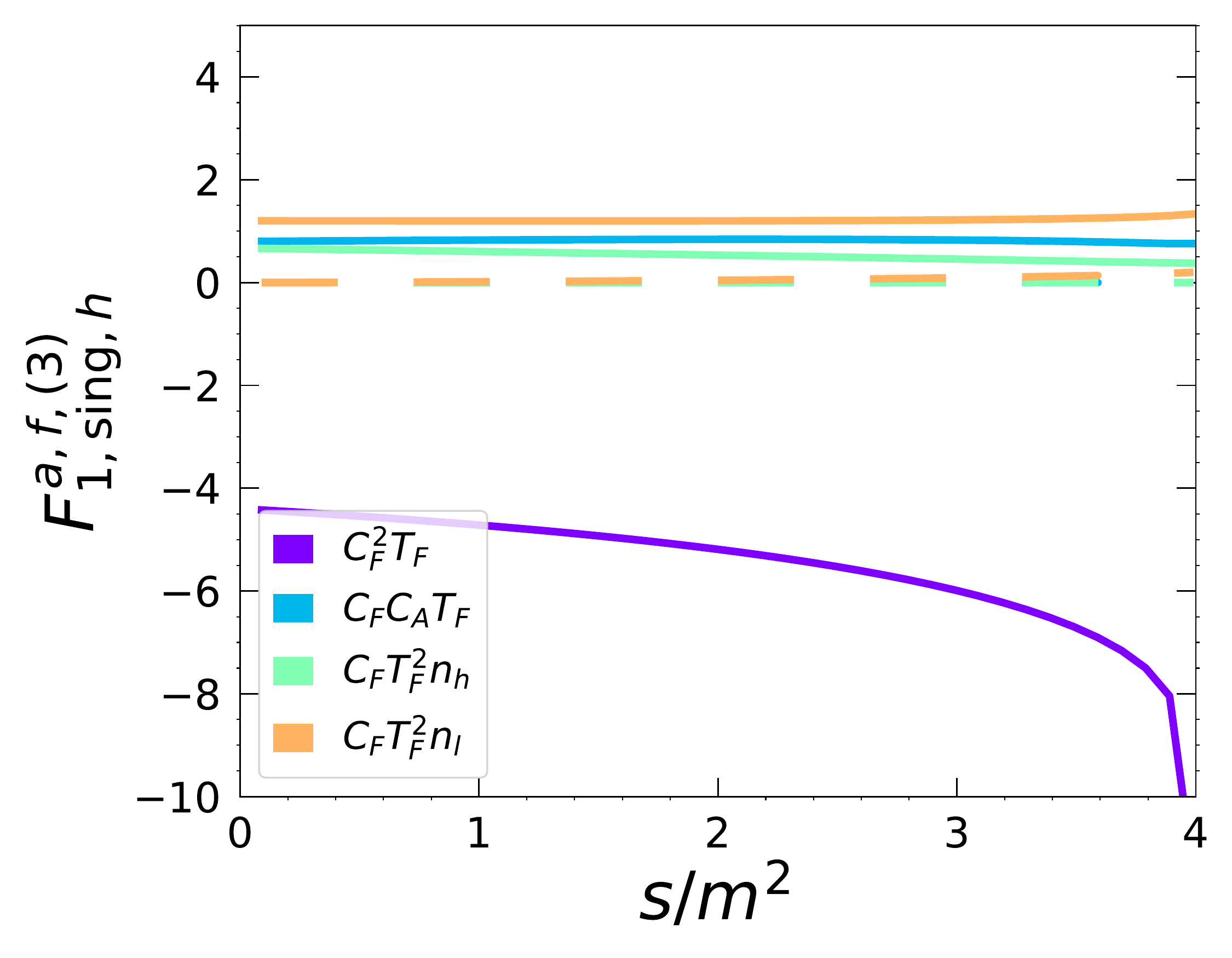} &
      \includegraphics[width=0.3\textwidth]{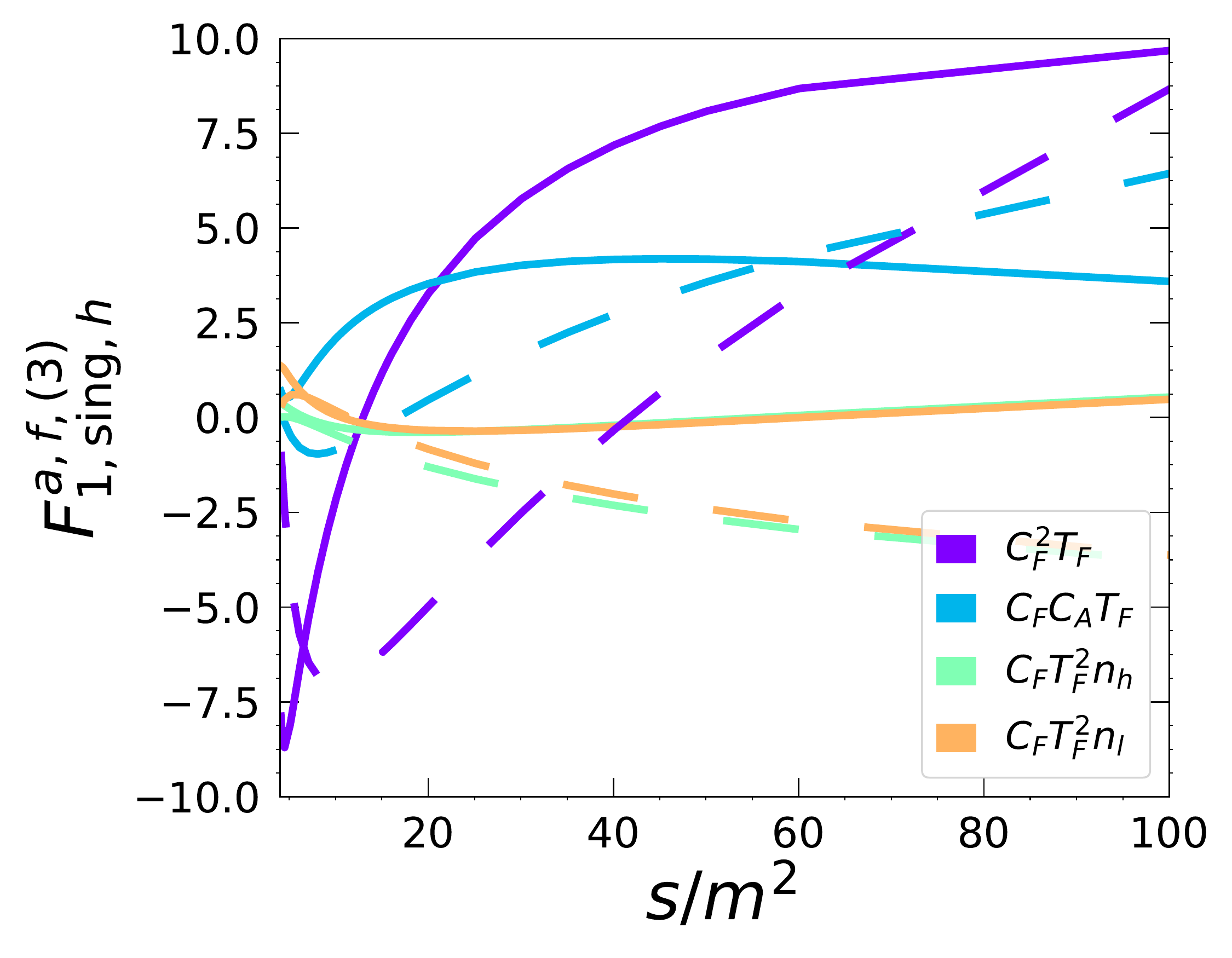}
      \\ (g) & (h) & (i) \\
      \includegraphics[width=0.3\textwidth]{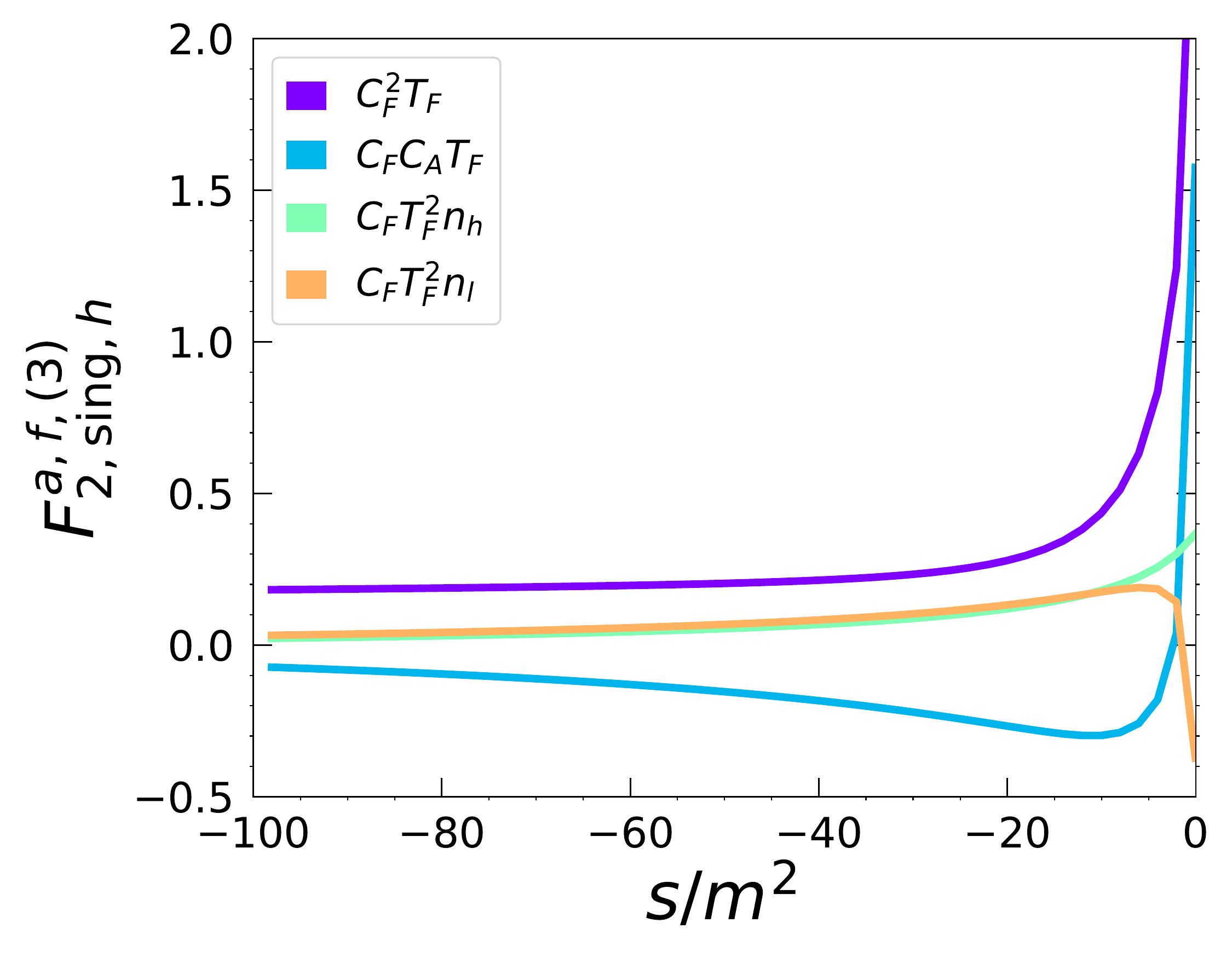} &
      \includegraphics[width=0.3\textwidth]{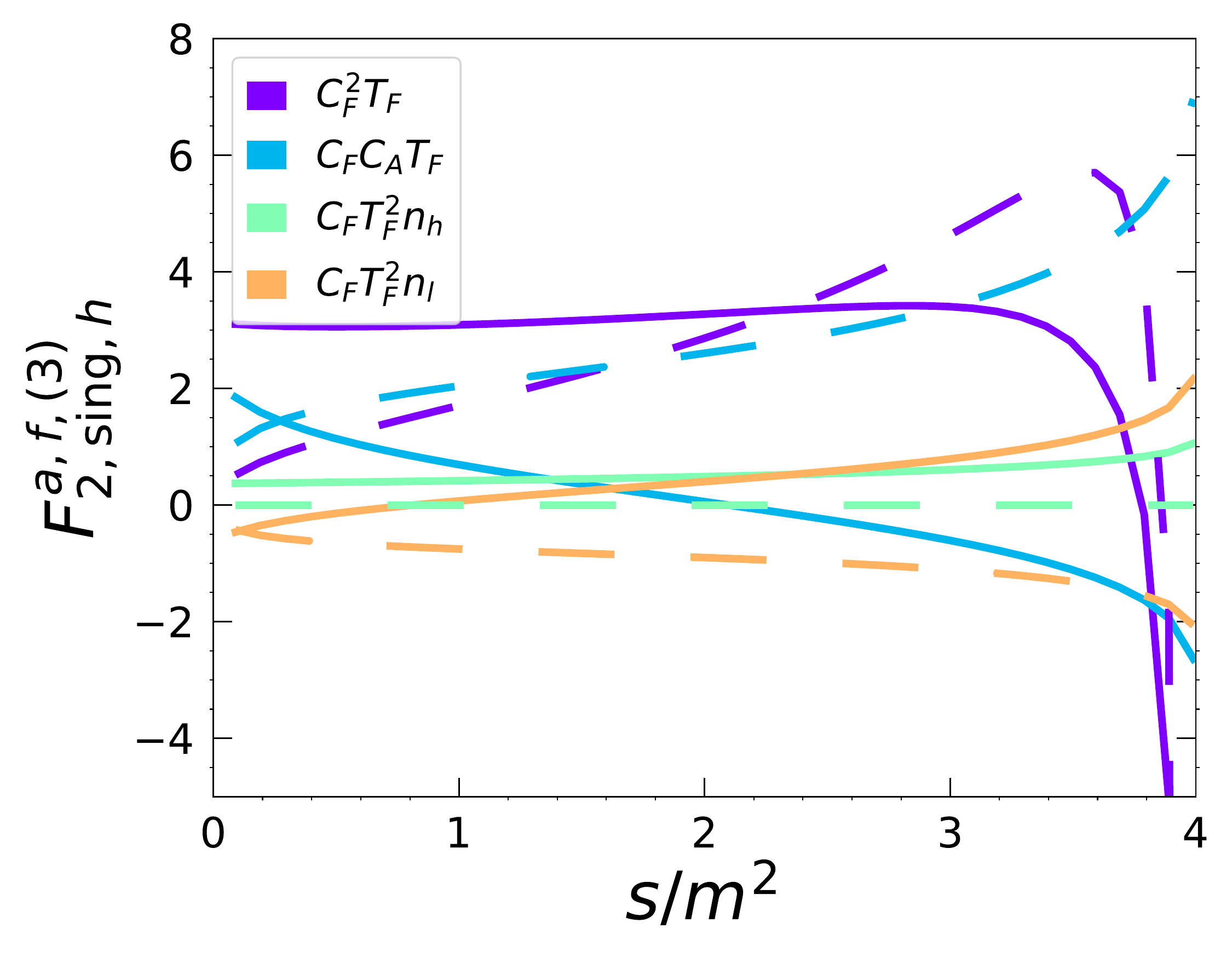} &
      \includegraphics[width=0.3\textwidth]{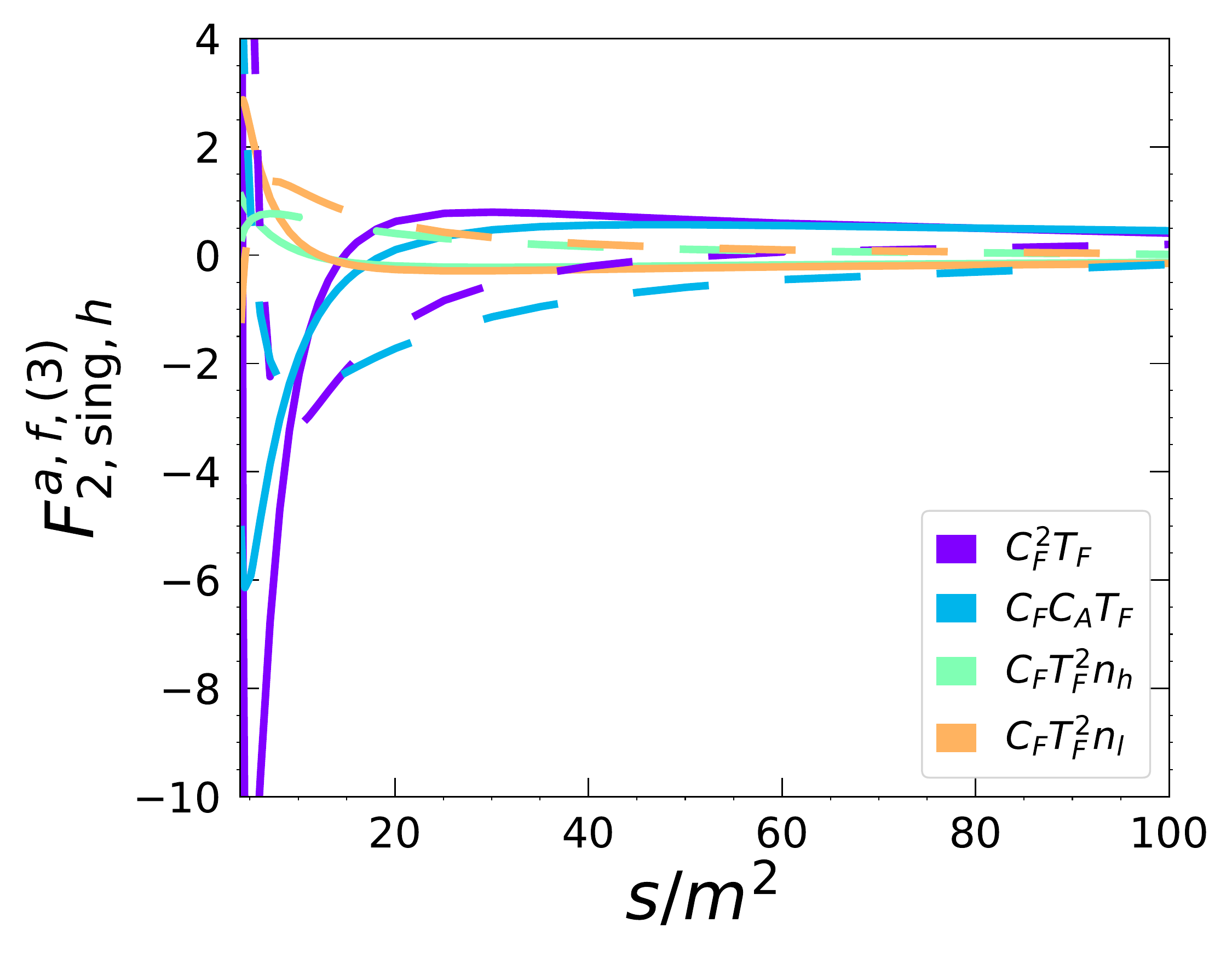}
      \\ (j) & (k) & (l)
    \end{tabular}
    \caption{\label{fig::FF_nhsing_1} Massive singlet vector and axial-vector
      form factors as a function of $s$ for $\mu^2=m^2$.}
  \end{center}
\end{figure}

\begin{figure}[h]
  \begin{center}
    \begin{tabular}{ccc}
      \includegraphics[width=0.3\textwidth]{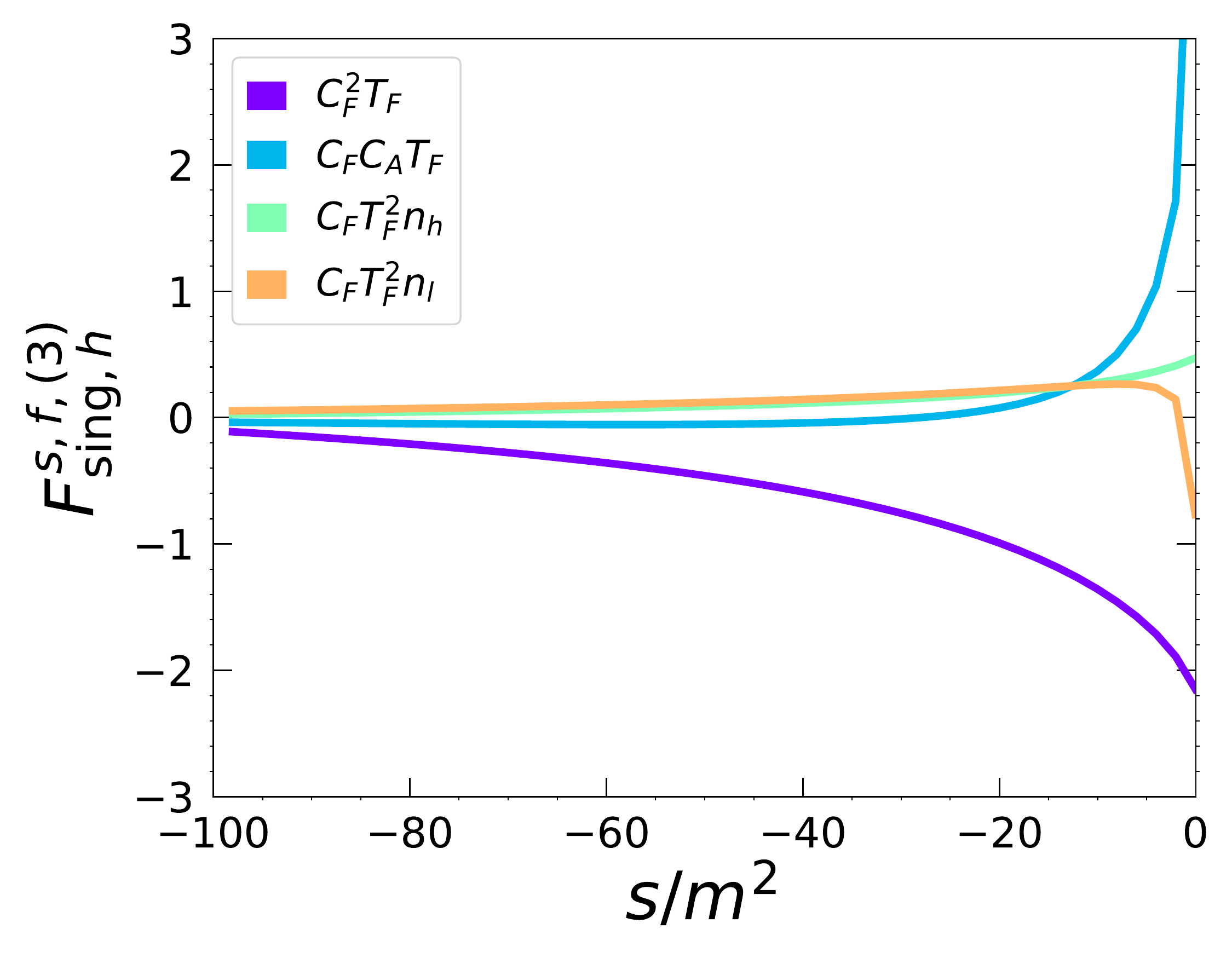} &
      \includegraphics[width=0.3\textwidth]{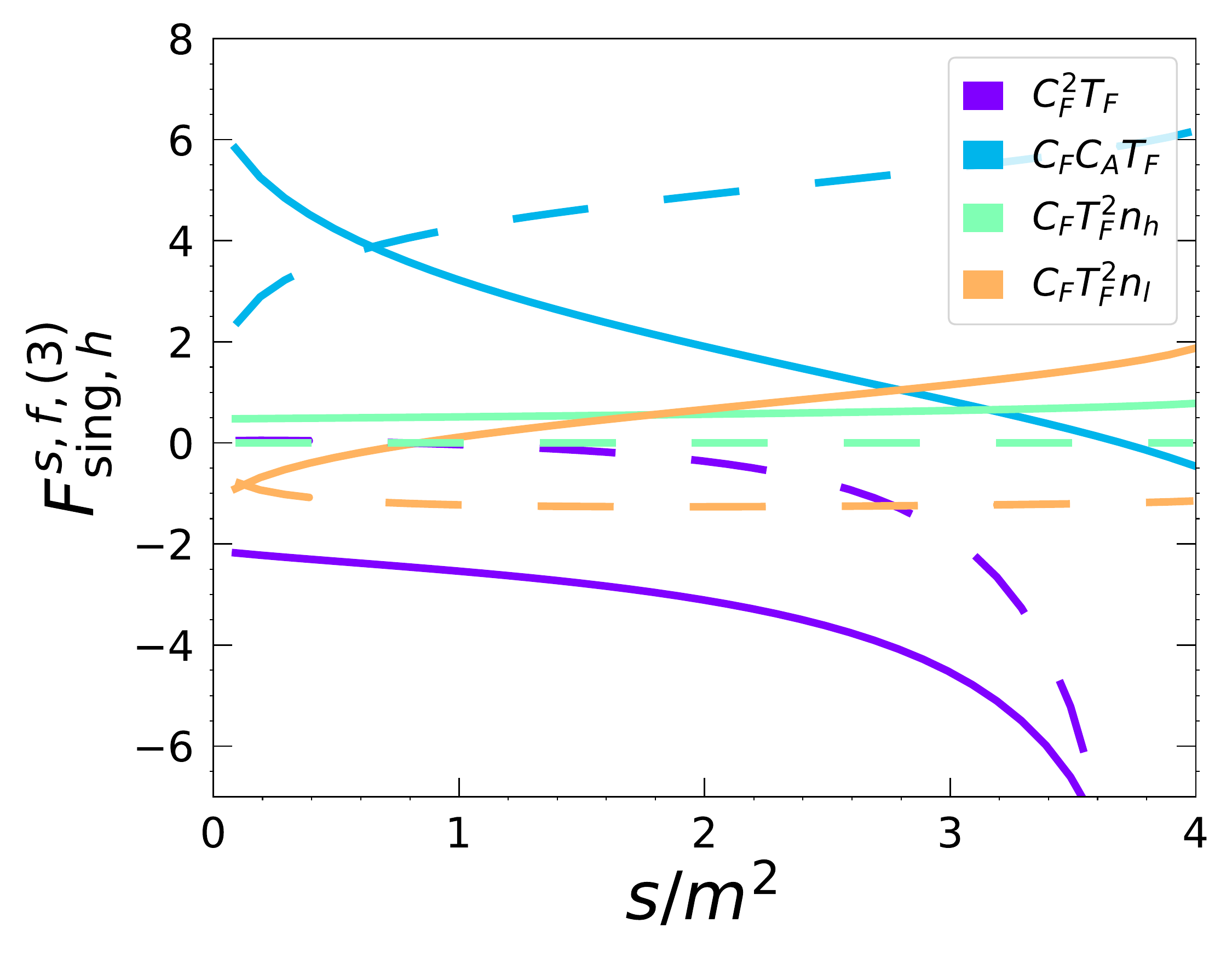} &
      \includegraphics[width=0.3\textwidth]{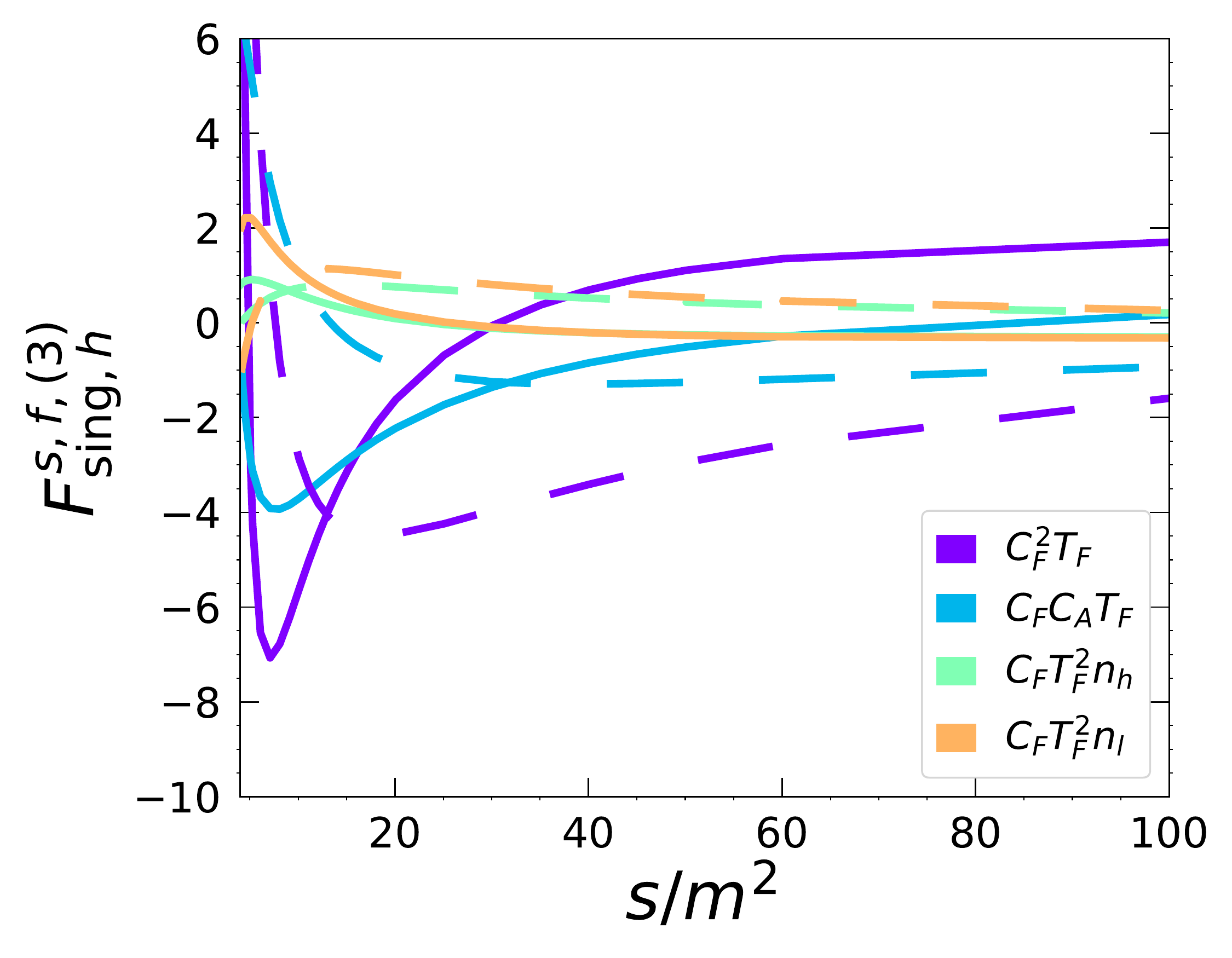}
      \\ (a) & (b) & (c) \\
      \includegraphics[width=0.3\textwidth]{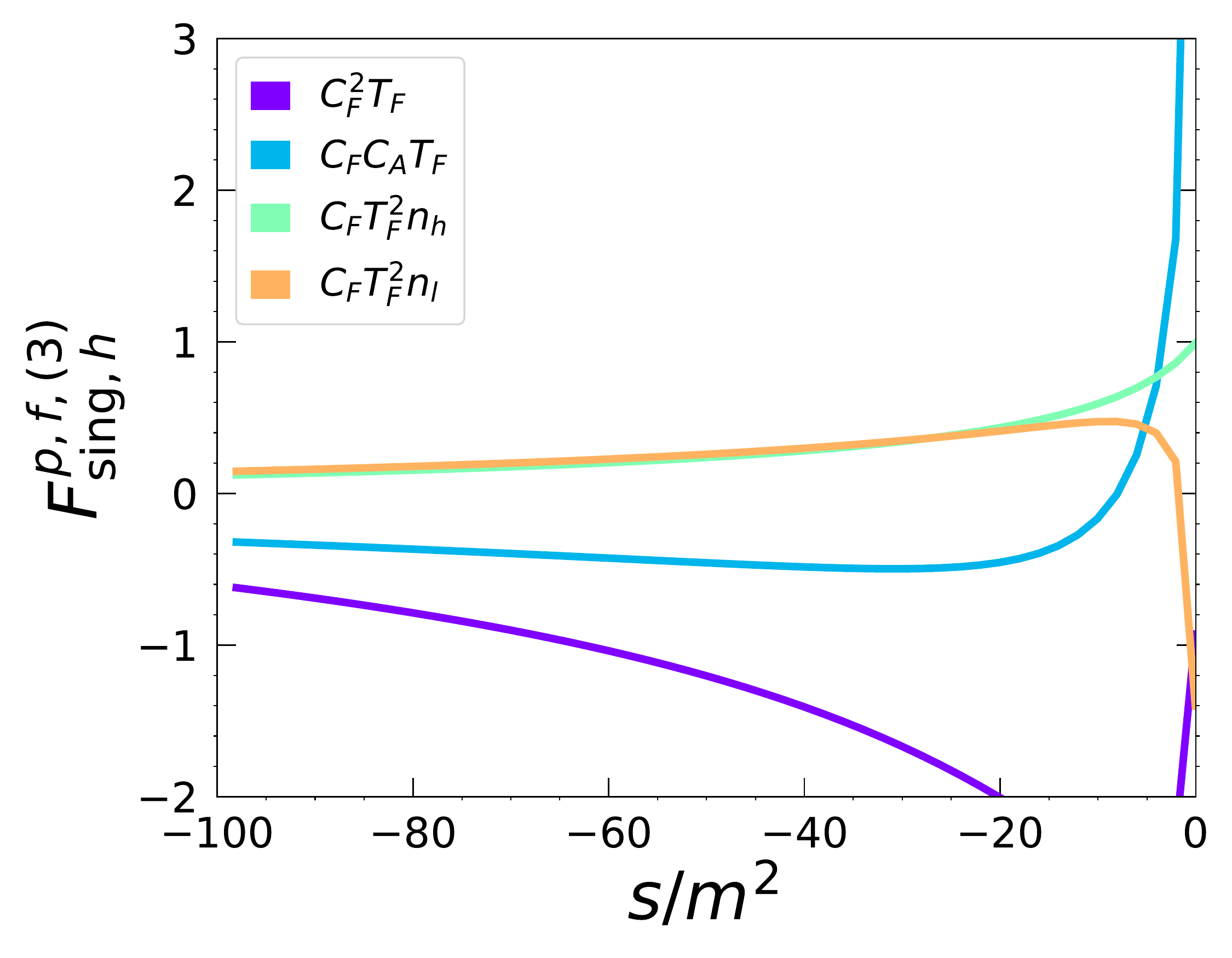} &
      \includegraphics[width=0.3\textwidth]{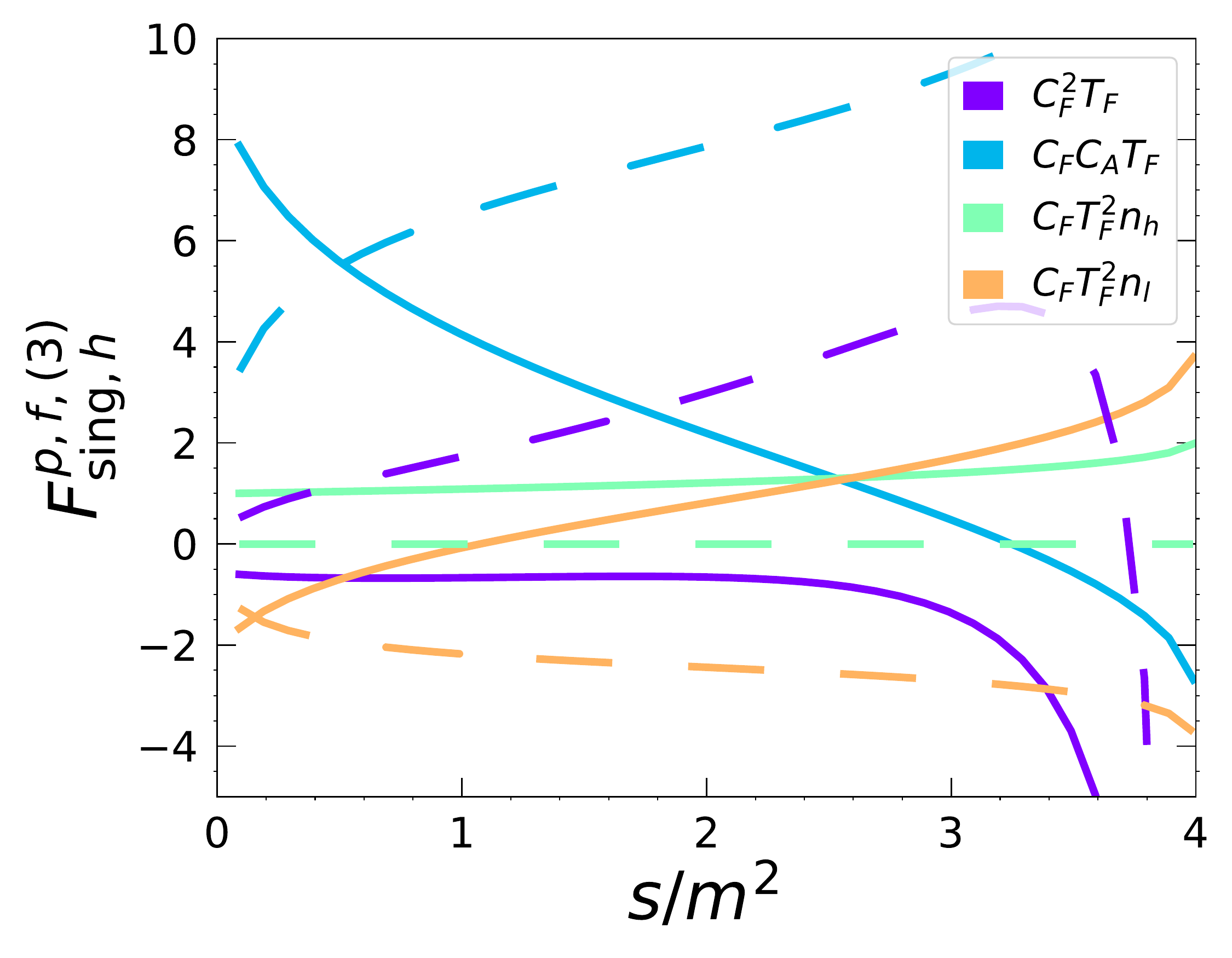} &
      \includegraphics[width=0.3\textwidth]{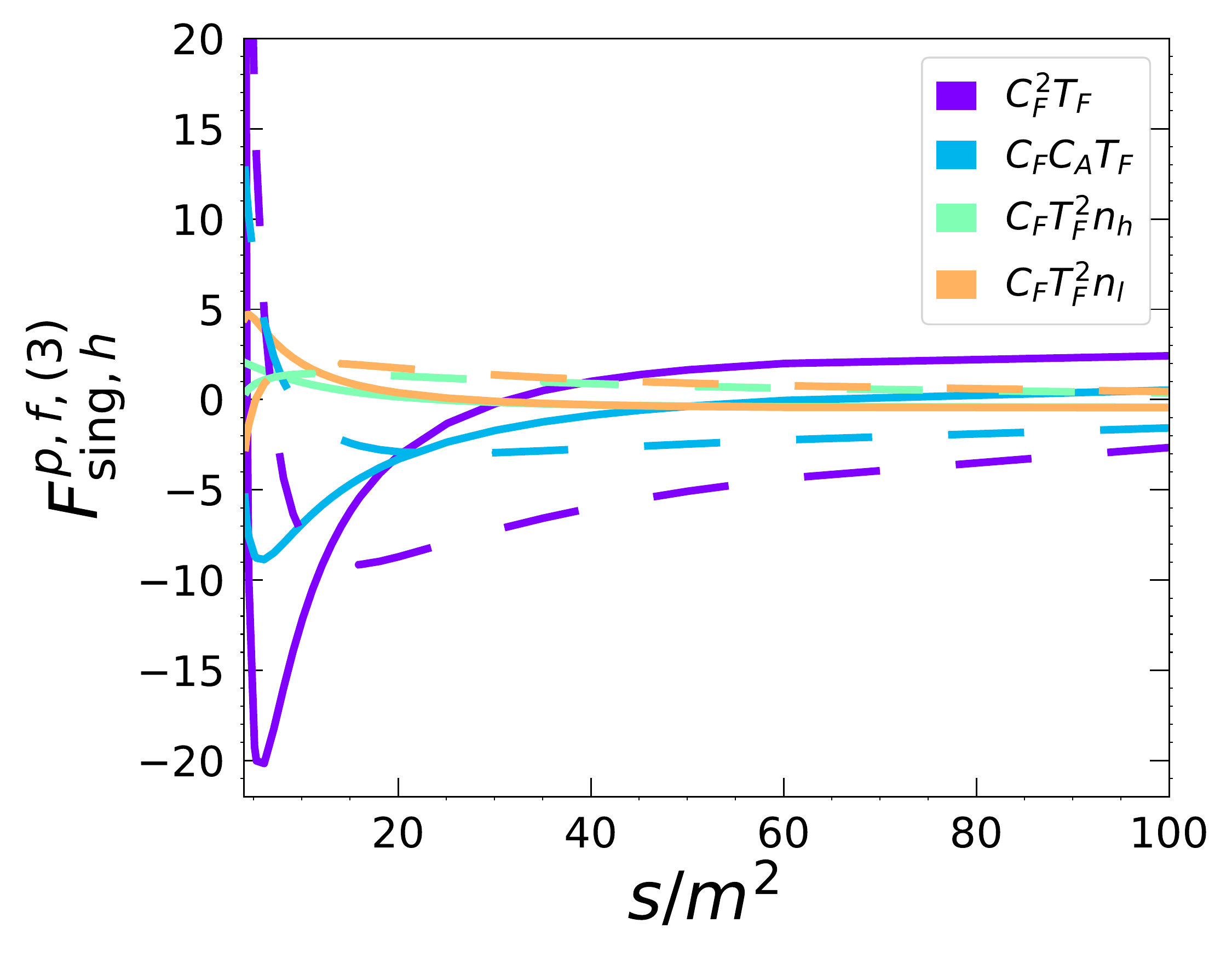}
      \\ (d) & (e) & (f)
    \end{tabular}
    \caption{\label{fig::FF_nhsing_2} Massive singlet scalar and pseudoscalar
      form factors as a function of $s$ for $\mu^2=m^2$.}
  \end{center}
\end{figure}

In Figs.~\ref{fig::FF_nhsing_1} and~\ref{fig::FF_nhsing_2} we show
the finite parts of the massive singlet form factors as a function of $s$.


\section{\label{app::GGtil}One- and two-loop result for $F_{G\tilde{G}}^f$}

Our one- and two-loop results for $F_{G\tilde{G}}^f$ are given by
\begin{align}
  F_{G\tilde{G}}^{f,(1)} &=
  -3 C_F L_m
  +C_F \biggl(
        -7
        +\frac{2 \pi ^2 (1-x)}{3 (1+x)}
        +\frac{(1-x) H_0^2}{2 (1+x)}
        +\frac{2 (1-x) H_{0,1}}{1+x}
  \biggr)
  ~,
  \\
  F_{G\tilde{G}}^{f,(2)} &=
  L_m^2 
  \biggl\{
  	-\frac{11}{4} C_F C_A 
  	+ C_F n_l T_F
        + C_F^2
        \biggl(
                \frac{3}{2}
                +\frac{3 \big(1+x^2\big) H_0}{2 (1-x) (1+x)}
        \biggr) 
  \biggr\}
  \nonumber \\ &
  + L_m \biggl\{
        C_F T_F n_h
        \biggl(
           	\frac{8}{3}
		-\frac{2 \pi ^2 (1-x)}{9 (1+x)}
		-\frac{(1-x) H_0^2}{6 (1+x)}
		-\frac{2 (1-x) H_{0,1}}{3 (1+x)}
	\biggr)
	\nonumber \\ &
	+C_F T_F n_l 
	\biggl(
		5
		-\frac{4 \pi ^2 (1-x)}{9 (1+x)}
		-\frac{(1-x) H_0^2}{3 (1+x)}
		-\frac{4 (1-x) H_{0,1}}{3 (1+x)}
	\biggr)
	+ C_F C_A 
	\biggl(
		-\frac{75}{4}
		\nonumber \\ &
		+\frac{11 \pi ^2 (1-x)}{9 (1+x)}
		+\frac{11 (1-x) H_0^2}{12 (1+x)}
		+\frac{11 (1-x) H_{0,1}}{3 (1+x)}
   	\biggr)
        +C_F^2 
        \biggl(
                \frac{29}{4}
                -\pi ^2 
                \biggl[
                        \frac{7-8 x+7 x^2}{12 (1-x) (1+x)}
                        \nonumber \\ &
                        +\frac{\big(1+x^2\big) H_0}{3 (1+x)^2}
                \biggr]
                +\biggl[
                         \frac{7 \big(1+x^2\big)}{2 (1-x) (1+x)}
                        -\frac{\big(1+x^2\big) H_{0,1}}{(1+x)^2}
                \biggr] H_0
                -\frac{3 \big(1+x^2\big) H_{-1} H_0}{(1-x) (1+x)}
                \nonumber \\ &
                +\frac{\big(1+x+x^2\big) H_0^2}{2 (1-x) (1+x)}
                -\frac{\big(1+x^2\big) H_0^3}{4 (1+x)^2}
                -\frac{(1-x) H_{0,1}}{1+x}
                +\frac{3 \big(1+x^2\big) H_{0,-1}}{(1-x) (1+x)}
        \biggr)
  \biggr\}
  \nonumber \\ &
  + C_F T_F n_h
  \biggl(
  	\frac{385}{36}
	-\pi ^2 
	\biggl[
		 \frac{3-8 x+3 x^2}{3 (1-x)^2}
		+\frac{2 \big(1-4 x-4 x^3+x^4\big) H_0}{9 (1-x)^3 (1+x)}
	\biggr]
	\nonumber \\ &
	-\frac{\big(19+8 x+19 x^2\big) H_0^2}{18 (1-x)^2}
	-\frac{2 \big(1-4 x-4 x^3+x^4\big) H_0^3}{9 (1-x)^3 (1+x)}
  \biggr)
  + C_F T_F n_l
  \biggl(
  	\frac{293}{36}
  	\nonumber \\ &
	-\pi ^2 
	\biggl[
		 \frac{4 (5-14 x)}{27 (1+x)}
		+\frac{2 (1-x) H_0}{9 (1+x)}
		+\frac{8 (1-x) H_1}{9 (1+x)}
	\biggr]
	-\biggl[
		 \frac{19 (1-x)}{18 (1+x)}
		+\frac{2 (1-x) H_1}{3 (1+x)}
	\biggr] H_0^2
	\nonumber \\ &
	-\frac{2 (1-x) H_0^3}{9 (1+x)}
	-\frac{38 (1-x) H_{0,1}}{9 (1+x)}
	-\frac{8 (1-x) H_1 H_{0,1}}{3 (1+x)}
	-\frac{4 (1-x) H_{0,0,1}}{3 (1+x)}
	\nonumber \\ &
	+\frac{8 (1-x) H_{0,1,1}}{3 (1+x)}
	-\frac{4 (1-x) \zeta_3}{3 (1+x)}
  \biggr)
  +C_F C_A
  \biggl(
	-\frac{4963}{144}
	+\zeta_3
	\biggl[
		\frac{49-75 x+21 x^2+17 x^3}{6 (1-x)^2 (1+x)}
		\nonumber \\ &
		+\frac{\big(1-4 x+8 x^2-4 x^3+x^4\big) H_0}{(1-x)^3 (1+x)}
	\biggr]
	-\frac{\pi ^4 \big(29-212 x+24 x^2-212 x^3+29 x^4\big)}{360 (1-x)^3 (1+x)}
	\nonumber \\ &
	+\pi ^2 
	\biggl[
		 \frac{601-1803 x+993 x^2-547 x^3}{108 (1-x)^2 (1+x)}
		-\biggl(
			 \frac{-10+13 x-12 x^2+7 x^3+11 x^4}{9 (1-x)^3 (1+x)}
			 \nonumber \\ &
			+\frac{2 (1-x) H_1}{3 (1+x)}
		\biggr) H_0
		+\frac{\big(1+2 x+18 x^2+2 x^3+x^4\big) H_0^2}{12 (1-x)^3 (1+x)}
		-\frac{\big(5-8 x+5 x^2\big) H_{-1}}{2 (1-x)^2}
		\nonumber \\ &
		-\frac{\big(-29+159 x-69 x^2+47 x^3\big) H_1}{18 (1-x)^2 (1+x)}
		-\frac{2 x \big(1+x+x^2\big) H_{0,1}}{(1-x)^3 (1+x)}
		\nonumber \\ &
		-\frac{\big(1-4 x+8 x^2-4 x^3+x^4\big) H_{0,-1}}{(1-x)^3 (1+x)}
	\biggr]
	+\biggl[
		 \frac{3}{2 (1-x)}
		-\frac{\big(5+x-5 x^2+7 x^3\big) H_{0,1}}{2 (1-x)^2 (1+x)}
		\nonumber \\ &
		+\frac{\big(5-8 x+5 x^2\big) H_{0,-1}}{(1-x)^2}
		-\frac{4 \big(1-4 x+8 x^2-4 x^3+x^4\big) H_{0,0,1}}{(1-x)^3 (1+x)}
		\nonumber \\ &
		+\frac{4 \big(1-4 x+8 x^2-4 x^3+x^4\big) H_{0,0,-1}}{(1-x)^3 (1+x)}
	\biggr] H_0
	+\biggl[
		-\frac{-305+681 x-717 x^2+269 x^3}{72 (1-x)^2 (1+x)}
		\nonumber \\ &
		-\frac{\big(-37+63 x-51 x^2+x^3\big) H_1}{12 (1-x)^2 (1+x)}
		+\frac{2 \big(1-4 x+7 x^2-4 x^3+x^4\big) H_{0,1}}{(1-x)^3 (1+x)}
		\nonumber \\ &
		-\frac{\big(1-4 x+8 x^2-4 x^3+x^4\big) H_{0,-1}}{(1-x)^3 (1+x)}
	\biggr] H_0^2
	-\frac{\big(5-8 x+5 x^2\big) H_{-1} H_0^2}{2 (1-x)^2}
	\nonumber \\ &
	-\biggl[
		 \frac{-22+46 x-39 x^2+22 x^3+29 x^4}{36 (1-x)^3 (1+x)}
		+\frac{2 (1-x) H_1}{3 (1+x)}
	\biggr] H_0^3
	-\frac{\big(5-8 x+5 x^2\big) H_{0,0,-1}}{(1-x)^2}
	\nonumber \\ &
	-\frac{\big(1-10 x-8 x^2-10 x^3+x^4\big) H_0^4}{48 (1-x)^3 (1+x)}
	+\biggl[
		\frac{3}{2}
		+\frac{19 (1-x) H_{0,1}}{3 (1+x)}
	\biggr] H_1
	\nonumber \\ &
	+\frac{\big(287-412 x+287 x^2\big) H_{0,1}}{18 (1-x) (1+x)}
	+\frac{\big(37-3 x+93 x^2+5 x^3\big) H_{0,0,1}}{6 (1-x)^2 (1+x)}
	-\frac{16 (1-x) H_{0,1,1}}{3 (1+x)}
	\nonumber \\ &
	+\frac{\big(5-14 x+48 x^2-14 x^3+5 x^4\big) H_{0,0,0,1}}{(1-x)^3 (1+x)}
	-\frac{6 \big(1-4 x+8 x^2-4 x^3+x^4\big) H_{0,0,0,-1}}{(1-x)^3 (1+x)}
  \biggr)
  \nonumber \\ &
  +C_F^2 
  \biggl(
        \frac{125}{16}
        -\frac{\pi ^4 \big(-4+12 x-16 x^2+124 x^3-93 x^4+31 x^5\big)}{90 (1-x)^3 (1+x)^2}
        \nonumber \\ &
        -\zeta_3 
        \biggl[
                \frac{3-3 x+3 x^2+x^3}{(1-x)^2 (1+x)}
                +\frac{2 \big(1-4 x+8 x^2-4 x^3+x^4\big) H_0}{(1-x)^3 (1+x)}
        \biggr]
        \nonumber \\ &
        +\pi ^2 
        \biggl[
                -\frac{25-42 x+x^2+6 x^3}{6 (1-x)^2 (1+x)}
                +\frac{\big(-7+12 x+22 x^2-44 x^3+27 x^4\big) H_0}{6 (1-x)^3 (1+x)}
                \nonumber \\ &
                -\frac{\big(7-21 x+28 x^2+68 x^3-51 x^4+17 x^5\big) H_0^2}{24 (1-x)^3 (1+x)^2}
                -\frac{2 (1-x) H_1}{3 (1+x)}
                \nonumber \\ &
                +\biggl(
                        \frac{5-8 x+5 x^2}{(1-x)^2}
                        +\frac{2 \big(1+x^2\big) H_0}{3 (1+x)^2}
                \biggr) H_{-1}
                -\frac{\big(1+x^2\big) H_{0,1}}{2 (1+x)^2}
                \nonumber \\ &
                +\frac{4 \big(1-3 x+4 x^2+8 x^3-6 x^4+2 x^5\big) H_{0,-1}}{3 (1-x)^3 (1+x)^2}
        \biggr]
        +\biggl[
                \frac{3 (1+x)}{2 (-1+x)}
                +\frac{4 \big(3-5 x+3 x^2\big) H_{0,1}}{(1-x)^2}
                \nonumber \\ &
                -\frac{2 \big(5-8 x+5 x^2\big) H_{0,-1}}{(1-x)^2}
                +\frac{8 \big(1-4 x+8 x^2-4 x^3+x^4\big) H_{0,0,1}}{(1-x)^3 (1+x)}
                \nonumber \\ &
                -\frac{8 \big(1-4 x+8 x^2-4 x^3+x^4\big) H_{0,0,-1}}{(1-x)^3 (1+x)}
        \biggr] H_0
        +\biggl[
                -\frac{1-3 x+5 x^2+17 x^3}{8 (1-x)^2 (1+x)}
                \nonumber \\ &
                -\frac{\big(13-11 x-5 x^2+11 x^3\big) H_1}{2 (1-x)^2 (1+x)}
                -\frac{\big(3-9 x+12 x^2+4 x^3-3 x^4+x^5\big) H_{0,1}}{(1-x)^3 (1+x)^2}
                \nonumber \\ &
                +\frac{\big(3-9 x+12 x^2+20 x^3-15 x^4+5 x^5\big) H_{0,-1}}{2 (1-x)^3 (1+x)^2}
        \biggr] H_0^2
        \nonumber \\ &
        +\frac{\big(-3+6 x+44 x^2-70 x^3+43 x^4\big) H_0^3}{12 (1-x)^3 (1+x)}
        -\frac{3 (1-x) H_{0,1}}{1+x}
        -\frac{2 (1-x) H_1 H_{0,1}}{1+x}
        \nonumber \\ &
        -\frac{\big(5-15 x+20 x^2+44 x^3-33 x^4+11 x^5\big) H_0^4}{48 (1-x)^3 (1+x)^2}
        +\frac{2 (1-x) H_{0,1,1}}{1+x}
        \nonumber \\ &
        +\biggl[
                \biggl(
                        -\frac{7 \big(1+x^2\big)}{(1-x) (1+x)}
                        +\frac{2 \big(1+x^2\big) H_{0,1}}{(1+x)^2}
                \biggr) H_0
                +\frac{\big(5-8 x+5 x^2\big) H_0^2}{(1-x)^2}
                \nonumber \\ &
                +\frac{\big(1+x^2\big) H_0^3}{2 (1+x)^2}
        \biggr] H_{-1}
        -\frac{\big(1+x^2\big) H_{0,1}^2}{(1+x)^2}
        +\biggl[
                 \frac{7 \big(1+x^2\big)}{(1-x) (1+x)}
                -\frac{2 \big(1+x^2\big) H_{0,1}}{(1+x)^2}
        \biggr] H_{0,-1}
        \nonumber \\ &
        -\frac{2 \big(7-6 x-3 x^2+6 x^3\big) H_{0,0,1}}{(1-x)^2 (1+x)}
        +\frac{2 \big(5-8 x+5 x^2\big) H_{0,0,-1}}{(1-x)^2}   
        \nonumber \\ &    
        -\frac{12 \big(1-4 x+8 x^2-4 x^3+x^4\big) H_{0,0,0,1}}{(1-x)^3 (1+x)}
        +\frac{12 \big(1-4 x+8 x^2-4 x^3+x^4\big) H_{0,0,0,-1}}{(1-x)^3 (1+x)}
  \biggr)        
         ~,
\end{align}
with $L_m=\ln(\mu^2/m^2)$ and we dropped the arguments of the
harmonic polylogarithms $H_{\vec{w}} \equiv H_{\vec{w}}(x)$~\cite{Remiddi:1999ew}.

The one-loop result agrees with Ref.~\cite{Bernreuther:2005rw},
the two-loop expression is new.



\section{\label{app:formfactors3l}The Fortran library \texttt{FF3l}}

In this appendix we present the Fortran library \texttt{FF3l} for the
numerical evaluation of the third-order corrections to the form factors.  We
implement the ultraviolet renormalized form factors, but we do not perform the infrared
subtraction.  In this way, any infrared subtraction scheme can be applied and it is
the task of the user to implement it.  The code is available at
\begin{verbatim}
https://gitlab.com/formfactors3l/ff3l
\end{verbatim}
where a documentation and sample programs can be found.  The code provides
interpolation grids and series expansion which can be used for instance in a
Monte Carlo program.  For the non-singlet contributions interpolation grids
are used in the ranges
\begin{itemize}
    \item $-40 < s/m^2 < 3.75$,
    \item $4.25 < s/m^2 < 16$,
    \item $16< s/m^2 < 60$.
\end{itemize}
In the remaining regions we implemented the series expansion around
$s=\pm\infty$ and $s=4m^2$.  We do not implement the expansion around
$s=16m^2$ since at this point the form factors are continuous functions (but
not holomorphic). For the massive singlet contributions
interpolation grids are used for
\begin{itemize}
    \item $-40 < s/m^2 < -1$,
    \item $1< s/m^2 < 3.75$,
    \item $4.25 < s/m^2 < 16$,
    \item $16< s/m^2 < 60$,
\end{itemize}
and for the massless singlet contributions interpolation grids are used in the ranges
\begin{itemize}
  \item $-40 < s/m^2 < -0.125$,
  \item $0.125< s/m^2 < 3.75$,
    \item $4.25 < s/m^2 < 16$,
    \item $16< s/m^2 < 60$.
\end{itemize}
In the remaining regions we implemented the series expansion around
$s=\pm\infty$, $s=0$, and $s=4m^2$.

A copy of \texttt{FF3l} can be obtained with
\begin{verbatim}
$ git clone https://gitlab.com/formfactors3l/ff3l.git
\end{verbatim}
A Fortran compiler such as \verb|gfortran| is needed. The
library can be compiled by running
\begin{verbatim}
$ ./configure
make
\end{verbatim}
The command \verb|make| will generate the static library \verb|libff3l.a|
which can be linked to the user’s program.  The module files
are located in the directory \texttt{modules} which must
be also passed to the compiler.  This gives access to the public functions and
subroutines. The names of all subroutines start with the suffix \verb|ff3l_|.

It is instructive to look at a program that uses \texttt{FF3l}.  We evaluate
the vector form factor $F_1^{v,(3)} (s)$ at $s/m^2 = 10$ at order
$\epsilon=-3, \dots, 0$ in the $\epsilon$ expansion.  The fortran program
looks as follows:
\begin{verbatim}
program example1
  use ff3l
  implicit none

  double complex :: f1v
  double precision :: s = 10
  integer :: eporder

  do eporder = -3,0
    f1v = ff3l_veF1(s,eporder)
    print *,"F1( s = ",s,", ep = ",eporder," ) = ", f1v
  enddo
end program example1
\end{verbatim}
In the preamble of the program, \texttt{use ff3l} loads the respective module.
The function \verb|ff3l_veF1| returns the sum of non-singlet, massive, and
massless singlet contributions to the ultraviolet renormalized vector form factor $F_1^v$
at three loops and receives two input parameters:
\begin{verbatim}
    double precision :: s
    integer :: eporder
\end{verbatim}
The variable \verb|s|$=s/m^2$ is the squared momentum transferred normalized w.r.t.\ the quark mass.
The order in the $\epsilon$ is set by the integer \verb|eporder|.
Only the values \verb|eporder|$=-3,-2,-1,0$ are valid.
The returned values is a \verb|double complex|, corresponding to the form factor value at third order as an expansion in $\alpha_s^{(n_l+n_h)}(m)$.
We assume that the strong coupling constant is renormalized in the $\overline{\mathrm{MS}}$ scheme with the renormalization scale $\mu=m$.
The choice whether to use interpolation grids or series expansion is handled internally.

The other types of form factors can be evaluated in a similar way with the functions \verb|ff3l_type|
where \verb|type| can be \verb|veF1, veF2, axF1, axF2, scF1, psF1|.
These six routines are implemented for the QCD group SU$(3)$.
We implemented also the abelian form factors. The corresponding functions come with the suffix \verb|_qed|, e.g.\ \verb|ff3l_veF1_qed(s,eporder)|.

By default, if not set explicitly, the library assumes the number of massive
and massless quarks to be $n_l = 4$
and $n_h = 1$, respectively.  However the user can
chose other values, for instance $n_l=3$ and $n_h=1$, in the following
way:
\begin{verbatim}
call ff3l_set_nl(3)
call ff3l_set_nh(1)
\end{verbatim}
Also by default all contributions from non-singlet and singlet diagrams are included.
They can be turned off with
\begin{verbatim}
call ff3l_nonsinglet_off()
call ff3l_nhsinglet_off()
call ff3l_nlsinglet_off()
\end{verbatim}
and turned on with
\begin{verbatim}
call ff3l_nonsinglet_on()
call ff3l_nhsinglet_on()
call ff3l_nlsinglet_on()
\end{verbatim}
In that case the output is the sum of the non-singlet and massive and massless
singlet contributions. In case a different linear combination is needed (see,
e.g., Eq.~(\ref{eq::FaNS})), the individual contributions have to be computed
individually using \texttt{FF3l} and the combination has to be done
afterwards.

It is useful to interface the library to Mathematica for simple and fast numerical evaluation and cross checks.
To this end, we provide also a Mathematica interface by making use of Wolfram’s MathLink interface (for details on the set up see Ref.~\cite{Hahn:2011gf}).
The interface is complied with
\begin{verbatim}
$ make mathlink
\end{verbatim}
To use the library within Mathematica, the interface must be loaded
\begin{verbatim}
In[] := Install["PATH/ff3l"]
\end{verbatim}
where \verb|PATH| is the directory where the mathlink executable \verb|ff3l| is saved.
Form factors in QCD are evaluated with one of the following: \verb|FF3lveF1|, \verb|FF3lveF2|, \verb|FF3laxF1|, \verb|FF3laxF2|, \verb|FF3lscF1|, \verb|FF3lpsF1|.
For example, the $\epsilon^0$ term of the vector form factor $F_1$ at third order in $\alpha_s$ is evaluated in the following way:
\begin{verbatim}
In[] := s = 10;
In[] := eporder = 0;
In[] := FF3lveF1[s,eporder]
Out[]:= 60.1219 - 172.027 I
\end{verbatim}
The number of massless and massive quarks can be set with \verb|FF3lSetNl| and \verb|FF3lSetNh|. The contribution from non-singlet, $n_l$- and $n_h$-singlet diagrams can be switched on and off with the following commands:
\begin{verbatim}
In[] := FF3lNonSingletOff[]
In[] := FF3lNonSingletOn[]
In[] := FF3lNhSingletOff[]
In[] := FF3lNhSingletOn[]
In[] := FF3lNlSingletOff[]
In[] := FF3lNlSingletOn[]
\end{verbatim}

The standalone \texttt{Mathematica} package \texttt{formfactors3l},
which evalutes the bare and finite form factors, can be found in Ref.~\cite{gitmath}.


\end{appendix}


\end{document}